\documentclass[preprint,review,12pt]{elsarticle}

\usepackage{algorithm}
\usepackage{algorithmicx}
\usepackage{algpseudocode}
\usepackage{amsmath}
\usepackage{amssymb}

\usepackage{color}
\usepackage{cancel} 

\usepackage{dsfont}

\usepackage{graphics}
\usepackage{graphicx}

\usepackage{hyperref}

\usepackage{overpic}

\usepackage{subfigure}

\usepackage{ulem}

\newtheorem{proposition}{Proposition}

\newtheorem{corollary}{Corollary}

\makeatletter
\newenvironment{breakablealgorithm}
  {
   \begin{center}
     \refstepcounter{algorithm}
     \hrule height.8pt depth0pt \kern2pt
     \renewcommand{\caption}[2][\relax]{
       {\raggedright\textbf{\ALG@name~\thealgorithm} ##2\par}%
       \ifx\relax##1\relax 
         \addcontentsline{loa}{algorithm}{\protect\numberline{\thealgorithm}##2}%
       \else 
         \addcontentsline{loa}{algorithm}{\protect\numberline{\thealgorithm}##1}%
       \fi
       \kern2pt\hrule\kern2pt
     }
  }{
     \kern2pt\hrule\relax
   \end{center}
  }
\makeatother

\journal{ArXiv}

\begin{document}

\begin{frontmatter}



\title{Riemannian information gradient methods for the parameter estimation of ECD\,: Some applications in image processing}


\author[label]{Jialun~Zhou,
        Salem~Said,
        Yannick Berthoumieu}

\address[label]{ University Bordeaux, CNRS, IMS, UMR 5218, Groupe Signal et Image, F-33405 Talence, France }

\begin{abstract}
Elliptically-contoured distributions (ECD) play a significant role, in computer vision, image processing, radar, and biomedical signal processing. Maximum likelihood estimation (MLE) of ECD leads to a system of non-linear equations, most-often addressed using fixed-point (FP) methods. Unfortunately, the computation time required for these methods is unacceptably long, for large-scale or high-dimensional datasets. To overcome this difficulty, the present work introduces a Riemannian optimisation method, the information stochastic gradient (ISG). The ISG is an online (recursive) method, which achieves the same performance as MLE, for large-scale datasets, while requiring modest memory and time resources.  To develop the ISG method, the Riemannian information gradient is derived taking into account the product manifold associated to the underlying parameter space of the ECD.  From this information gradient definition, we define also, the information deterministic gradient (IDG), an offline (batch) method, which is an alternative, for moderate-sized datasets. The present work formulates these two methods, and demonstrates their performance through numerical simulations. Two applications, to image re-colorization, and to texture classification, are also worked out. 
\end{abstract}



\begin{keyword}
elliptically-contoured distribution \sep Riemannian information gradient \sep large-scale dataset \sep image re-colorization \sep texture classification.
\end{keyword}
\end{frontmatter}


\section{Introduction}\label{intro}
The family of Elliptically-contoured distributions (ECD) was originally introduced in~\cite{kelker1970distribution}, and investigated in~\cite{fang1990generalized,fang2018symmetric}. It contains many widely-used statistical distributions, such as elliptical Gamma, Pearson type II, and elliptical multivariate logistic distributions. 
In terms of applications, the most popular classes of ECD are multivariate generalized Gaussian distributions (MGGD), and multivariate Student-T distributions~\cite{gomez1998multivariate,sanchez2002matrix,kotz2004multivariate}. These are location-scale distributions, and are further parameterised by a shape parameter, or a degrees of freedom parameter.

MGGD are used in image processing, as models for wavelet and curvelet coefficients, and as models for three-channel color vectors, in image denoising, context-based image retrieval, image thresholding, texture classification, and image quality assessment  ~\cite{boubchir2005multivariate,cho2009image,verdoolaege2008multiscale,bazi2007image,scharcanski2006wavelet,GUPTA201887}. MGGD are also used in video coding and denoising, radar signal processing, and biomedical signal processing~\cite{desai2003robust,elguebaly2010bayesian}. 

Some applications of Student-T distributions are presented in \cite{laus2019multivariate}, involving image denoising. In radar imaging, the Student-T distribution, so-called $\mathcal{G}^0$ model within  the family of spherically invariant random vectors (SIRVs), is largely exploited in the context of SAR or PolSAR imaging, for tasks such as despeckling, classification, segmentation or detection \cite{5570982,7887730,8930947}.

Because ECD have been successful in real-world image and signal processing applications, much attention has been devoted to developing efficient methods for estimating their parameters. 
The vast majority of works, dedicated to this estimation problem, are  focused on estimating the scatter matrix, considering the other parameters, i.e. location and shape parameters, as known. 

In terms of maximum-likelihood estimation, two main classes of algorithms have been studied. Fixed-point (FP) algorithms, and gradient descent algorithms have been proposed, based on the geometric properties of the manifold of positive definite matrices~\cite{tyler1987,6263313,sra2013geometric,zhang2013multivariate,sra2015conic}. For MGGD, when the location parameter is equal to zero and the shape parameter is given, the uniqueness of the maximum-likelihood estimator has been shown, under a restriction on the value of the shape parameter \cite{pascal2013parameter}. In this case, a method of moments has also been developed \cite{verdoolaege2011geodesics}. For Student-T distributions, with a known degrees of freedom parameter, a fixed-point method for parameter estimation is given in~\cite{laus2019multivariate}, where the existence and uniqueness, of location and scatter maximum-likelihood estimates, is shown for a fixed degrees of freedom parameter, superior to 1.

There is a shared drawback, in all of the maximum-likelihood estimation methods, just mentioned~\cite{sra2013geometric,zhang2013multivariate,laus2019multivariate,verdoolaege2011geodesics,pascal2013parameter}. Specifically, these methods work well for datasets of moderate size and dimension, but require excessive resources in memory and time, for large-scale datasets, e.g. with a few millions of samples, an order of magnitude commonly encountered in optical or SAR image processing. This issue can be so severe as to make any of these methods inapplicable.

Mainly, this is due to the fact that all of these methods are off-line, or batch, estimation methods. They require access to the whole dataset, at once, for each iteration, and therefore consume increasing time and memory resources, in order to converge to a useful estimate, as the dataset grows large. 

In order to overcome this drawback, the present work builds on the ideas from Riemannian stochastic optimisation, proposed in~\cite{bonnabel2013stochastic,tripuraneni2018averaging}. The problem of estimating the parameters of an ECD is viewed as the problem of minimising the Kullback-Leibler divergence, between the true (unknown) and estimate distributions. When this problem is addressed using Riemannian stochastic optimisation, each iteration of a stochastic optimisation method requires access to only one datapoint (one sample), instead of the whole dataset. In this way, the present work proposes a recursive method, for estimating the parameters of ECD (each time a new sample is processed, this sample is used to update the current estimate).  

The proposed method will be called the information stochastic gradient (ISG). In its simplest form, it is an improvement of a previous method, used to estimate the scatter matrix, when the location and shape parameters are known~\cite{zhou2019fast}. In this paper, we consider also the general case where the scatter matrix, the location and shape parameters are unknown. The ISG method relies on two main ideas\,:
\begin{itemize}
  \item The greatest difficulty, in using recursive methods, is that they may require a careful choice of step-sizes. The standard Riemannian stochastic gradient method (as in~\cite{bonnabel2013stochastic}), is very sensitive to the choice of step-sizes. However, using the information gradient (also called the natural gradient~\cite{amari2016information,amari1998natural}) leads to an automatic choice of step-sizes, which guarantees optimal performance. The ISG method implements the information gradient, relying on the Fisher information metric (or matrix), of the ECD model.
\item The parameter space of an ECD model does not only contain the scatter matrix, but also location and shape parameters. In the case of MGGD or Student-T models, this parameter space is a product space, made up of triplets: (scatter matrix, location parameter, shape/degrees of freedom parameter). Since the geodesic curves of this space do not have a tractable expression~\cite{verdoolaege2012geometry}, an intuitive idea is to update each one of the three parameters, in its own turn, in an alternating fashion.
\end{itemize}
To understand the benefit of combining these two ideas, consider the special case of MGGD. In this case, a method of moments (MM) was used for the joint estimation of all three parameters~\cite{verdoolaege2011geodesics}, while their maximum-likelihood esitmation (MLE) was studied in \cite{pascal2013parameter}. It is well known that MLE performs better than MM in most scenarios~\cite{fuhrer1995estimating}. However, as mentioned above, MLE cannot be applied to large scale datasets, due to its computational requirements. The ISG method strikes a balance between the low complexity of MM, and the stronger performance of MLE. For example, in the case where the scatter matrix and the location parameter are unknown, the complexity of ISG is comparable to that of the MM, while its performance is similar to that of the MLE, when the number of available samples is sufficiently large. In other words, the size of the dataset is leveraged as a source of information, rather than as a computational burden. 

The two ideas which underly the ISG method (discussed above), are also implemented  in an offline (batch) method, called the information deterministic gradient (IDG) method. While its complexity (and therefore computation time) is much higher than the ISG method, the IDG method consistently outperforms other methods, even in the case where all three parameters are unknown. 

The main results of the present work are given in Section \ref{sec:mainresults}. A detailed comparison of various estimation methods (MM, FP, ISG, IDG), based on computer experiments, is carried out in Section \ref{sec:Computer_Experiments}. Two image processing applications with real datasets are presented in Section \ref{sec:App_Real}.

First, Sections \ref{sec:Formulation_of_Pb} and \ref{sec:geo_concepts} define the general estimation problem for ECD models, and introduce some necessary geometric concepts.

\section{The estimation problem}\label{sec:Formulation_of_Pb}
\subsection{The ECD family}
ECD is a general family of probability distributions that contains many important sub-families. The name ECD comes from the fact that when an ECD has a probability density function, the contours (level surfaces) of this function are ellipsoids.


The location, or expectation, parameter $\mu$ of an ECD determines the centre of these ellipsoids, while the axes of these ellipsoids are proportional to the eigenvalues of the inverse of the scatter matrix $\Sigma$. The shape parameter $\beta$ determines the factor for this proportionality ($\beta$ is the degrees of freedom parameter, for Student-T distributions).

Let $X$ be a $m$-dimensional random vector that follows a ECD model. Denote $\theta=(\mu,\Sigma,\beta)$ the parameters of this ECD, and $\Theta = \mathbb{R}^m \times \mathcal{P}_m \times \mathbb{R}_+$ its parametric space, where $\mathcal{P}_m$ is the set of all symmetric positive definite matrices of size $m \times m$. If $X$ has a probability density function, then this takes on the following form
\begin{equation}\label{eq:density}
p(x;\theta) = c(\beta) \, |\Sigma|^{-1/2} \, g \left[ \delta_x(\mu,\Sigma), \beta \right]
\end{equation}
where $c(\beta)$ is a normalizing factor which depends only on $\beta$, and $\delta_x(\mu,\Sigma) = (x-\mu)^{\dagger}\Sigma^{-1}(x-\mu)$. The density generator $g$ depends on the specific sub-family of ECD distributions, for example
\begin{equation*}
g \left[ \delta_x(\mu,\Sigma),\beta \right] = \exp \left( -\frac{1}{2} \delta_x^{\beta} \right) \text{ for MGGD}
\end{equation*}
\begin{equation*}
g \left[ \delta_x(\mu,\Sigma),\beta \right] = \left( 1  + \frac{1}{\beta} \delta_x \right)^{-\frac{\beta+m}{2}} \text{ for Student-T}
\end{equation*}
\subsection{Problem formulation}
Parameter estimation will be formulated as the problem of minimising the Kullback-Leibler divergence $D(\theta^*||\theta)$, denoted $D(\theta)$ for short. That is to say, the estimator $\hat{\theta}$ is sought which is the solution of the following minimisation problem
\begin{equation}\label{eq:pb}
\hat{\theta} = \mathop{\arg\min}_{\theta \in \Theta} \, D(\theta)
\end{equation}
Recall the definition of the KL divergence
\begin{equation}\label{eq:KL_D}
\begin{aligned}
D(\theta) & = \int_{\mathbb{R}^p} p(x;\theta^*) \ln \left( \frac{p(x;\theta^*)}{p(x;\theta)} \right) \mathrm{d}x \\
& = \mathbb{E}_{\theta^*} \left[ \ell(\theta^*;x) \right] - \mathbb{E}_{\theta^*} \left[ \ell(\theta;x) \right]
\end{aligned}
\end{equation}
Where $\ell(\theta;x) = \log p(x;\theta)$ is the log-likelihood, 
\begin{equation}\label{eq:ell}
\ell(\theta;x) = \alpha( \beta ) - \frac{1}{2}\log\det(\Sigma) + h(\delta_x,\beta) 
\end{equation}
with $\alpha(\beta) = \log c(\beta)$ and $h = \log g$. In the following, the KL divergence (\ref{eq:KL_D}) will be minimised using Riemannian information gradient descent. Some Riemannian Information geometry concepts are recalled in the following section.

\section{Necessary geometric concepts}\label{sec:geo_concepts}
The gradient descent method on Riemannian manifolds is based on the following update rule~\cite{absil2009optimization}
\begin{equation}\label{eq:general_Retraction}
\theta_{n+1} = R_{\theta_n} ( \alpha_{n+1} u(\theta_n) )
\end{equation}
Here, the smooth mapping $R_{\theta}$  from the tangent space $T_{\theta}\Theta$ to $\Theta$ is required to be a retraction, in the sense that it verifies
\begin{subequations}\label{eq:R_prop}
\begin{equation}\label{eq:R_prop1}
R_{\theta} (0_{\theta}) = \theta
\end{equation}
\begin{equation}\label{eq:R_prop2}
\mathrm{D} R_{\theta}(0_{\theta}) = \mathrm{Id}_{T_{\theta}\Theta}
\end{equation}
where $0_{\theta}$ denotes the zero element in $T_{\theta} \Theta$, and $\mathrm{Id}_{T_{\theta}\Theta}$ denotes the identity mapping on $T_{\theta}\Theta$.
\end{subequations}
Each vector $u(\theta_n)$ belongs to the tangent space $T_{\theta_n} \Theta$, and provides the direction of descent. In the present work, $-u(\theta_n)$ is the Riemannian information gradient, derived using the Fisher information metric. The positive scalar $\alpha_{n}$ is the step-size. The aim of equation (\ref{eq:general_Retraction}) is to generate a sequence $(\theta_n)_{n\geq 0} \in \Theta$ that converges to a stationary point $\theta^*$ of the cost function (under some restrictions on $u$ and $\alpha$).

For our estimation problem, the model has three different parameters, which belong to three different Riemannian manifolds. Precisely, the parameter space is the product manifold $\Theta = \mathbb{R}^m \times \mathcal{P}_m \times \mathbb{R}_+$. Therefore, tractable expressions of the Fisher information metric, and of the intrinsic geodesic map, on this product manifold, are needed. However, $\Theta$ does not support any such expressions~\cite{verdoolaege2012geometry}. As the global Fisher information metric has not a closed form, we propose to use the product metric 
\begin{equation}\label{eq:product_metric}
\left< u,v \right>_{\theta} = \left< u_{\mu},v_{\mu} \right>_{\mu} + \left< u_{\scriptscriptstyle \Sigma},v_{\scriptscriptstyle \Sigma} \right>_{\scriptscriptstyle \Sigma} + \left< u_{\beta},v_{\beta} \right>_{\beta}
\end{equation}
where $u = (u_{\mu},u_{\scriptscriptstyle \Sigma},u_{\beta})$ and $v = (v_{\mu},v_{\scriptscriptstyle \Sigma},v_{\beta})$ are tangent vectors at the point $\theta = (\mu,\Sigma,\beta)$. The metrics $\left<\cdot,\cdot\right>_{\scriptscriptstyle \mu}$, $\left<\cdot,\cdot\right>_{\scriptscriptstyle \Sigma}$ and $\left<\cdot,\cdot\right>_{\scriptscriptstyle \beta}$ are respectively the intrinsic Fisher information metrics of their corresponding sub-spaces. 
For the location parameter $\mu$ in $\mathbb{R}^m$, its information metric is expressed in terms of the usual Euclidean metric, 
\begin{equation}
\left< u_{\mu},v_{\mu} \right>_{\mu} = I_{\mu} \,  u_{\mu}^{\dagger} \, \Sigma^{-1} \, v_{\mu}
\end{equation}
where $\left<\cdot,\cdot\right>$ denotes the scalar product in Euclidean space. The information constant $I_{\mu}$ is
\begin{equation}\label{eq:I_mu}
I_{\mu} = -\frac{4}{m} \mathbb{E} \left[ \frac{\partial^2 h(\delta_x,\beta)}{\partial \delta_x^2} \, \delta_x \right] - 2 \mathbb{E} \left[ \frac{\partial h(\delta_x,\beta)}{\partial \delta_x} \right]
\end{equation}
with $h = \log g$. As for $\Sigma \in \mathcal{P}_m$, the Fisher information metric $\left<\cdot,\cdot\right>_{\Sigma}$ for the ECD model is defined by the Riemannian geometry of $\mathcal{P}_m$~\cite{berkane1997geodesic}. 
\begin{equation}\label{eq:info_metric_Sigma}
\left< U_{\scriptscriptstyle \Sigma},V_{\scriptscriptstyle \Sigma} \right>_{\scriptscriptstyle \Sigma} = I_1 \mathrm{tr}( \Sigma^{-1} U_{\scriptscriptstyle \Sigma} \Sigma^{-1} V_{\scriptscriptstyle \Sigma}) + I_2 \mathrm{tr}(\Sigma^{-1} U_{\scriptscriptstyle \Sigma} ) \mathrm{tr}(\Sigma^{-1} V_{\scriptscriptstyle \Sigma})
\end{equation}
Here the constants $I_1>0$ and $I_2 \geqslant 0$ depend on the particular model under consideration, as follows
\begin{equation}\label{eq:I1_I2}
\begin{aligned}
I_1 = \frac{2\mathcal{A}}{m (m+2)} \qquad I_2 = \frac{\mathcal{A}}{m(m+2)} - \frac{1}{4}\\
\mathcal{A} = \mathbb{E} \left[ \left( \frac{\partial h(\delta_x,\beta)}{\partial \delta_x} \delta_x \right)^2 \right] \hspace{1.15cm}
\end{aligned}
\end{equation}
The shape factor $\beta$ belongs to $\mathbb{R}_+$, so the Fisher information metric is given by
\begin{equation}
\left< u_{\beta},v_{\beta} \right>_{\beta} = I_{\beta} u_{\beta} v_{\beta}
\end{equation}   
with the information constant
\begin{equation}\label{eq:I_beta}
I_{\beta} = -\mathbb{E} \left[ \frac{\partial^2 \alpha(\beta)}{\partial \beta^2} + \frac{\partial^2 h(\delta_x,\beta)}{\partial \beta^2} \right]
\end{equation}

\begin{figure}[!htbp]
\centering
\includegraphics[scale=0.3]{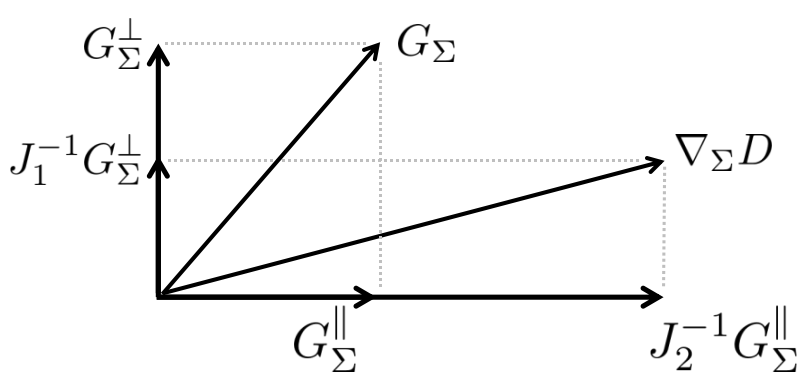}
\label{fig:fimsigma}
\end{figure}

Now, the information gradient $\nabla_{\theta} D(\theta)$ with respect to the product metric (\ref{eq:product_metric}) is obtained by solving the following equation,
\begin{equation}\label{eq:diff_eq}
\mathrm{d}\,D(\theta)\,v\,=\,\left<\nabla_{\theta} D(\theta), v \right>_{\theta}
\end{equation}
where the scalar product on the right-hand side is given by (\ref{eq:product_metric}), and $\mathrm{d}$ is the differential form of $D$. Precisely, this product information gradient has the following form
\begin{equation}\label{eq:igrad_theta}
\nabla_{\theta} D(\theta) = \left( \nabla_{\mu} D(\theta), \nabla_{\Sigma} D(\theta), \nabla_{\beta} D(\theta) \right)
\end{equation}
The first component $\nabla_{\mu} D(\theta)$ is expressed as
\begin{equation}\label{eq:igrad_mu}
\nabla_{\mu} D(\theta) = - I_{\mu}^{-1} \ \Sigma \ \mathbb{E} \left[ \mathrm{G}_{\mu}(\theta;x) \right]
\end{equation}
where $I_{\mu}$ is given in equation (\ref{eq:I_mu}), and vector $G_{\mu}(\theta;x)$ is actually the gradient in the classic Euclidean sense
\begin{equation}\label{eq:g_mu}
G_{\mu}(\theta;x) = 2 \frac{\partial h(\delta_x,\beta)}{\partial \delta_x}   \Sigma^{-1} (x-\mu)
\end{equation}
The second component $\nabla_{\Sigma} D(\theta)$ is a bit more complicated
(see Figure \ref{fig:fimsigma}, for an illustration of the following computations)
\begin{equation}\label{eq:igrad_Sigma}
\begin{aligned}
\nabla_{\Sigma} D(\theta) = -J_1^{-1} \, \mathbb{E} \left\{ \left[ \mathrm{G}_{\Sigma} (\Sigma;x) \right]^{\bot} \right\} - J_2^{-1} \, \mathbb{E} \left\{ \left[ \mathrm{G}_{\Sigma} (\theta;x) \right]^{\parallel} \right\}
\end{aligned}
\end{equation}
where $J_1 = I_1$ and $J_2 = I_1 + m I_2$, in terms of $I_1$ and $I_2$ given in (\ref{eq:I1_I2}), and where $\bot$ and $\parallel$ denote the following decomposition of $\mathrm{G}_{\Sigma} (\theta;x)$, 
\begin{equation}
\left[ \mathrm{G}_{\Sigma} (\theta;x) \right]^{\parallel} = \frac{1}{m} \mathrm{tr} \left[ \Sigma^{-1} \mathrm{G}_{\Sigma}(\theta;x) \right] \Sigma
\end{equation}
\begin{equation}
\left[ \mathrm{G}_{\Sigma}(\theta;x) \right]^{\bot} = \mathrm{G}_{\Sigma} (\theta;x) - \left[ \mathrm{G}_{\Sigma} (\theta;x) \right]^{\parallel}
\end{equation}
in terms of
\begin{equation}\label{eq:g_Sigma}
G_{\Sigma}(\theta;x) = -\frac{1}{2}\Sigma - \frac{\partial h(\delta_x,\beta)}{\partial \delta_x} S_x \ \text{ with } \ S_x = (x-\mu)(x-\mu)^{\dagger}
\end{equation}
Finally, for the third component, 
\begin{equation}\label{eq:igrad_beta}
\nabla_{\beta} D(\theta) = -I_{\beta}^{-1} \mathbb{E} \left[ \mathrm{G}_{\beta} (\beta;x) \right]
\end{equation}
where $I_{\beta}$ was given in (\ref{eq:I_beta}), and
\begin{equation}\label{eq:g_beta}
\mathrm{G}_{\beta} (\theta;x) = \frac{\partial \alpha(\beta)}{\partial \beta} - \frac{\partial h(\delta_x,\beta)}{\partial \beta}
\end{equation}
With regard to the retraction $R_{\theta}$, it will be defined as the product Riemannian exponential map, 
\begin{equation}\label{eq:prod_retraction}
\begin{array}{c c c c}
R_{\theta} : & T_{\theta} \Theta & \longrightarrow & \Theta \\[0.1cm]
& u = \left( \begin{matrix} u_{\mu}\\ u_{\Sigma}\\ u_{\beta} \end{matrix} \right) & \longmapsto & \left( \begin{matrix} \mathrm{Exp}_{\mu} ( u_{\mu}) \\ \mathrm{Exp}_{\Sigma} ( u_{\Sigma} ) \\ \mathrm{Exp}_{\beta} ( u_{\beta} ) \end{matrix} \right)
\end{array}
\end{equation}
where $u$ is the direction of descent. The exponential map on $\mathbb{R}^p$ (a Euclidean space) reduces to vector addition
\begin{equation}\label{eq:geod_mu}
\mathrm{Exp}_{\mu} (  u_{\mu}) = \mu + u_{\mu}
\end{equation}
The exponential map on $\mathcal{P}_m$ is defined as follows~\cite{pennec2006riemannian}:
\begin{equation}\label{eq:geod_Sigma}
\mathrm{Exp}_{\Sigma} ( u_{\scriptscriptstyle \Sigma}) = \Sigma \exp \left( \Sigma^{-1} u_{\scriptscriptstyle \Sigma} \right)
\end{equation}
As for $\beta$, since it belongs to $\mathbb{R}_+$, the corresponding exponential map is a $1-$dimensional version of (\ref{eq:geod_Sigma})
\begin{equation}\label{eq:geod_beta}
\mathrm{Exp}_{\beta}(v_{\scriptscriptstyle \beta}) = \beta \exp( \beta^{-1} v_{\scriptscriptstyle \beta} )
\end{equation}
All these three exponential map $\mathrm{Exp}$ verify the properties (\ref{eq:R_prop}), therefore the their direct product (\ref{eq:prod_retraction}) also verifies these properties, and is a well-defined retraction. 
Finally, the Riemannian distance associated to the metric (\ref{eq:product_metric}) is given by
\begin{equation}\label{eq:inf_d_prod}
\mathrm{d}^2 (\theta_1,\theta_2) = \mathrm{d}^2_{\mathbb{R}^m} (\mu_1,\mu_2) + \mathrm{d}^2_{\mathcal{P}_m} (\Sigma_1,\Sigma_2) + \mathrm{d}^2_{\mathbb{R}_+} (\beta_1,\beta_2)
\end{equation}
For $\mu$, the information distance is proportional to the Euclidean distance in $\mathbb{R}^p$
\begin{subequations}\label{eq:inf_d}
\begin{equation}\label{eq:inf_d_mu}
\mathrm{d}^2_{\mathbb{R}^m}(\mu_1,\mu_2) = I_{\mu} \ (\mu_1-\mu_2)^{\dagger} (\mu_1-\mu_2)
\end{equation}
for $\mu_1,\mu_2 \in \mathbb{R}^m$, with the constant $I_{\mu}$ given by (\ref{eq:I_mu}). For $\Sigma$, the information distance is defined as in~ \cite{cyrus}
\begin{equation}\label{eq:inf_d_Sigma}
\mathrm{d}^2_{\mathcal{P}_m} (\Sigma_1,\Sigma_2) = I_1 \ \mathrm{tr} \left[ \log (\Sigma_1^{-1} \ \Sigma_2 ) \right]^2 + I_2 \ \mathrm{tr}^2 \left[ \log ( \Sigma_1^{-1} \Sigma_2 ) \right]
\end{equation}
for $\Sigma_1, \Sigma_2 \in \mathcal{P}_m$, where the constants $I_1$ and $I_2$ are given by (\ref{eq:I1_I2}), and the function $\log$ denotes the symmetric matrix logarithm. Finally, for $\beta$, the information distance is given by
\begin{equation}\label{eq:inf_d_beta}
\mathrm{d}^2_{\mathbb{R}_+} (\beta_1, \beta_2) = I_{\beta} \ \log^2 \left( \beta_1^{-1} \ \beta_2 \right)
\end{equation}
for $\beta_1, \beta_2 \in \mathbb{R}_+$, where $I_{\beta}$ is given in (\ref{eq:I_beta}).
\end{subequations}
With the necessary geometric concepts now in place, the next section will introduce our estimation algorithms.

\section{The IDG and ISG methods} \label{sec:mainresults}
This section will describe the IDG and ISG methods, and discuss their main properties.  The IDG method (information deterministic gradient) is a deterministic gradient method, and the ISG method (information stochastic gradient) is a stochastic gradient method.

When the direction of descent is chosen according to (\ref{eq:igrad_theta}), the updated estimates $\theta_{k+1}=(\mu_{k+1},\Sigma_{k+1},\beta_{k+1})$ rely on the current estimates $\theta_k=(\mu_{k},\Sigma_{k},\beta_{k})$, through the following alternating optimisation scheme
\begin{equation}\label{eq:alternate_update}
\begin{aligned}
\text{ step 1 : } & \mu_{k+1} \leftarrow (\mu_{k},\Sigma_{k},\beta_{k}) \\
\text{ step 2 : } & \Sigma_{k+1} \leftarrow (\mu_{k+1},\Sigma_{k},\beta_{k}) \\
\text{ step 3 : } & \beta_{k+1} \leftarrow (\mu_{k+1},\Sigma_{k+1},\beta_{k}) \\
\end{aligned}
\end{equation}
 
\subsection{Deterministic gradient}
The IDG method is a second-order offline method, somewhat similar to a Newton method. In the Newton method, the direction of descent is found by solving the Newton equation~\cite{absil2009optimization}. In the IDG method, the Hessian in the Newton equation is approximated by the Fisher information metric (or matrix) $\mathcal{I}(\theta)$.

Since IDG is an offline method, it choses a direction of descent which depends on the complete dataset. The cost function (\ref{eq:pb}) is reformulated, by replacing the KL divergence, with the empirical average of $- \ell(\theta,x_n)$. This empirical average is denoted by $\hat{D}(\theta)$,
\begin{equation}\label{eq:pb_idg}
\hat{D}(\theta) = -\frac{1}{N} \sum_{n=1}^N \ell(\theta_n;x_n)
\end{equation}
If the current estimate is $\theta_k$, the direction of descent is given by its three components
\begin{subequations}\label{eq:compts_igrad_theta_idg}
\begin{equation}\label{eq:igrad_mu_idg}
\nabla_{\mu} \hat{D}(\theta_k) = - I_{\mu}^{-1} \Sigma \frac{1}{N} \sum_{n=1}^N  \mathrm{G}_{\mu}(\theta_k;x_n) 
\end{equation}
\begin{equation}\label{eq:igrad_Sigma_idg}
\nabla_{\Sigma} \hat{D}(\theta_k) = - J_1^{-1} \, \frac{1}{N} \sum_{n=1}^N  \left[ \mathrm{G}_{\Sigma} (\theta_k;x_n) \right]^{\bot} - J_2^{-1} \, \frac{1}{N} \sum_{n=1}^N \left[ \mathrm{G}_{\Sigma} (\theta_k;x_n) \right]^{\parallel}
\end{equation}
\begin{equation}\label{eq:igrad_beta_idg}
\nabla_{\beta} \hat{D}(\theta_k) = -I_{\beta}^{-1} \frac{1}{N} \sum_{n=1}^N  \mathrm{G}_{\beta} (\theta_k;x_n)
\end{equation}
\end{subequations}
which are the same as (\ref{eq:igrad_mu}), (\ref{eq:igrad_Sigma}), (\ref{eq:igrad_beta}), but with expectations replaced by empirical averages.
Using the expressions (\ref{eq:igrad_mu_idg}), (\ref{eq:igrad_Sigma_idg}), (\ref{eq:igrad_beta_idg}), the IDG algorithm can now be stated as follows.
\begin{breakablealgorithm}
\caption{\label{algo:idg}Information Deterministic Gradient (IDG) algorithm}
\begin{algorithmic}[1]
\Require A dataset $\mathcal{X}=(x_1,\cdots,x_N)$, an initialization $\theta_0 \in S_0 \subset \Theta$;
\Ensure The estimate $\hat{\theta}$;
\For{ $k = 0,1,2 \cdots, K $ }
	\State $\mu_{\scriptscriptstyle k+1} \gets \mu_{\scriptscriptstyle k}  - \alpha_{\scriptscriptstyle \mu} \nabla_{\scriptscriptstyle \mu} \hat{D}( \mu_{\scriptscriptstyle k}, \Sigma_{\scriptscriptstyle k}, \beta_{\scriptscriptstyle k} )$;
	\State $\Sigma_{\scriptscriptstyle k+1} \gets \Sigma_{\scriptscriptstyle k} \exp \left( - \Sigma_{\scriptscriptstyle k}^{\scriptscriptstyle -1} \alpha_{\scriptscriptstyle \Sigma} \nabla_{\Sigma} \hat{D}( \mu_{\scriptscriptstyle k+1}, \Sigma_{\scriptscriptstyle k}, \beta_{\scriptscriptstyle k} ) \right)$;
	\State $\beta_{\scriptscriptstyle k+1} \gets \beta_{\scriptscriptstyle k} \exp \left( - \beta_{\scriptscriptstyle k}^{\scriptscriptstyle -1} \alpha_{\scriptscriptstyle \beta} \nabla_{\beta} \hat{D}( \mu_{\scriptscriptstyle k+1}, \Sigma_{\scriptscriptstyle k+1}, \beta_{\scriptscriptstyle k} ) \right)$;
\EndFor
\State $\hat{\theta} \gets (\mu_{\scriptscriptstyle K+1},\Sigma_{\scriptscriptstyle K+1},\beta_{\scriptscriptstyle K+1})$
\end{algorithmic}
\end{breakablealgorithm}
In this algorithm, $\alpha$ denotes the step-size, which is selected according to the Armijo-Goldstein rule, and $S_0$ denotes a neighborhood of the true parameter value $\theta^*$. The following Proposition \ref{prop:conv_idg} states the convergence of Algorithm \ref{algo:idg}.
\begin{proposition}\label{prop:conv_idg}
Assume the cost function (\ref{eq:pb_idg}) has an isolated stationary point $\theta=\theta^*$ in some neighborhood $S_0 \subset \Theta$, and that the estimates $(\theta_k)_{k\geq 0}$ remain within $S_0$. Then, for the sequence $(\theta_k)_{k\geq 0}$ generated by Algorithm \ref{algo:idg},
\begin{equation*}
\lim_{k \rightarrow \infty} \theta_k = \theta^*
\end{equation*}
\end{proposition}
\ref{prove:conv_idg} sketches a proof of this convergence. For the case $\theta=(\Sigma)$ or $\theta=(\mu,\Sigma)$, near the true value $\theta^*$, the Hessian of the function $\hat{D}(\theta)$ is approximated by the Fisher information metric. Therefore, one should expect the $\theta_k$ converge to $\theta^*$ with a superlinear rate of convergence, just like the Newton method dose. Precisely, if $\theta=(\Sigma)$ or $\theta=(\mu,\Sigma)$, with a fixed shape parameter $\beta^*$, then, under the assumptions of Proposition \ref{prop:conv_idg}, one should expect Algorithm \ref{algo:idg} to generate a sequence $(\theta_k)_{k\geq 0}$ converging superlinearly to $\theta^*$.
This is essentially due to Theorem 6.3.2 in~\cite{absil2009optimization}, and will be observed experimentally in Section \ref{sec:Computer_Experiments} (see Figure \ref{fig:Super_rate}). 

\subsection{Stochastic gradient}
The ISG method is an online quasi-Newton method. For each update, only one sample or one mini-batch is used. Here, the cost function remains the same as in equation (\ref{eq:KL_D}). For the current estimate $\theta_{\scriptscriptstyle k} = (\mu_{\scriptscriptstyle k}, \Sigma_{\scriptscriptstyle k}, \beta_{\scriptscriptstyle k})$ the stochastic information gradients are
\begin{subequations}\label{eq:compts_igrad_theta_isg}
\begin{equation}\label{eq:igrad_mu_isg}
\nabla_{\mu} \ell(\theta;x_n) = I_{\mu}^{-1} \ \Sigma \  \mathrm{G}_{\mu}(x_n;\theta) 
\end{equation}
\begin{equation}\label{eq:igrad_Sigma_isg}
\begin{aligned}
\nabla_{\Sigma} \ell(\theta;x_n) = & J_1^{-1} \, \left[ \mathrm{G}_{\Sigma} (x_n;\theta) \right]^{\bot} \\
& \quad + J_2^{-1} \, \left[ \mathrm{G}_{\Sigma} (x_n;\theta) \right]^{\parallel}
\end{aligned}
\end{equation}
\begin{equation}\label{eq:igrad_beta_isg}
\nabla_{\beta} \ell(\theta;x_n) = I_{\beta}^{-1} \ \mathrm{G}_{\beta} (x_n;\theta)
\end{equation}
\end{subequations}
Accordingly, the expected direction of descent is $\mathbb{E}_{\theta^*} [ \nabla_{\theta} \ell(\theta;x) ]$, which is equal to $0$ at the global minimum $\theta^*$. As in the classic stochastic gradient descent method, the step-size $\alpha_n = \frac{a}{n}$ is strictly positive, decreasing, and verifies the usual conditions
\begin{equation*}
\sum \alpha_n = \infty \qquad \sum \alpha_n^2 < \infty
\end{equation*}
Using the expressions (\ref{eq:igrad_mu_isg}), (\ref{eq:igrad_Sigma_isg}), (\ref{eq:igrad_beta_isg}), the ISG algorithm can now be stated as follows. 
\begin{breakablealgorithm}
\caption{\label{algo:isg}ISG algorithm}
\begin{algorithmic}[1]
\Require A dataset $\mathcal{X}=(x_1,\cdots,x_N)$, an initialization $\theta_0 \in S_0 \subset \Theta$, the coefficient $a>0$;
\Ensure The estimate $\hat{\theta}$;
\For{ $n = 0,1,2 \cdots, N $ }
	\State $\alpha_{n+1} \gets \frac{a}{n+1}$;
	\State $\mu_{\scriptscriptstyle n+1} \gets \mu_{\scriptscriptstyle k} + \alpha_{\scriptscriptstyle n+1} \nabla_{\mu} \ell( \mu_{\scriptscriptstyle n}, \Sigma_{\scriptscriptstyle n}, \beta_{\scriptscriptstyle n}, x_{\scriptscriptstyle n})$;
	\State $\Sigma_{\scriptscriptstyle n+1} \gets \Sigma_{\scriptscriptstyle n} \exp \left( \Sigma_{\scriptscriptstyle n}^{\scriptscriptstyle -1} \alpha_{\scriptscriptstyle n+1} \nabla_{\Sigma} \ell( \mu_{\scriptscriptstyle n+1}, \Sigma_{\scriptscriptstyle n}, \beta_{\scriptscriptstyle n}, x_{\scriptscriptstyle n} ) \right)$;
	\State $\beta_{\scriptscriptstyle n+1} \gets \beta_{\scriptscriptstyle n} \exp \left( \beta_{\scriptscriptstyle n}^{\scriptscriptstyle -1} \alpha_{\scriptscriptstyle n+1} \nabla_{\beta} \ell( \mu_{\scriptscriptstyle n+1}, \Sigma_{\scriptscriptstyle n+1}, \beta_{\scriptscriptstyle n}, x_{\scriptscriptstyle n} ) \right)$;
\EndFor
\State $\hat{\theta} \gets (\mu_{\scriptscriptstyle N+1}, \Sigma_{\scriptscriptstyle N+1}, \beta_{\scriptscriptstyle N+1})$;
\end{algorithmic}
\end{breakablealgorithm}
Remark that, the descending direction is $-\nabla_{\theta} \ell(\theta;x)$, and the double negative sign is simplified as positive in the algorithm. The compact and convex set $S_0$ is a neighborhood of $\theta^*$, in which the cost function $D(\theta)$ has an isolated stationary point $\theta = \theta^*$. The following proposition \ref{prop:conv_isg} states the convergence of Algorithm \ref{algo:isg}.
\begin{proposition}\label{prop:conv_isg}
Assume the function $D(\theta)$ has an isolated stationary point at $\theta=\theta^*$ in $S_0$, and that the estimates $(\theta_n)_{n\geq 0}$ remain within $S_0$. Then, $\lim \theta_n = \theta^*$ almost surely.
\end{proposition}
The proof of this convergence is discussed in \ref{prove:conv_isg}. Note that $S_0$ admits a system of normal coordinates $(\theta^{i}; i=1,\cdots, d)$ with origin at $\theta^*$, where $d$ is the dimension of the parameter space $\Theta$, $d=\frac{m(m+1)}{2} + m + 1$. Since $D(\theta)$ has an isolated stationary point at $\theta=\theta^*$, the Hessian at point $\theta=\theta^*$ can be expressed in normal coordinates
\begin{equation}\label{eq:Hessian_theta*}
\mathcal{H}_{ij} = \left. \frac{\partial^2 D}{\partial \theta^{i} \partial \theta^{j}} \right|_{\theta^{i}=0} 
\end{equation}
The matrix $\mathcal{H} = (\mathcal{H}_{ij})$ is positive definite~\cite{absil2009optimization}. With these notations, the rate of convergence is given by the following proposition.
\begin{proposition}\label{prop:mean_square_rate}
Under the assumptions of Proposition \ref{prop:conv_isg}, if $a>\frac{1}{2 \lambda}$, where $\lambda > 0$ is the smallest eigenvalue of $\mathcal{H}$,
\begin{equation}\label{eq:mean_square_rate}
\mathbb{E} [ \mathrm{d}^2(\theta^*,\theta_n) ] = \mathcal{O}(n^{-1})
\end{equation}
\end{proposition}
Here, $\mathrm{d}(\cdot,\cdot)$ stands for the product distance in (\ref{eq:inf_d_prod}), and the "big O" notation means that there exist $K>0$ and $n_0 > 0$ such that
\begin{equation*}
\forall n\geqslant n_0 \qquad \mathbb{E}[ \mathrm{d}^2(\theta^*,\theta_n)] \leqslant \frac{K}{n}
\end{equation*}
In terms of the normal coordinates $(\theta^{i})$, let the direction of descent $\nabla_{\theta} \ell(\theta^*;x)$ at the point $\theta=\theta^*$ have components $(u^{i}(\theta^*))$. Let $\mathcal{G}^* = (\mathcal{G}^*_{ij})$, be the matrix
\begin{equation}\label{eq:fiher_inf_matrix_at_theta*}
\mathcal{G}^*_{ij} = \mathbb{E}_{\theta^*} \left[ u^{i}(\theta^*) u^{j}(\theta^*) \right]
\end{equation}
Then, the following proposition gives the asymptotic normality of the ISG algorithm.
\begin{proposition}[asymptotic normality]\label{prop:asymptotic_normality}
Under the assumptions of Propositions \ref{prop:conv_isg} and \ref{prop:mean_square_rate}, the distribution of the re-scaled coordinates $(n^{\frac{1}{2}}\theta^{i})_{i \in \{1,\cdots, d \}}$ converges to a centred $d-$variate normal distribution, where $d$ is the dimension of $\Theta$, with covariance matrix $\mathcal{G}$ given by the following Lyabunov equation
\begin{equation}
A \mathcal{G} + \mathcal{G} A = -a^2 \mathcal{G}^*
\end{equation}
Here, $A = (A_{ij})$ with $A_{ij} = \frac{1}{2}\delta_{ij} - a \mathcal{H}_{ij}$ ($\delta$ denotes Kronecker's delta).
\end{proposition}
The proofs of Propositions \ref{prop:asymptotic_normality} and \ref{prop:mean_square_rate} are discussed in \ref{prove:msr_&_an}. For the case $\theta = (\Sigma)$ or $\theta = (\mu,\Sigma)$, the product metric (\ref{eq:product_metric}) coincides with the information metric of the ECD model. Then, the assumptions of Proposition 5 in~\cite{zhou2019fast} are satisfied, and the following corollary may be obtained.
\begin{corollary}\label{coro:ISG_for_mS}
For the ECD model, parameterised by $\theta = (\Sigma)$ or $\theta=(\mu,\Sigma)$, with a fixed $\beta^*$, the product metric (\ref{eq:product_metric}) coincides with the information metric.
\begin{enumerate}
\item the rate in equation (\ref{eq:mean_square_rate}) holds, whenever $a>1/2$.
\item\label{item:normality} if $a = 1$ the distribution of the re-scaled coordinates $(n^{1/2} \theta^{i})$ converges to a centred $d$-variate normal distribution, with covariance matrix equal to the identity $\mathcal{G}^{*} = I_d$, and the recursive estimates $\theta_n$ are asymptotically efficient.
\end{enumerate}
\end{corollary}
Note that, Item \ref{item:normality}) of Corollary \ref{coro:ISG_for_mS} implies that the distribution of $n \mathrm{d}^2(\theta^*,\theta_n)$ converges to a $\chi^2$-distribution with $d$ degrees of freedom. 
\begin{subequations}
\begin{equation}
n \mathrm{d}^2(\theta^*,\theta_n) \Rightarrow \chi^2\left(\frac{m(m+1)}{2}\right) \text{ for } \theta = (\Sigma)
\end{equation}
\begin{equation}
n \mathrm{d}^2 (\theta^*,\theta_n) \Rightarrow \chi^2\left(\frac{m(m+1)}{2} + m\right) \text{ for } \theta = (\mu,\Sigma)
\end{equation}
\end{subequations}
This provides a practical means of confirming the asymptotic normality of the estimators $\theta_n$. The function $d^2(\cdot,\cdot)$ denotes the square information distance, here the same as (\ref{eq:inf_d_prod}). 

\subsection{Global convergence analysis}
This section studies the global convergence of the IDG and ISG algorithms, for two specific families of distributions, MGGD and Student-T. The main results are stated in the following two tables.
\begin{table}[!htbp]
\caption{Convergence analysis: MGGD}
\label{tab:Cov_MGGD}
\centering
\begin{tabular}{|c|c|}
\hline
 & MGGD \\
\hline
$\theta = (\Sigma)$ & Globally for $\beta>0$\\
\hline
$\theta = (\mu,\Sigma)$ & Globally for $\beta > 0$\\
\hline
\end{tabular}
\end{table}
\begin{table}[!htbp]
\caption{Convergence analysis: Student-T}
\label{tab:Cov_studentT}
\centering
\begin{tabular}{|c|c|}
\hline
 & Student\\
\hline
$\theta = (\Sigma)$ & Globally for $\beta>-m$\\
\hline
$\theta = (\mu,\Sigma)$ & Globally for $\beta>0$\\
\hline
\end{tabular}
\end{table}
For the cases indicated in Tables \ref{tab:Cov_MGGD} and \ref{tab:Cov_studentT}, the cost function $D(\theta)$ (or $\hat{D}(\theta)$) has a unique stationary point, at $\theta^*$, which is the global minimizer. This will be obtained from the following development.\\

\noindent
First, for the case of $\theta=(\Sigma)$ with known $\mu^*$ and $\beta^*$, let
\begin{equation}\label{eq:f=inv_g}
f(\delta_x,\beta) = \frac{1}{g(\delta_x,\beta)}
\end{equation}
then, for the MGGD model
\begin{subequations}\label{eq:fct_f}
\begin{equation}\label{eq:fct_f_mggd}
f(\delta_x,\beta) = \exp \left( \frac{1}{2} \delta_x^{\beta} \right) \qquad \beta>0
\end{equation}
and for the Student-T model,
\begin{equation}\label{eq:fct_f_student}
f(\delta_x,\beta) = \left( 1+\frac{\delta_x}{\beta} \right)^{\frac{\beta+m}{2}} \qquad \beta>-m
\end{equation}
\end{subequations}
The following proposition introduces a sufficient condition for the KL divergence $D(\Sigma)$ and its empirical approximation $\hat{D}(\Sigma)$ to be geodesically strictly convex.
\begin{proposition}\label{prop:convex_Sigma}
assume that the function $f:\mathbb{R}_+ \rightarrow \mathbb{R}_+$ in (\ref{eq:f=inv_g}) verifies the following condition : for any $\varphi:\mathbb{R} \rightarrow \mathbb{R}_+$
\begin{equation}\label{eq:convex_condition_Sigma}
\varphi \text{ strictly log-convex } \Rightarrow f \circ \varphi \text{ strictly log-convex }
\end{equation}
Then, the KL divergence $D(\Sigma)$ (and its approximation $\hat{D}$) is geodesically strictly convex. 
\end{proposition}
In particular, the unique global minimum, and the unique stationary point, of $D(\Sigma)$ is at the true $\Sigma^*$. This proposition \ref{prop:convex_Sigma} directly yields the following corollary, for the specific MGGD model and Student-T model, by plugging (\ref{eq:fct_f_mggd}) and (\ref{eq:fct_f_student}) into (\ref{eq:convex_condition_Sigma}).
\begin{corollary}\label{coro:mggd_stud_convex_Sigma}
the KL divergence $D(\Sigma)$ and its empirical approximation $\hat{D}(\theta)$ are geodesically strictly convex, with unique global minimum (and unique stationary point), in both of the following cases.
\begin{enumerate}
\item $\mathcal{X}$ is distributed according to an MGGD model, with scatter matrix $\Sigma^*$ and with shape parameter $\beta > 0$.
\item $\mathcal{X}$ is distributed according to a Student-T model, with scatter matrix $\Sigma^*$ and degree of freedom $\beta>-m$.
\end{enumerate}
\end{corollary}
Thus, when $\Sigma$ is unknown and $\beta$ satisfies the conditions of Corollary \ref{coro:mggd_stud_convex_Sigma}, this corollary implies the global convergence of Algorithms \ref{algo:idg} and \ref{algo:isg}. Precisely, these algorithms will always converge to the true value $\theta^*$ of the parameter $\theta$.\\

\noindent
For the more complicated situation $\theta = (\mu, \Sigma)$, global convergence does not always hold. The cost function $D(\theta)$ is not geodesically convex, but may be reformulated, using a new matrix argument~\cite{laus2019multivariate}
\begin{equation}
S = \left[
\begin{matrix}
\Sigma + \mu \mu^{\dagger} & \mu \\
\mu^{\dagger} & 1
\end{matrix}
\right]
\end{equation}
If the new random vector $y$ is given by
\begin{equation}
y = \left( x^{\dagger}, 1 \right)^{\dagger}
\end{equation}
then the cost function can be reformulated as
\begin{equation}
\tilde{D}(\theta) = -\frac{1}{2}\log\det(S) - \log \tilde{f}(\delta_y)
\end{equation}
where 
\begin{equation}\label{eq:delta_y_delta_x}
\delta_y = y^{\dagger} S^{-1} y = (x-\mu)^{\dagger} \Sigma^{-1} (x-\mu) + 1
\end{equation}
and
\begin{subequations}\label{eq:reformed_f_mS}
\begin{equation}\label{eq:reformed_f_mS_mggd}
\tilde{f}(\delta_y) = \exp\left[ \frac{1}{2} \left( \delta_y - 1 \right)^{\beta} \right] \text{ for MGGD}
\end{equation}
Then for Student-T is
\begin{equation}\label{eq:reformed_f_mS_stud}
\tilde{f}(\delta_y) = \left( 1-\frac{1}{\beta} + \frac{\delta_y}{\beta} \right)^{\frac{\beta+m}{2}} \text{ for Student-T}
\end{equation}
\end{subequations}
In \cite{laus2019multivariate}, the minimization of $\tilde{D}(\theta)$ was proven to be equivalent to the minimization of $D(\theta)$. Replacing the new function $\tilde{f}$ into (\ref{eq:convex_condition_Sigma}), the following corollary is obtained.
\begin{corollary}\label{coro:mggd_stud_convex_mS}
the KL divergence $D(\mu,\Sigma)$ (and $\hat{D}(\mu,\Sigma)$) has a unique global minimum (and unique stationary point) at $(\mu^*,\Sigma^*)$, in both of the following cases.
\begin{enumerate}
\item $\mathcal{X}$ is distributed according to an MGGD model, with expectation and scatter matrix $(\mu^*,\Sigma^*)$ and with fixed shape parameter $\beta>0$.
\item $\mathcal{X}$ is distributed according to a Student-T model, with expectation and scatter matrix $(\mu^*,\Sigma^*)$ and with the fixed degree of freedom $\beta>0$.
\end{enumerate}
\end{corollary}
For these two cases, global convergence is then guaranteed.\\

\noindent
Finally, for the most complicated case, $\theta = (\mu,\Sigma,\beta)$, the cost function is always non-convex. Moreover, we have verified experimentally that it has multiple stationary points in $\Theta = \mathbb{R}^m \times \mathcal{P}_m \times \mathbb{R}_+$. Therefore, the correct estimation can only be guaranteed when the initial value $\theta_0$ is close enough to the global minimum $\theta^*$.

\section{Computer experiments}\label{sec:Computer_Experiments}
This section presents a set of computer experiments, which confirm the theoretical results of Section \ref{sec:mainresults}, and provide a detailed comparison of the ISG and IDG estimation methods, with the already existing MM and FP. For every experiment, 1000 Monte Carlo trials were carried out. For each trial, the dataset $\mathcal{X} = \{ x_1,\cdots,x_N \}$ is independent and identically distributed, according to true parameters $(\mu^*,\Sigma^*,\beta^*)$. The dimension $m$ of $x_n$ is taken equal to $10$. The true $\mu^*$ is randomly chosen from a multivariate normal distribution. The scatter $\Sigma^*$ is defined as $\Sigma(i,j) = \rho^{|i-j|}$ for $i,j \in \{1,m\}$, and $\rho \sim \mathcal{U}(0.2,0.8)$. The shape parameter $\beta^*$ is uniformly selected from the intervals $[0.2,5]$ for MGGD and for Student-T.

\begin{figure}[!htbp]
\centering
\subfigure[The case $\theta=(\Sigma)$]{
\label{fig:Super_rate_S}
\includegraphics[width=6.5cm]{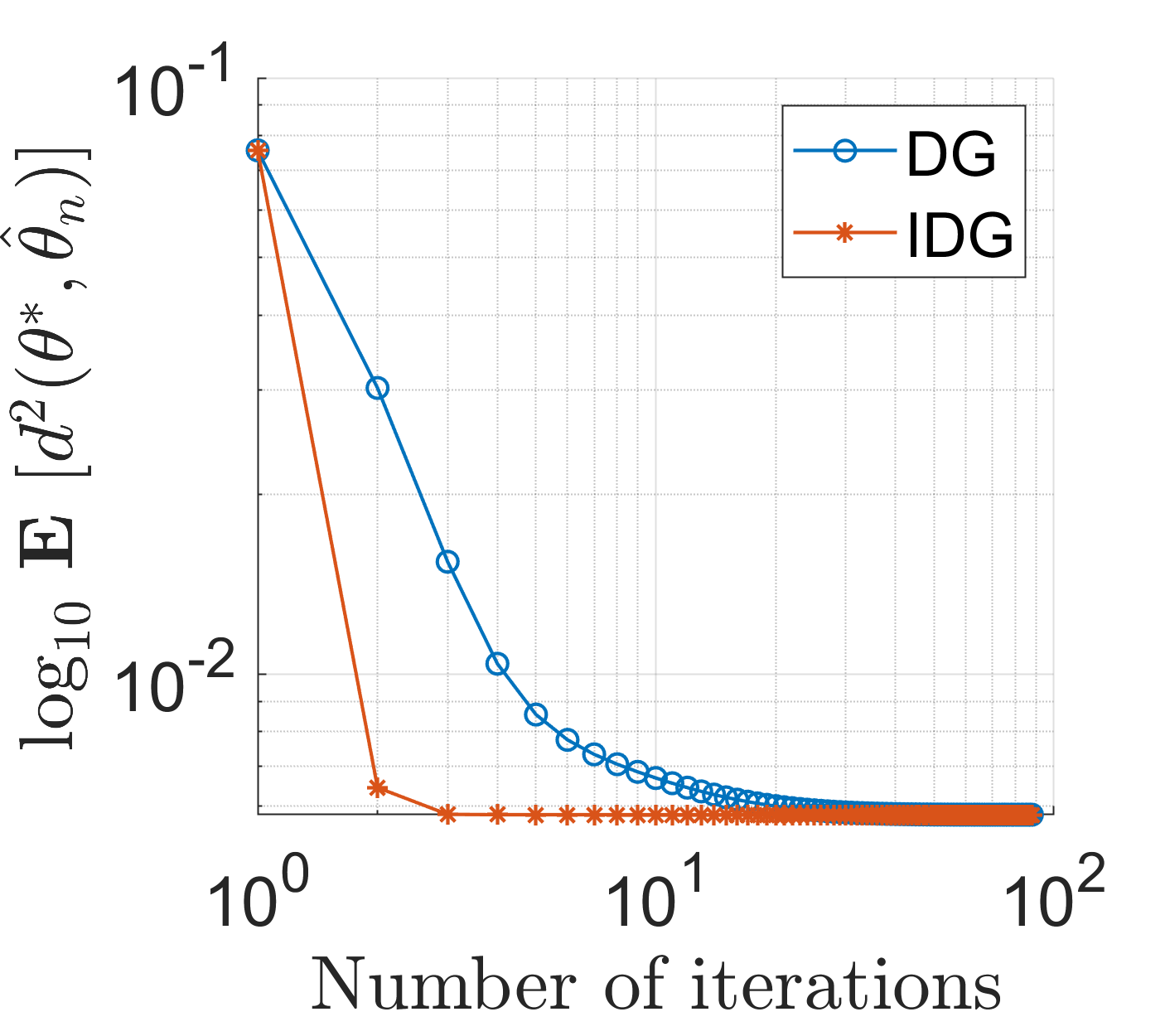}
}
\subfigure[The case $\theta=(\mu,\Sigma)$]{
\label{fig:Super_rate_mS}
\includegraphics[width=6.5cm]{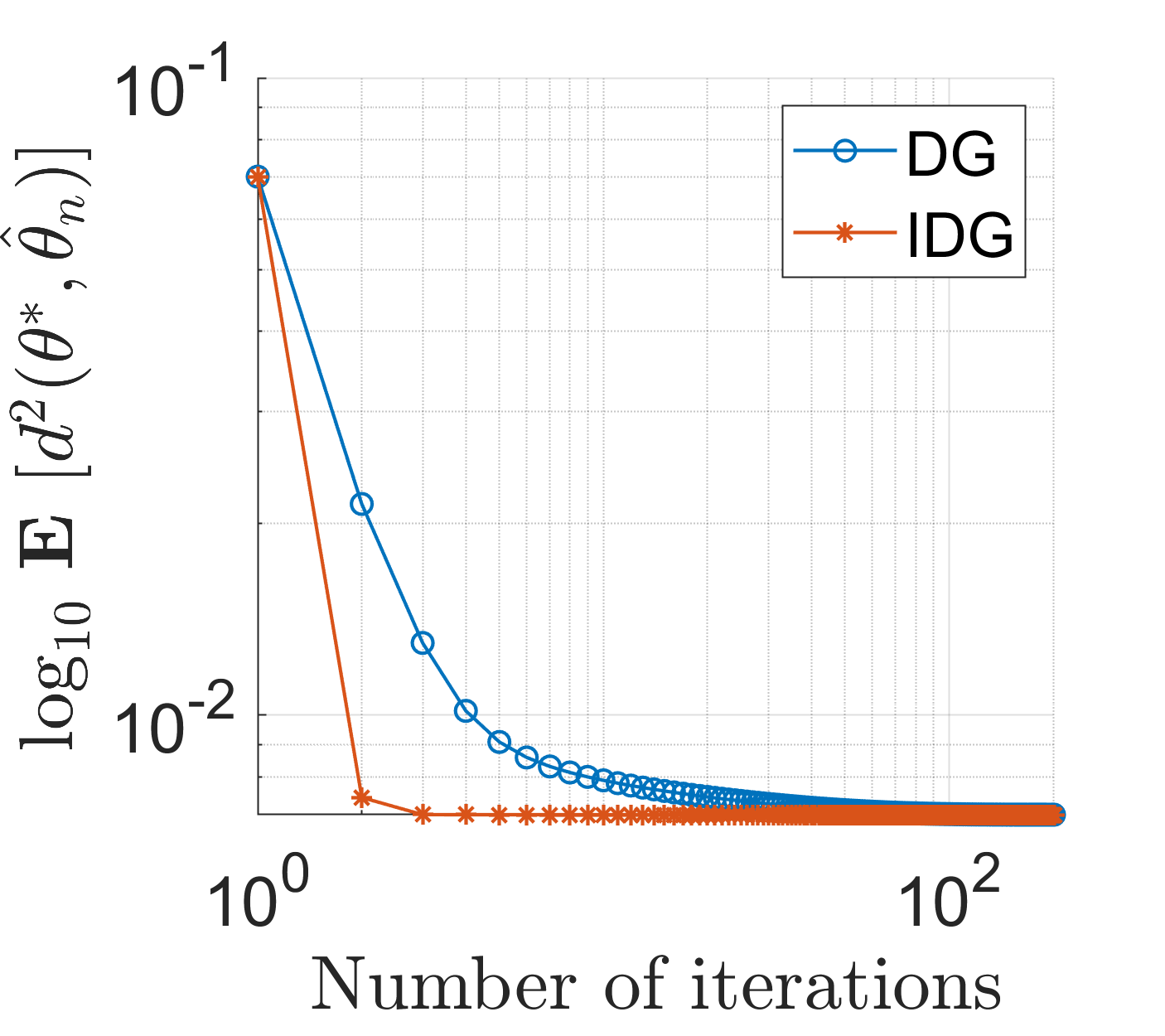}
}
\caption{The superlinear convergence rate for IDG}
\label{fig:Super_rate}
\end{figure}

The first experiment confirms the super-linear rate of convergence of IDG, for a dataset, distributed according to the MGGD model, which contains $N = 10^4$ samples. The initial value $\theta_0$ is defined as the MM estimate, using $10\%$ of the entire dataset. Figure \ref{fig:Super_rate_S} presents the case of $\theta=(\Sigma)$ with known $(\mu^*,\beta^*)$. The IDG method converges after only two iterations, and if the same accuracy needs to be achieved, the deterministic gradient method (not using the information gradient) requires at least $88$ iterations. For the case of $\theta=(\mu,\Sigma)$ with known $(\beta^*)$, things are similar. Figure \ref{fig:Super_rate_mS} shows that IDG, after two iterations, achieves the same accuracy as the traditional gradient method, after $200$ iterations.

\begin{figure}[!htbp]
\centering
\subfigure[The case $\theta=(\Sigma)$]{
\label{fig:rate_S}
\includegraphics[width=6.0cm,height=8.0cm]{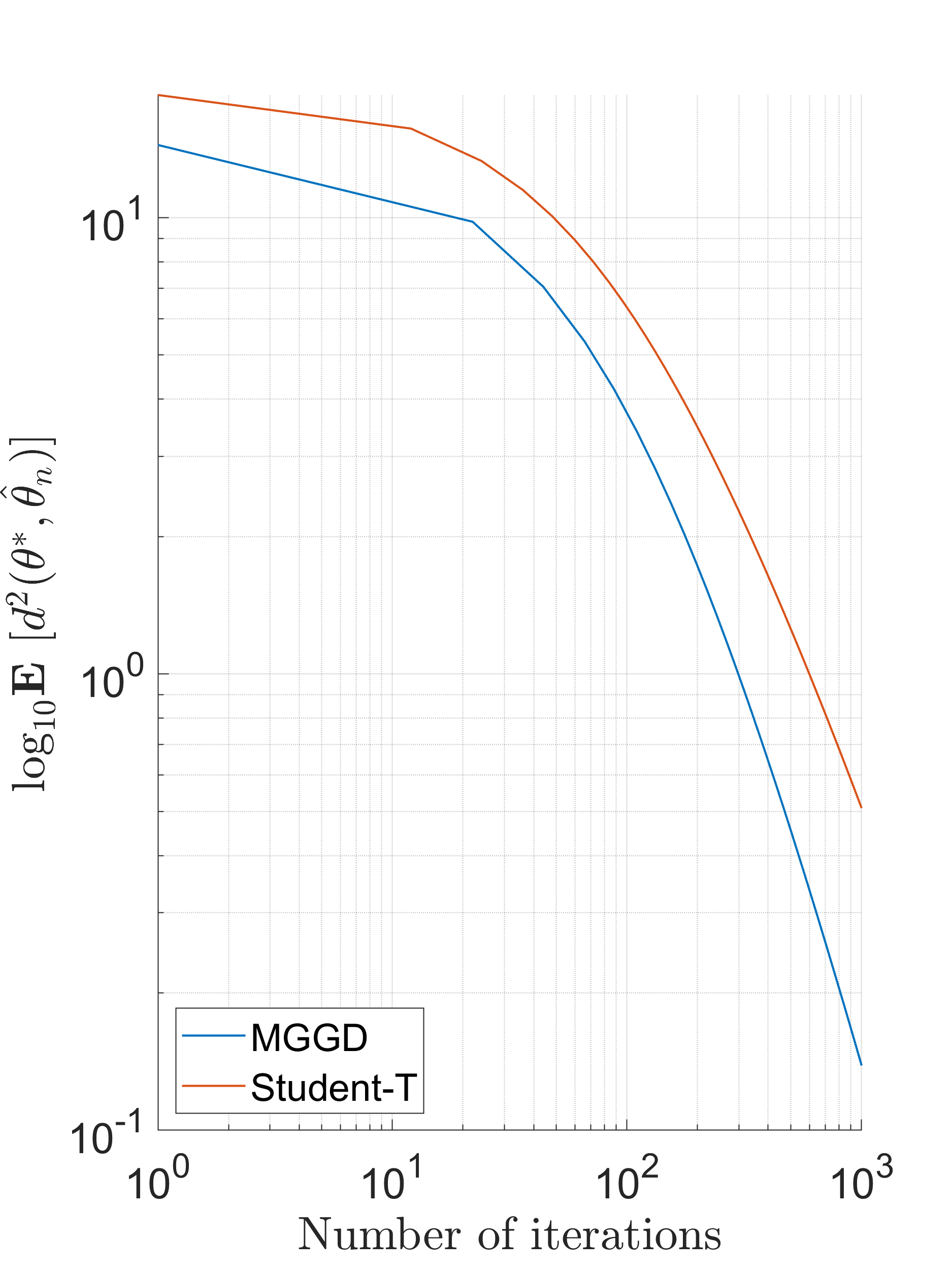}
}
\subfigure[The case $\theta=(\mu,\Sigma)$]{
\label{fig:rate_mS}
\includegraphics[width=6.0cm,height=8.0cm]{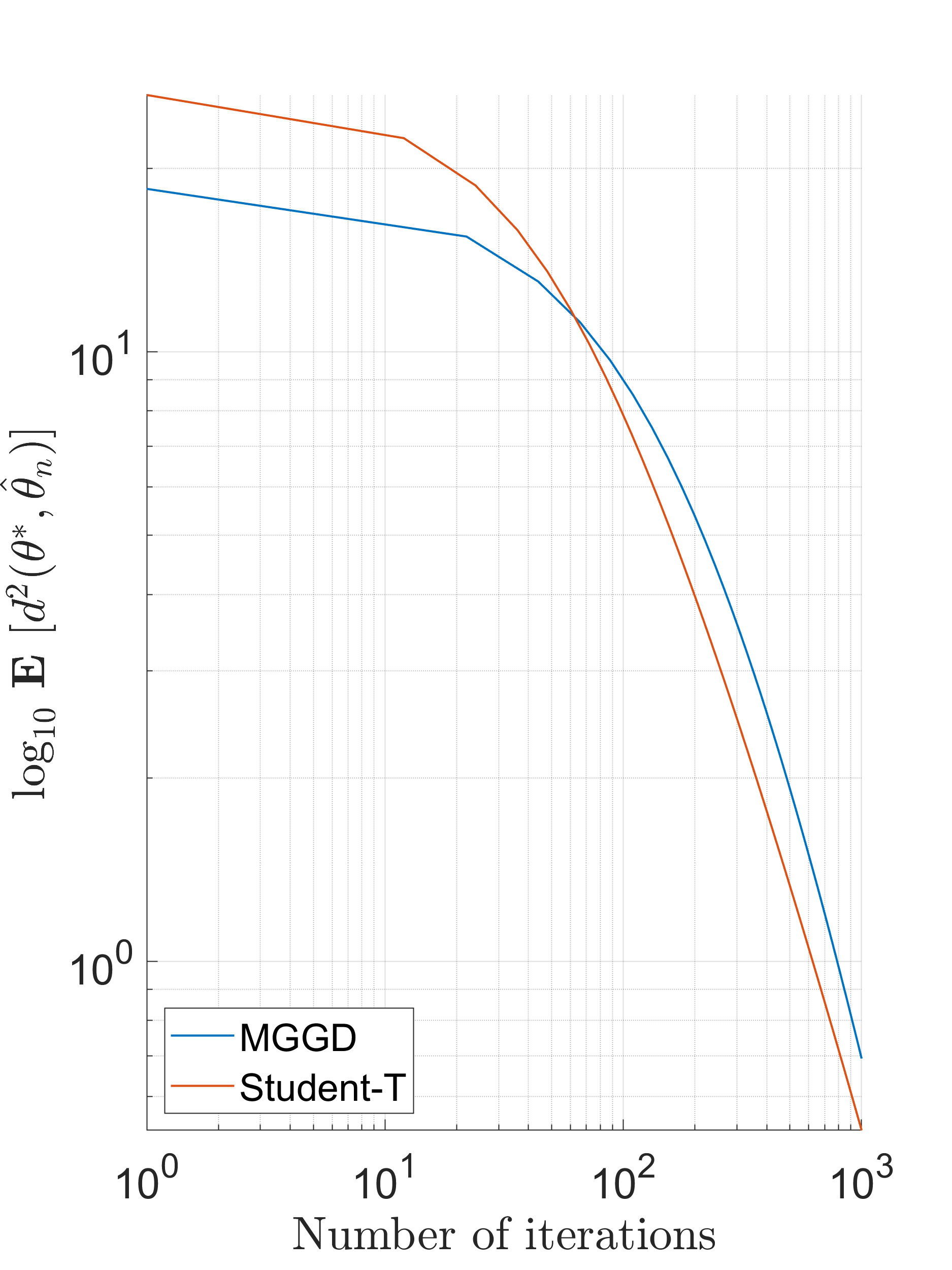}
}

\subfigure[The case $\theta=(\mu,\Sigma,\beta)$]{
\label{fig:rate_mSb}
\includegraphics[width=6.0cm,height=8.0cm]{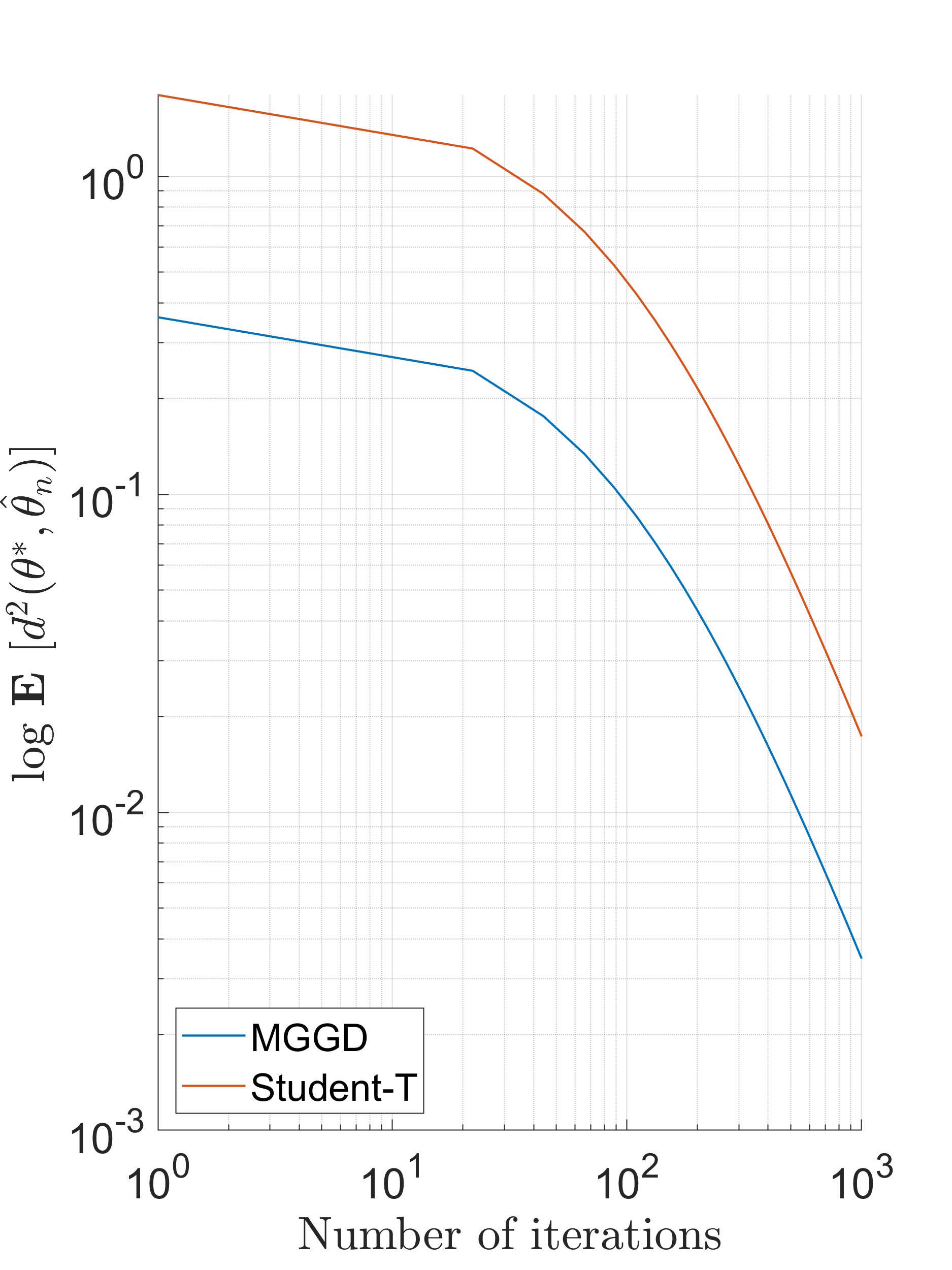}
}
\caption{Linearly convergence rate for ISG}
\end{figure}

The second experiment confirms the convergence rate of ISG. In this experiment, both MGGD and Student-T datasets are used. The initialization $\theta_0$ is randomly chosen. Figures \ref{fig:rate_S}, \ref{fig:rate_mS}, and \ref{fig:rate_mSb} confirm the rate of convergence stated in (\ref{eq:mean_square_rate}), in the neighborhood of $\theta^* = (\mu^*,\Sigma^*,\beta^*)$. In these log-log plots, the x-axis and y-axis represent the number of iterations and $\mathbb{E}[d^2(\theta^*,\hat{\theta}_n)]$, respectively, and $\mathbb{E}$ denotes the Monte Carlo approximation of the expectation, obtained by averaging over the 1000 trials. The slope of each curve approaches $-1$, while $\theta_n$ approaches the true value $\theta^*$. Note that, for the cases of $\theta = (\Sigma)$ and $\theta=(\mu,\Sigma)$, the initialization $\theta_0$ can be chosen far away from $\theta_0$ (e.g. $d^2(\theta^*,\theta_0) > 10$). However, when $\theta=(\mu,\Sigma,\beta)$, the initialization should be in a neighborhood of $\theta^*$ which satisfies the conditions in Proposition \ref{prop:mean_square_rate}. For the results obtained in Figures \ref{fig:rate_S} and \ref{fig:rate_mS} (that is to say, when $\beta^*$ is fixed), the step-size coefficient $a$ always equals $1$, which satisfies the condition in \ref{coro:ISG_for_mS}. For the case of unknown $\beta$ , the step-size coefficient $a$ is taken much larger, in order to meet the conditions of proposition \ref{prop:mean_square_rate}. In fact, here, $a=100$.

\begin{figure}[!htbp]
\centering
\subfigure[The case $\theta=(\Sigma)$]{
\label{fig:Chi2_S}
\includegraphics[width=6.5cm]{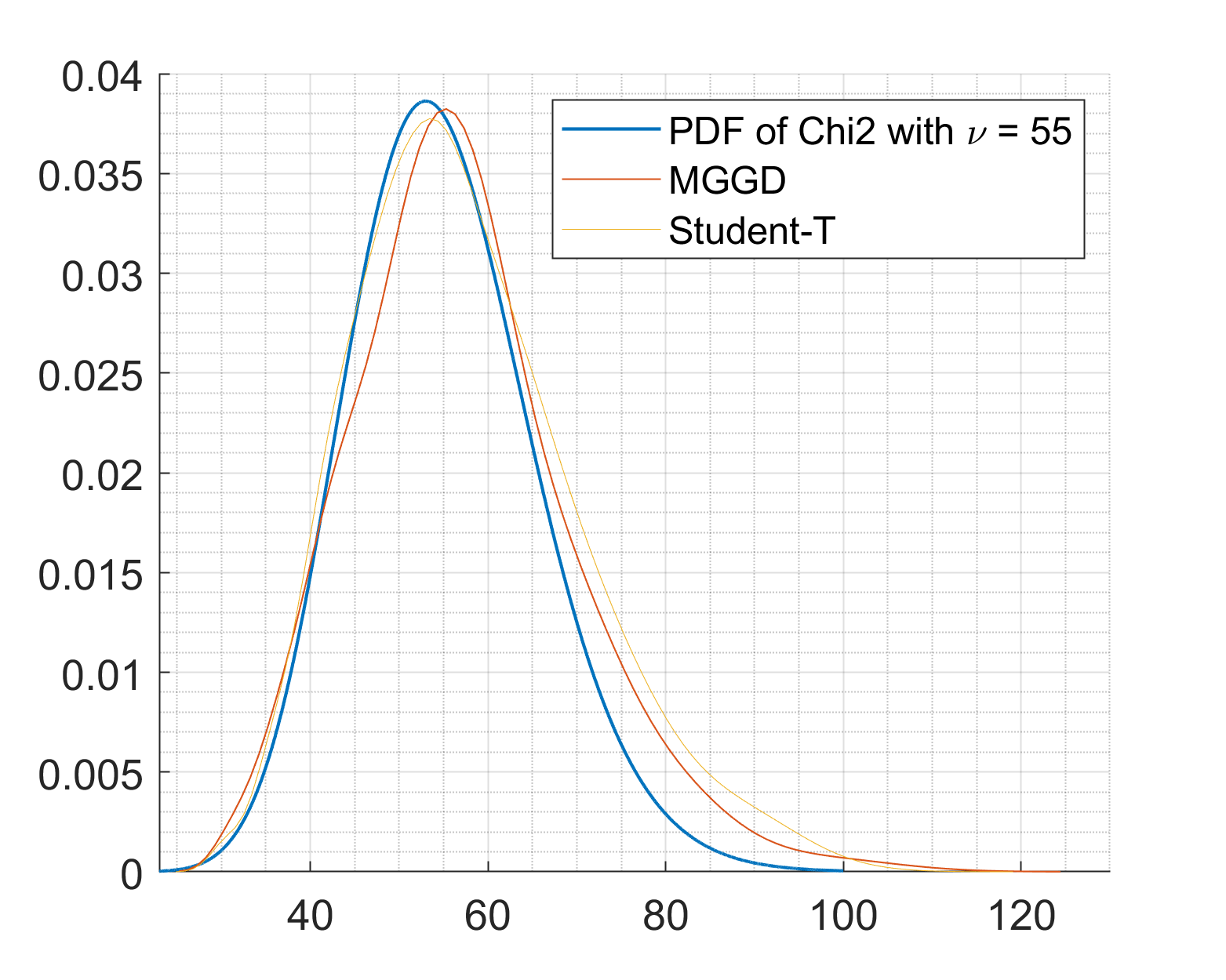}
}
\subfigure[The case $\theta=(\mu,\Sigma)$]{
\label{fig:Chi2_mS}
\includegraphics[width=6.5cm]{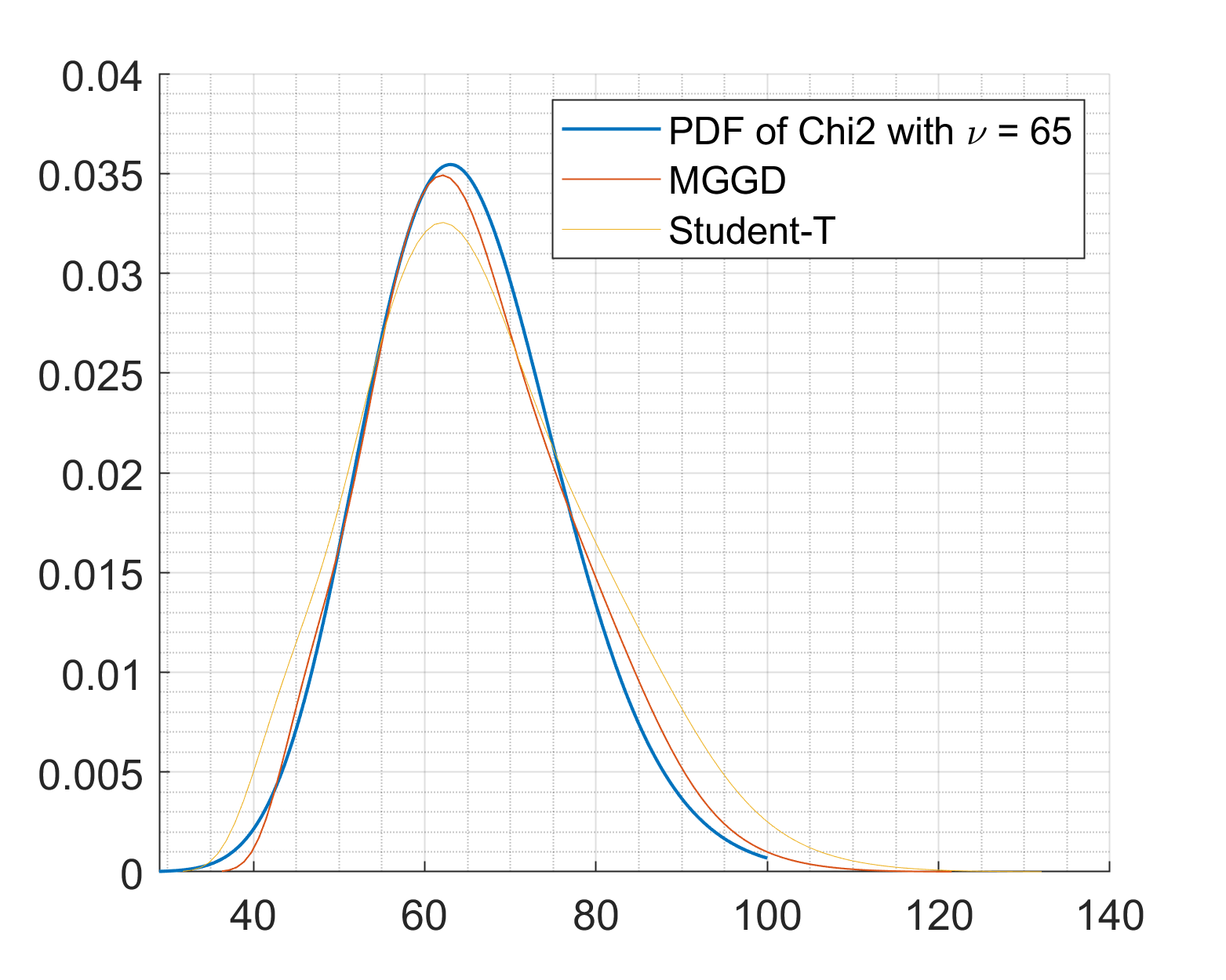}
}
\caption{Validation by fitting a Chi-2}
\end{figure}

For the case of $\theta = (\Sigma)$ and $\theta=(\mu,\Sigma)$, Figures \ref{fig:Chi2_S} and \ref{fig:Chi2_mS} confirm the chi-squared limit distribution in corollary (\ref{coro:ISG_for_mS}). 
The samples $x_n$ being matrices of size $m \times m$ with $m = 10$, the dashed blue curve is the probability density of a chi-
squared distribution with $55$ and $65$ degrees of freedom, for Figures \ref{fig:Chi2_S} and \ref{fig:Chi2_mS} respectively. The
solid lines are the smoothed histograms of $N d^2(\theta^*, \hat{\theta}_N)$ where $N = 10^5$. These "estimated p.d.f." coincide very closely with the theoretical chi-squared probability density.

\begin{figure}[!htbp]
\centering
\subfigure[The case $\theta=(\Sigma)$]{
\label{fig:eff_S}
\includegraphics[width=6.5cm]{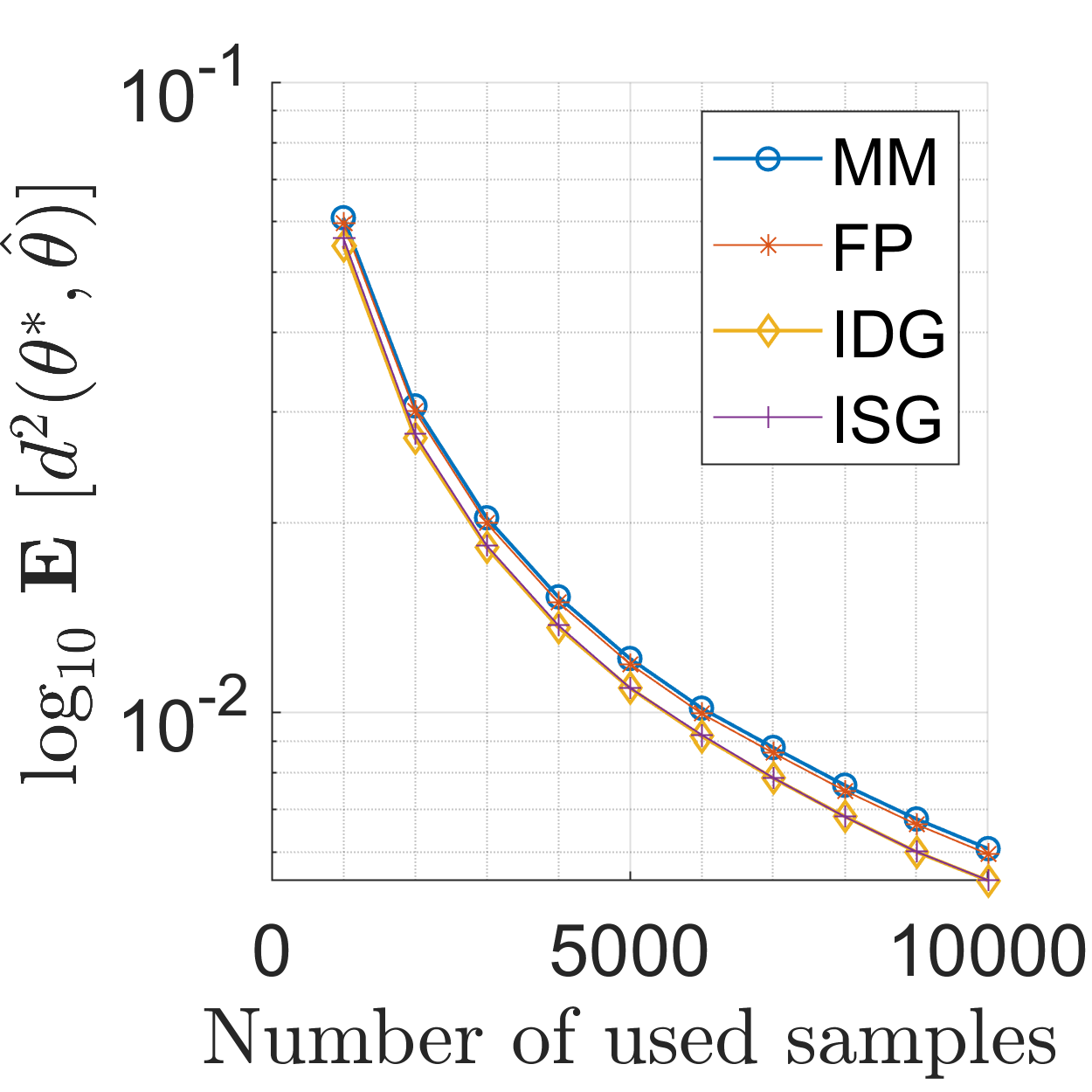}
}
\subfigure[The case $\theta=(\mu,\Sigma)$]{
\label{fig:eff_mS}
\includegraphics[width=6.5cm]{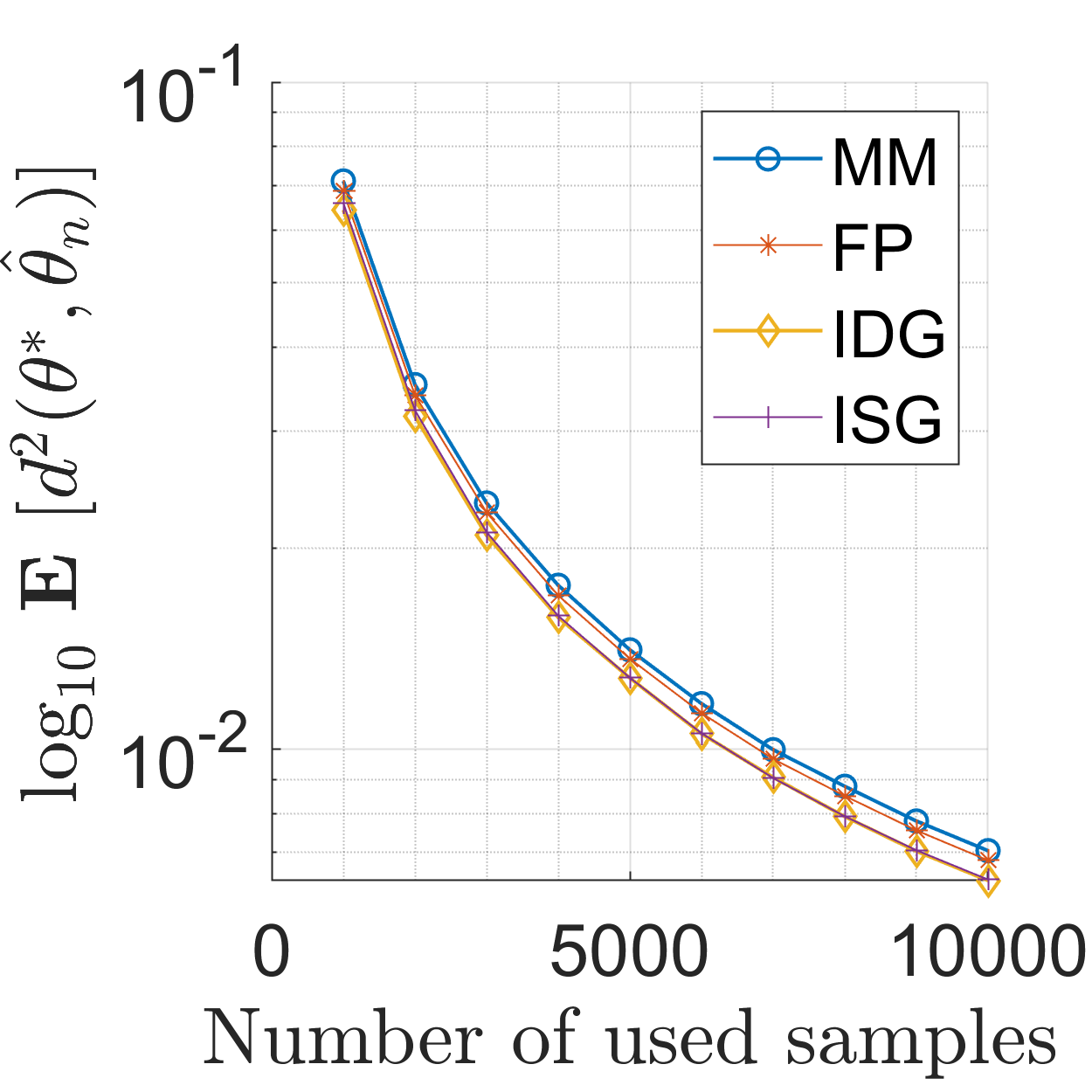}
}

\subfigure[The case $\theta=(\mu,\Sigma,\beta)$]{
\label{fig:eff_mSb}
\includegraphics[width=6.5cm]{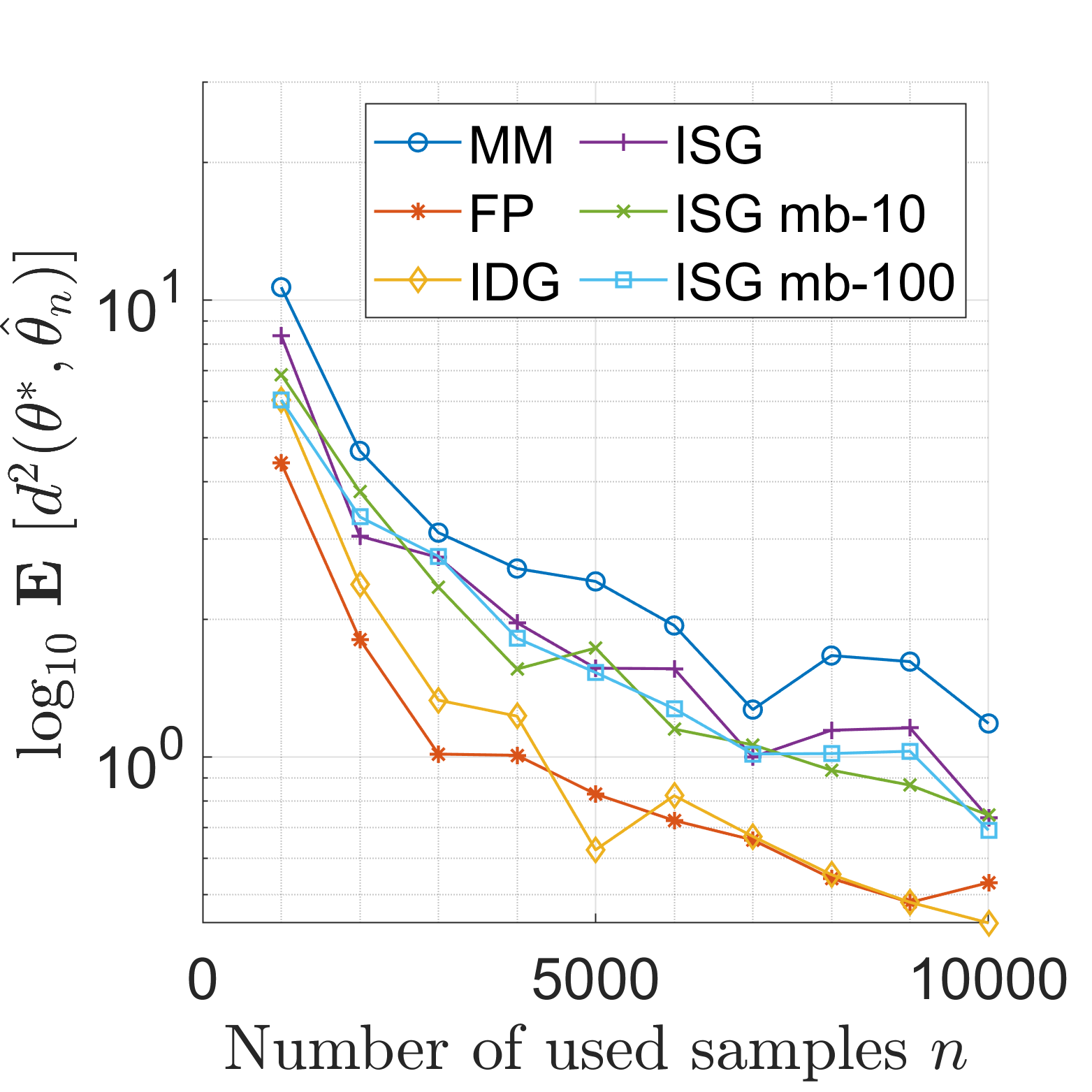}
}
\caption{Efficiency comparison}
\end{figure}

In the third experiment, we compare the efficiency of the IDG and ISG methods with other common estimation methods, MM and FP. In each trial, the dataset is generated from an MGGD model, and contains $N = 10^4$ datapoints. For MGGD, the MM  was given in~\cite{verdoolaege2011geodesics}, and the FP method in \cite{pascal2013parameter}. In Figures \ref{fig:eff_S}, \ref{fig:eff_mS} and \ref{fig:eff_mSb}, the x-axis denotes the size of the dataset, and the y-axis denotes the expectation of the square distance between $\theta^*$ and the estimated $\hat{\theta}$. This expectation is approximated by the average of $10^3$ Monte Carlo trials. For the cases $\theta = (\Sigma)$ and $\theta=(\mu,\Sigma)$, the IDG and ISG algorithms show a better accuracy. When $\theta=(\mu,\Sigma,\beta)$, the accuracy of the MLE method is still significantly better than MM, and the accuracies of IDG and FP coincide. However, the accuracy of ISG is not as good as as FP or IDG. This phenomenon may be explained theoretically. Indeed, when $\theta=(\mu,\Sigma,\beta)$, the product metric does not coincide with the information metric of the ECD model, and this leads to a less efficient estimation. The fluctuations of the curves in Figure \ref{fig:eff_mSb} are quite significant. This means the variance of the final estimate $\hat{\theta}$ is significant.

\begin{figure}[!htbp]
\centering
\subfigure[Gradient $\rightarrow 0$.]{
\label{fig:Grad2Zero}
\includegraphics[width=13cm]{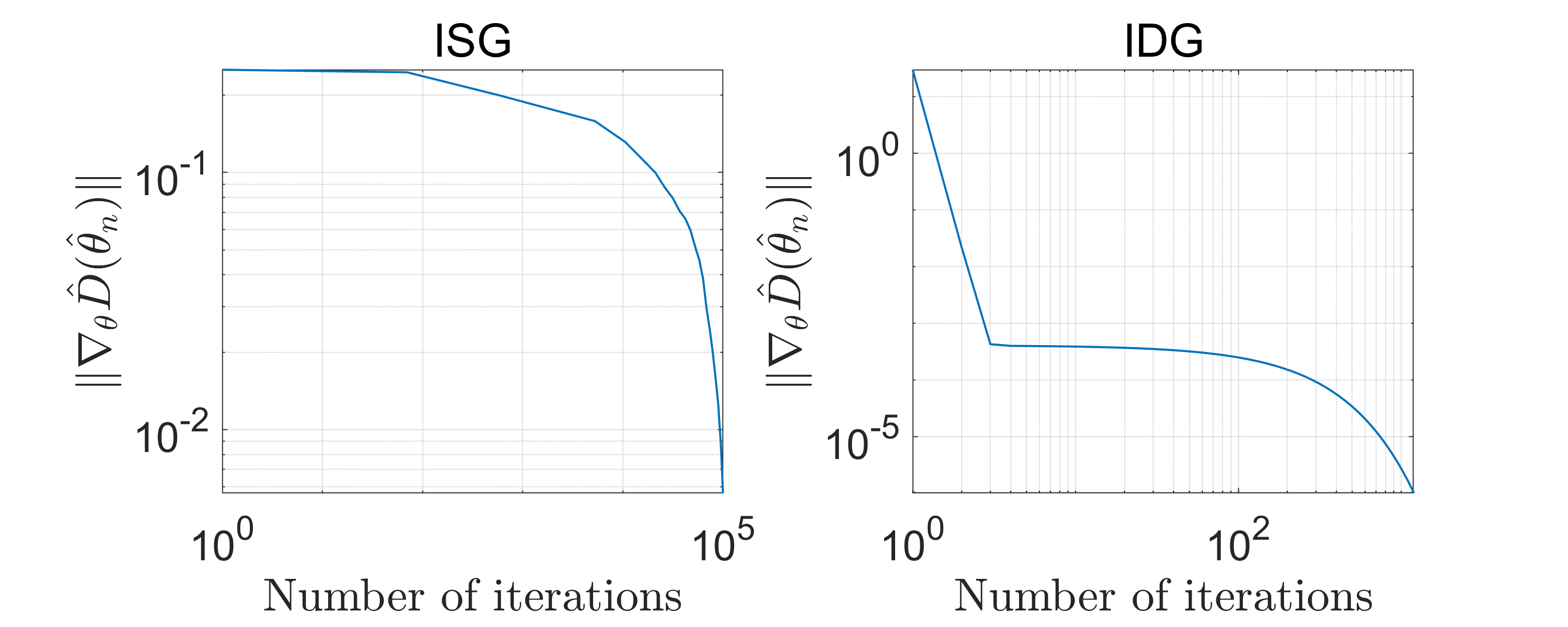}
}
\subfigure[Other stationary point]{
\label{fig:LGCov}
\includegraphics[width=6.5cm]{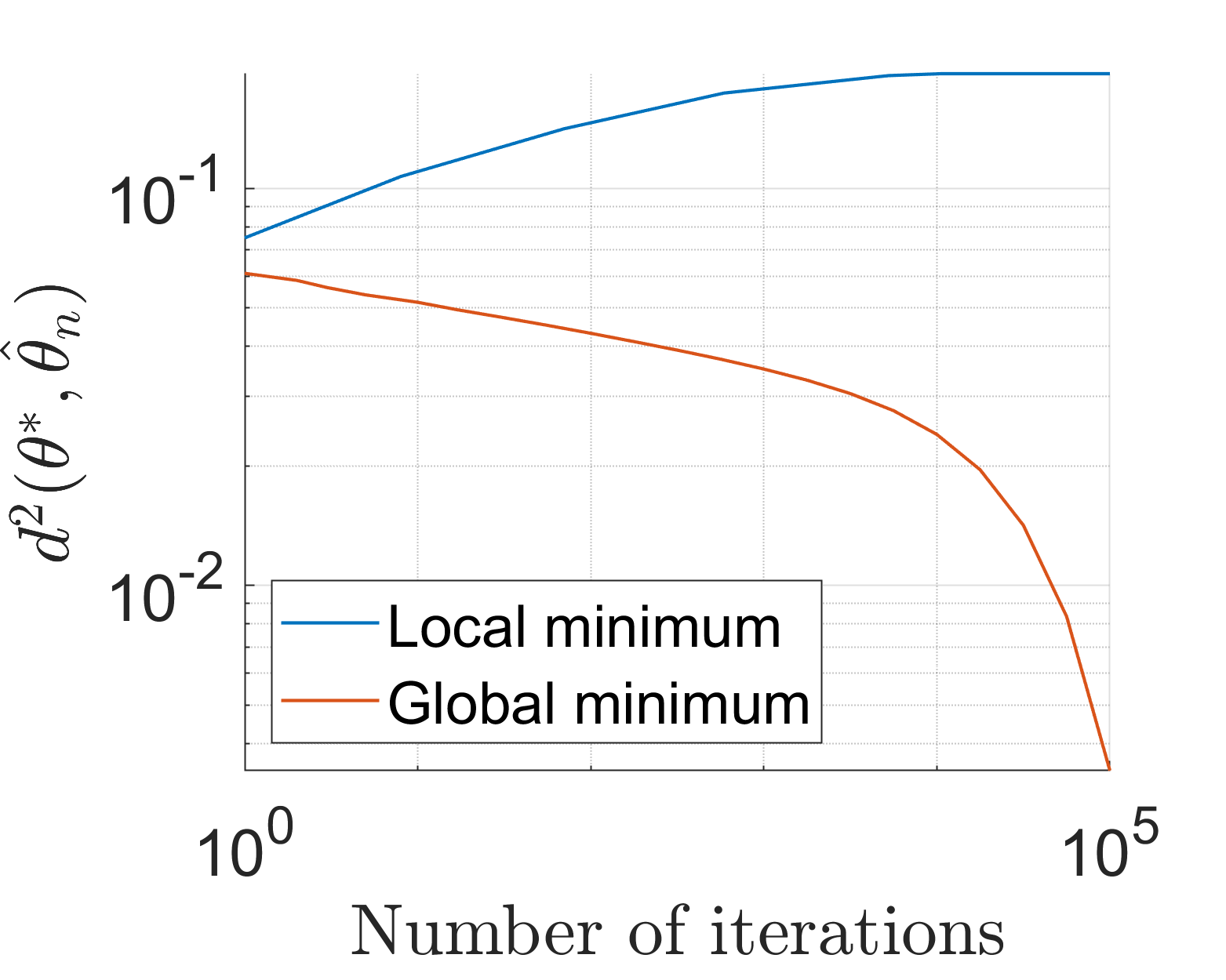}
}
\caption{Additional experiments}
\end{figure}

Two additional experiments were done, to explain these fluctuations. The first additional experiment shows that both IDG and ISG eventually converge to a stationary point (but not necessarily a global minimum). The variations of the norms of the gradients appear in Figure \ref{fig:Grad2Zero}. As the number of iterations increases, the norm of the gradient approaches $0$, for both IDG and ISG. The second additional experiment proved the existence of stationary points other than the true value $\theta^*$. For the same dataset, two different initial values $\theta_0$ were used for the ISG method. In Figure \ref{fig:LGCov}, the initial value $\theta_0$ of the red curve is close to the global minimum $\theta^*$, and its $\theta_n$ finally converge to $\theta^*$. The blue curve has $\theta_0$ farther away, and its $\mathrm{d}^2(\theta^*, \hat{\theta}_n)$ converges to a non-zero constant. In conclusion, for $\theta=(\mu,\Sigma,\beta)$, the convergence to global minimum $\theta^*$ can only be guaranteed locally. If the initial value $\theta_0$ is chosen in a neighborhood $S_0$, then FP and IDG can converge to the true point by virtue of their stability, where $S_0$ should always satisfy the conditions in proposition \ref{prop:conv_idg} and \ref{prop:conv_isg}. Due to its stochastic nature, ISG  may jump out of the neighborhood $S_0$ during the first few iterations. This leads to convergence to local minimum, different from $\theta^*$. Then, the final averaged accuracy of ISG is not as good as the other two MLE methods, and the variance of the ISG estimator is relatively important. As a possible remedy to this problem, the mini-batch ISG was also tested, and compared with other methods, in the Figure \ref{fig:eff_mSb}. Two sizes of the mini-batch, $10$ and $100$, were considered. However, the experimental results show that the mini-batch has no significant effect on the accuracy of ISG.

\begin{table}[!htbp]
\caption{Percentage of 'correct' estimates}
\label{tab:correct_percentage}
\centering
\begin{tabular}{|c|c|c|}
\hline
 & correct estimates & incorrect estimates\\
 & $\mathrm{d}^2(\theta^*,\theta_n) \searrow 0$ & $\mathrm{d}^2(\theta^*,\theta_n) \to c \gg 0$  \\
 & and $\nabla_{\theta}D(\theta_n) \searrow 0$ & and $\nabla_{\theta}D(\theta_n) \searrow 0$ \\
\hline
$\theta=(\mu,\Sigma,\beta)$ & $73\%$ & $27\%$\\
\hline
\end{tabular}
\end{table}

\begin{figure}[!htbp]
\centering
\subfigure[The case $\theta=(\Sigma)$]{
\label{fig:time_S}
\includegraphics[width=6.5cm]{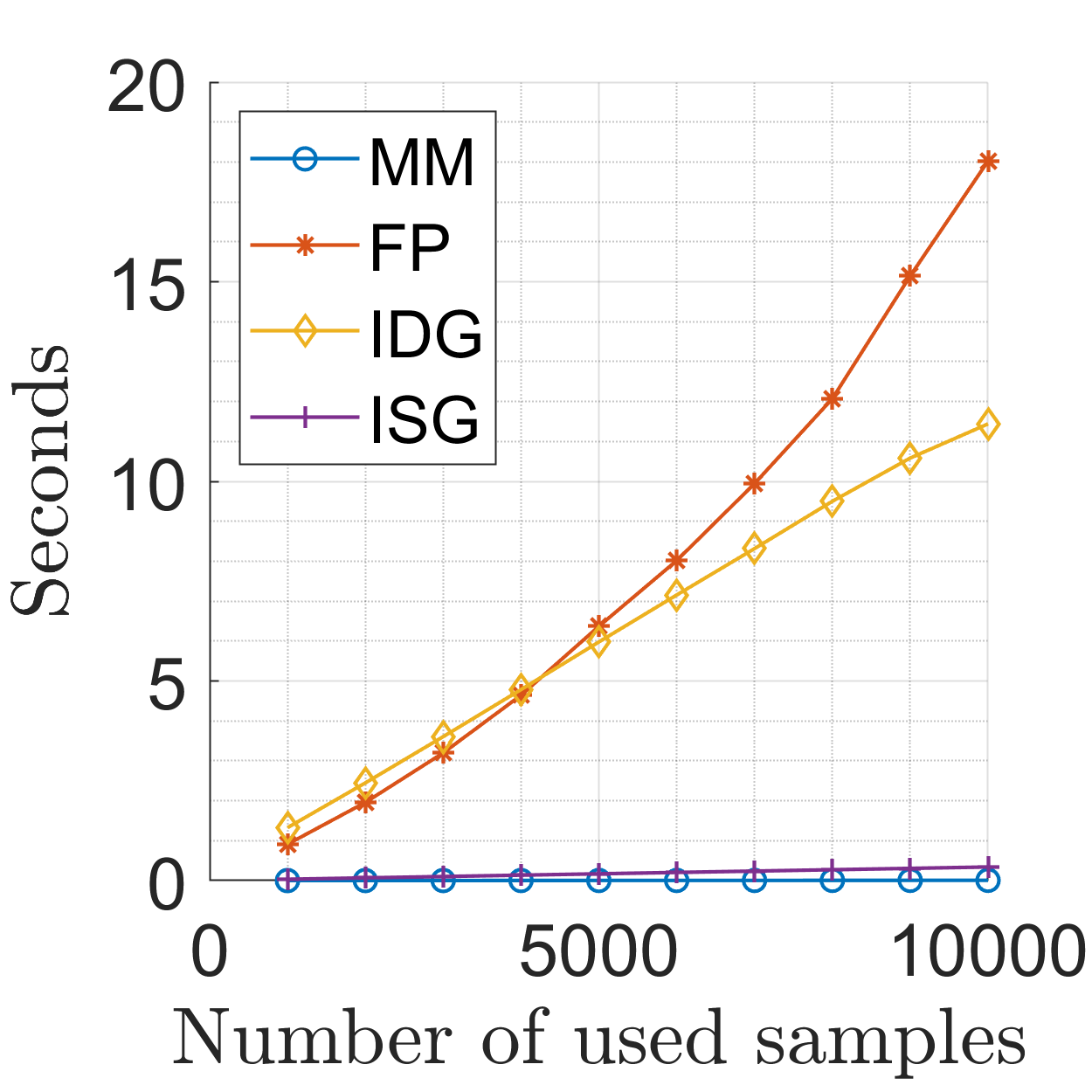}
}
\subfigure[The case $\theta=(\mu,\Sigma)$]{
\label{fig:time_mS}
\includegraphics[width=6.5cm]{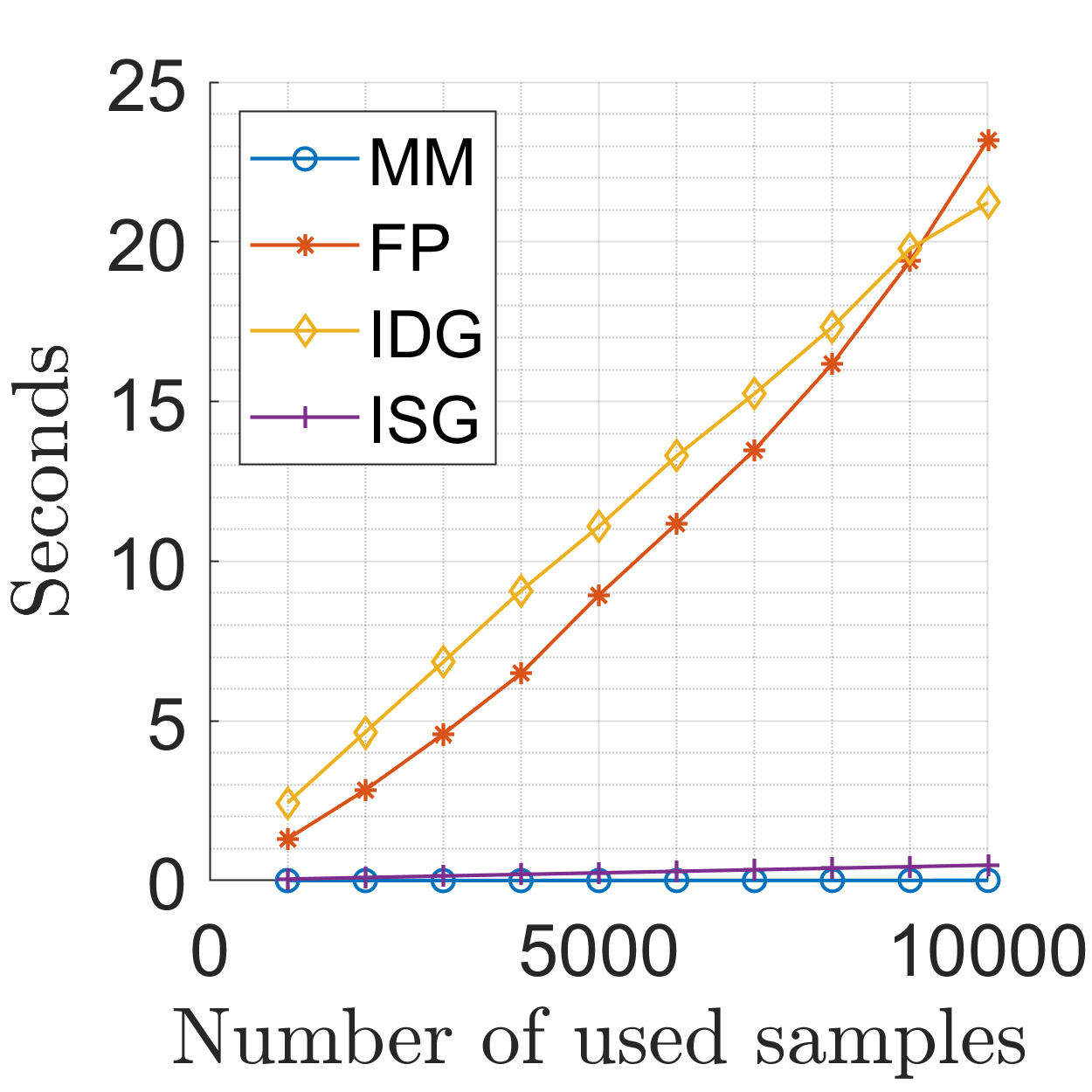}
}
\subfigure[The case $\theta=(\mu,\Sigma,\beta)$]{
\label{fig:time_mSb}
\includegraphics[width=6.5cm]{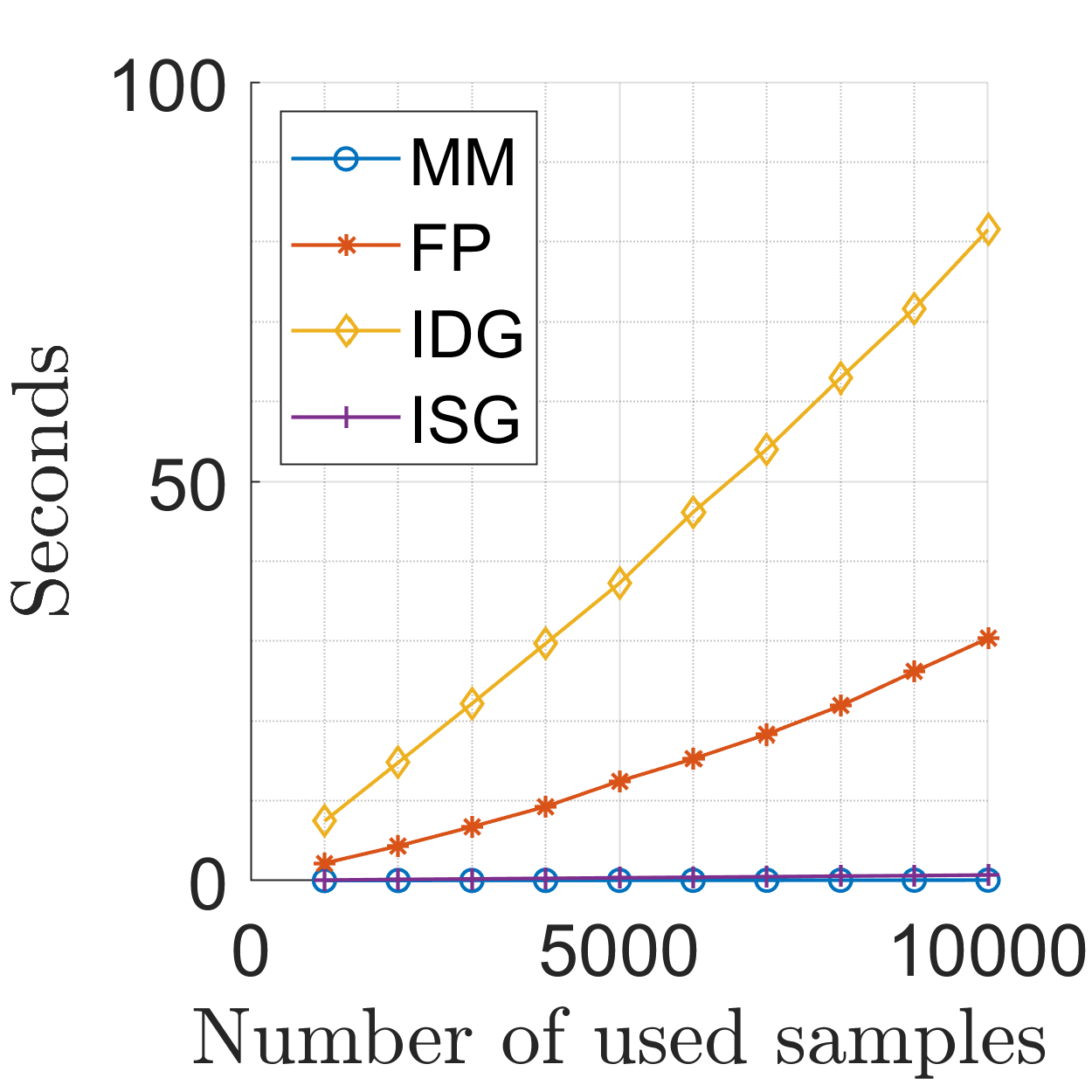}
}
\caption{Time consumption}
\end{figure}
As for computational time, information gradient methods have a significant advantage. The computational time of the ISG algorithm is similar to that of MM, and is significantly less than that of FP. Meanwhile, its accuracy is significantly better than that of MM. In most experiments, the accuracy of ISG is similar to, or even better than, FP. Although the computational time of IDG is greater than that of ISG, it is comparable to that of FP, while, in most cases, IDG can achieve the best accuracy, among the four estimation methods considered.

\section{Application with real dataset}\label{sec:App_Real}
In addition to experimental simulations, we also applied our methods to real datasets.
\subsection{Color transformation}
The first application is to color transformation for image editing with MGGD models, which was investigated in~\cite{hristova2017transformation}. Its goal is to replace the color distribution of the input image by that one of a target image. The main idea is to fit the input and the target distributions, with two different MGGD models. Then, the transformation between these two MGGDs is implemented by a linear Monge-Kantorovich transformation for $\Sigma$, and a stochastic transformation for $\beta$. Specifically, this conversion can be three-dimensional (3D), for RGB images, or five-dimensional (5D), when spatial gradient-field information is included.

\begin{figure}[!htbp]
    \centering
    \begin{overpic}[width=4.4cm]{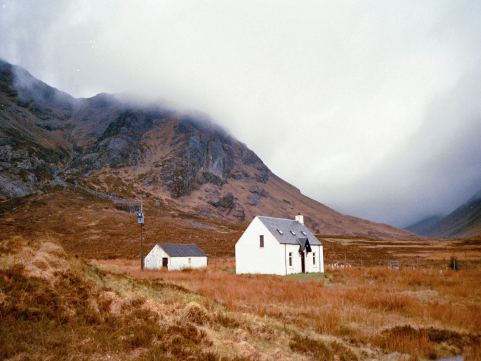}
    \put(60,5){\normalsize \color{white}{\bf Input}}
    \end{overpic}
    \begin{overpic}[width=4.4cm]{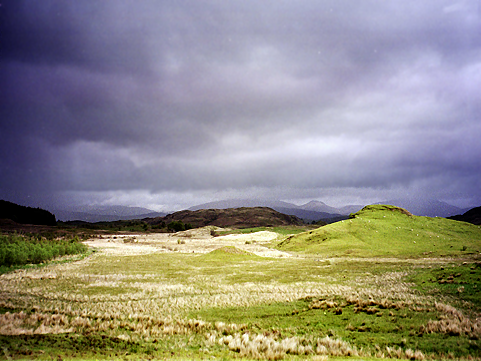}
    \put(55,5){\normalsize \color{white}{\bf Target}}
    \end{overpic}
    \begin{overpic}[width=4.4cm]{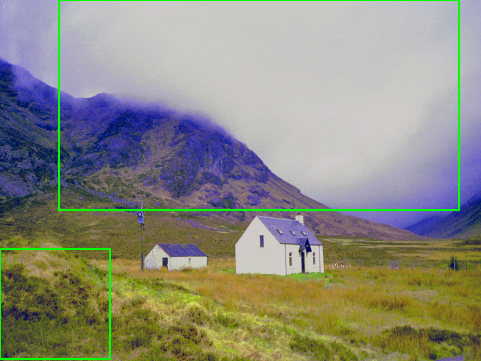}
    \put(72,5){\normalsize \color{white}{MM}}
    \put(2,14){\small \color{white}{\bf b1}}
    \put(15,64){\small \color{white}{\bf a1}}
    \end{overpic}
    
    \vspace*{0.08cm}
    \begin{overpic}[width=4.4cm]{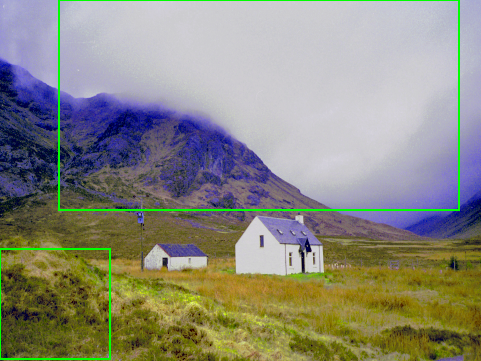}
    \put(78,5){\normalsize \color{white}{FP}}
    \put(2,14){\small \color{white}{\bf b2}}
    \put(15,64){\small \color{white}{\bf a2}}
    \end{overpic}
    \begin{overpic}[width=4.4cm]{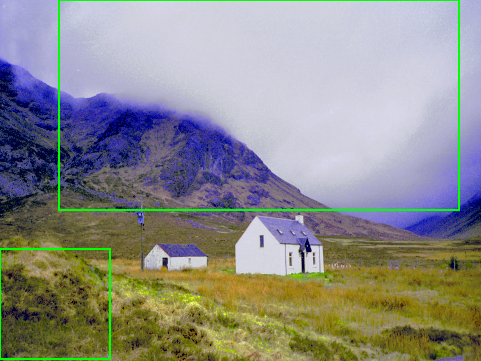}
    \put(72,5){\normalsize \color{white}{IDG}}
    \put(2,14){\small \color{white}{\bf b3}}
    \put(15,64){\small \color{white}{\bf a3}}
    \end{overpic}
    \begin{overpic}[width=4.4cm]{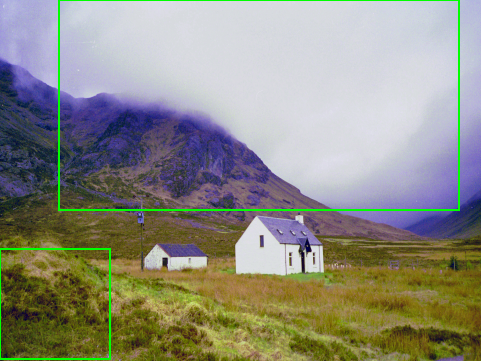}
    \put(75,5){\normalsize \color{white}{ISG}}
    \put(2,14){\small \color{white}{\bf b4}}
    \put(15,64){\small \color{white}{\bf a4}}
    \end{overpic}
    
    \vspace*{0.08cm}
    \begin{overpic}[width=6.65cm]{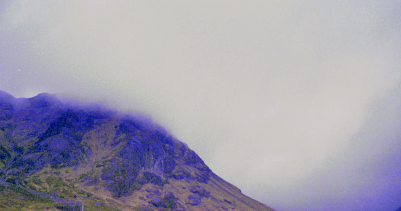}
    \put(2,46){\small \color{white}{\bf a1}}
    \end{overpic}
    \begin{overpic}[width=6.65cm]{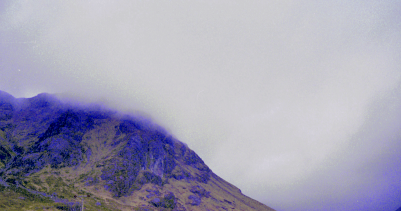}
    \put(2,46){\small \color{white}{\bf a2}}
    \end{overpic}
    
    \vspace*{0.08cm}
	\begin{overpic}[width=6.65cm]{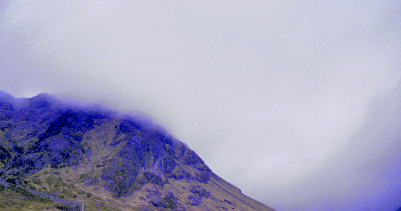}
    \put(2,46){\small \color{white}{\bf a3}}
    \end{overpic}
    \begin{overpic}[width=6.65cm]{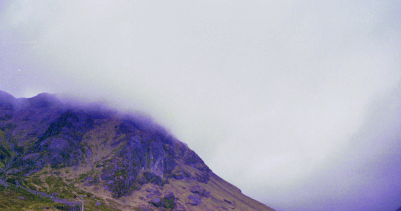}
    \put(2,46){\small \color{white}{\bf a4}}
    \end{overpic}
    
    \vspace*{0.08cm}
    \begin{overpic}[width=3.3cm]{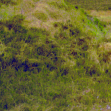}
    \put(3,87){\small \color{white}{\bf b1}}
    \end{overpic}
    \begin{overpic}[width=3.3cm]{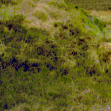}
    \put(3,87){\small \color{white}{\bf b2}}
    \end{overpic}
    \begin{overpic}[width=3.3cm]{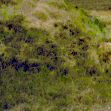}
    \put(3,87){\small \color{white}{\bf b3}}
    \end{overpic}
    \begin{overpic}[width=3.3cm]{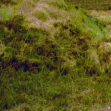}
    \put(3,87){\small \color{white}{\bf b4}}
    \end{overpic}
	\caption{3D transformation}
    \label{fig:scotland}
\end{figure}
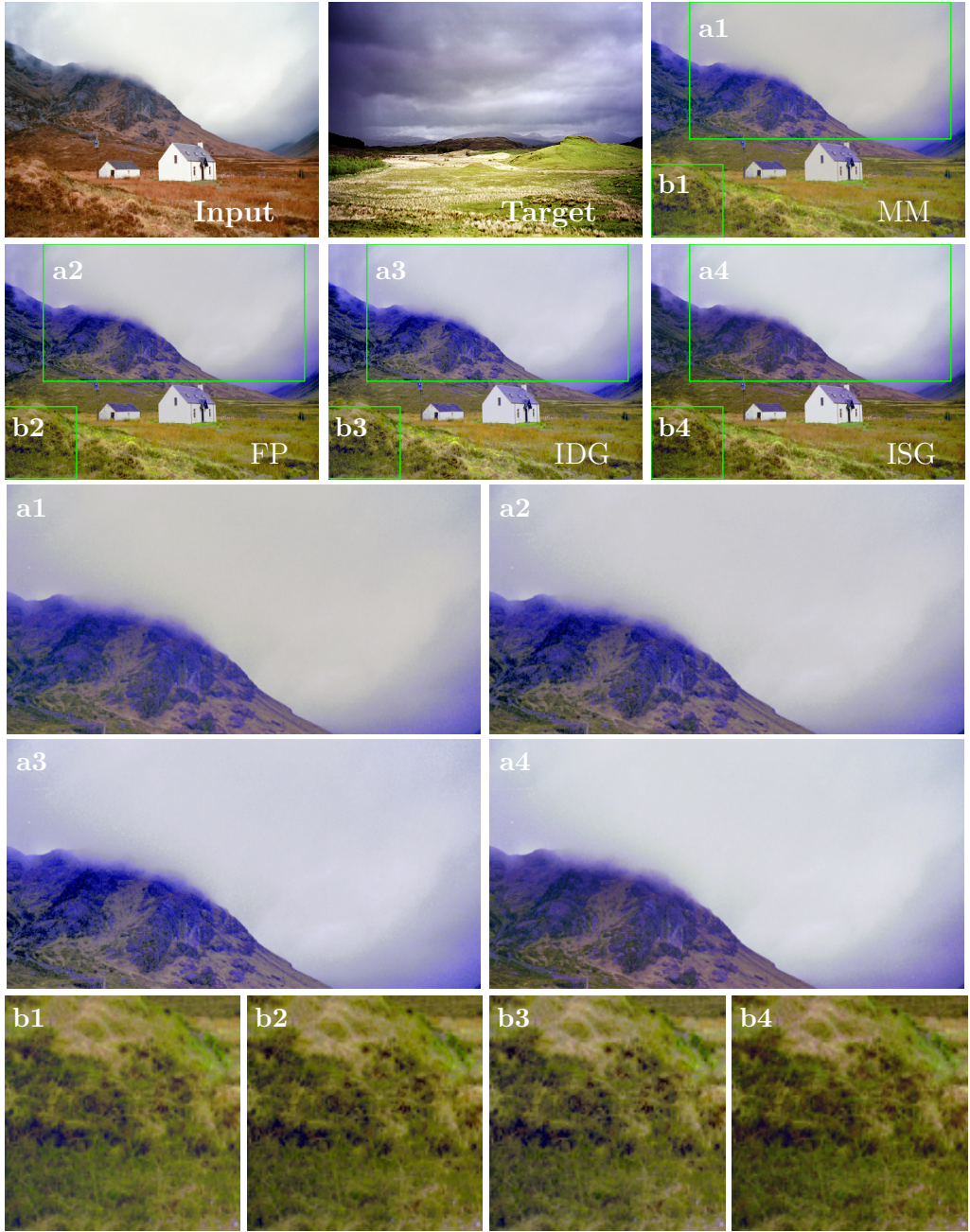

Starting with the 3D rgb case, Figure \ref{fig:scotland} presents the transformed images and some of their details. The detail (a1) clearly shows that the cloud 'drawn' by MM appears too green. Similarly, FP also presents a green appearance, in detail (a2). On the contrary, the two gradient methods, i.e. IDG and ISG methods,  show pure white cloud color in (a3) and (a4). Note also the difference in the amount of blue in the shadows on the grass. Too much blue is mixed with the shadow, in MM's output detail (b1). In details (b2),(b3),(b4), the results of MLE methods lead to a more natural appearance.

From the point of view of the present work, the most interesting aspect of this application is in term of computational time. The recursive (online) ISG method takes about $10$ seconds for two images (input and output). In contrast, FP and IDG each require more than two hours. In other words, ISG has a decisive advantage, in terms of time consumption. 

\begin{figure}[!htbp]
    \centering
    \begin{overpic}[width=4.4cm]{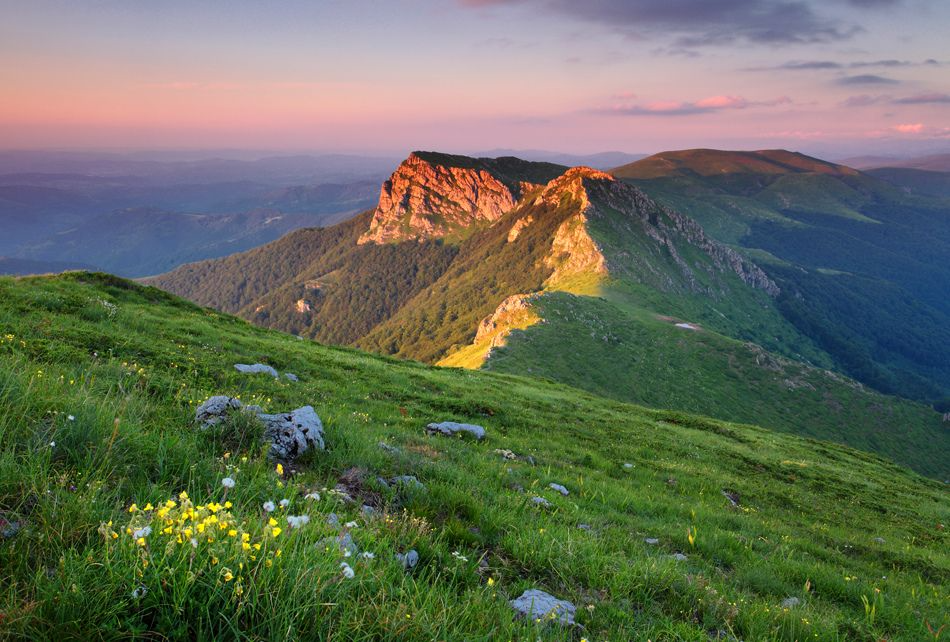}
    \put(5,5){\normalsize \color{white}{\bf Input}}
    \end{overpic}
    \begin{overpic}[width=4.4cm]{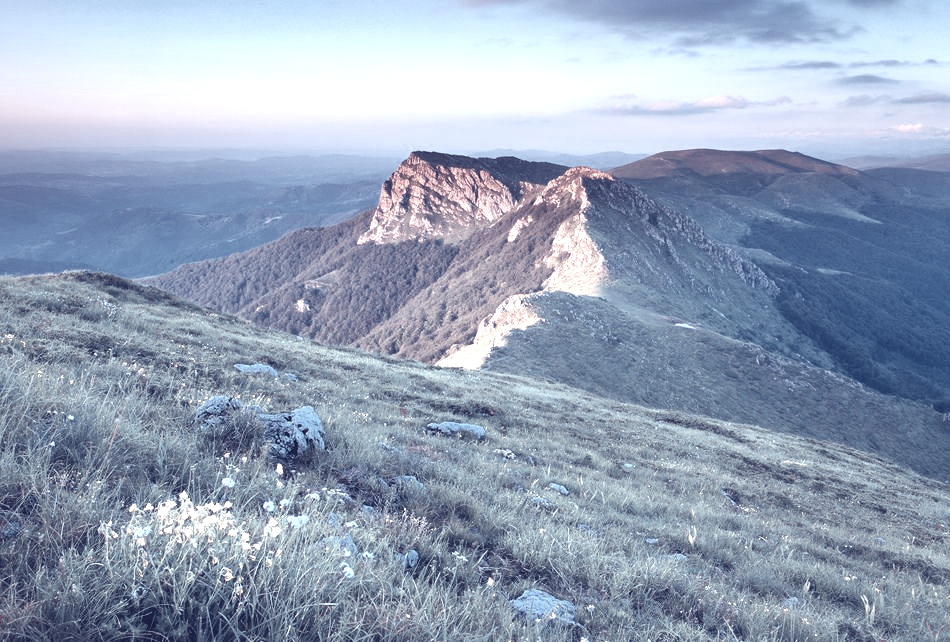}
    \put(5,5){\normalsize \color{white}{\bf MM}}
    \end{overpic}
    \begin{overpic}[width=4.4cm]{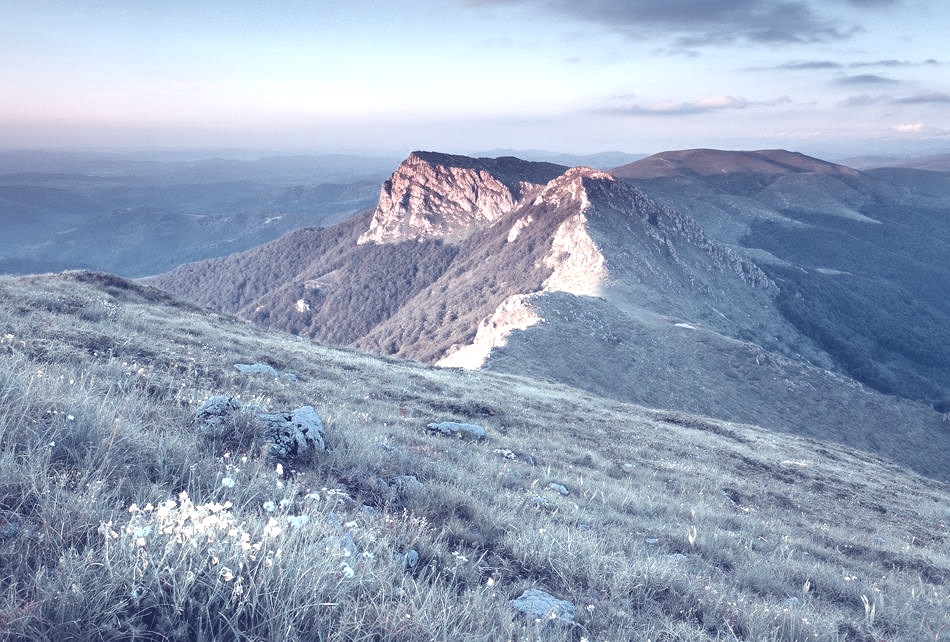}
    \put(5,5){\normalsize \color{white}{\bf FP}}
    \end{overpic}
    
	\vspace*{0.08cm}
	\begin{overpic}[width=4.4cm]{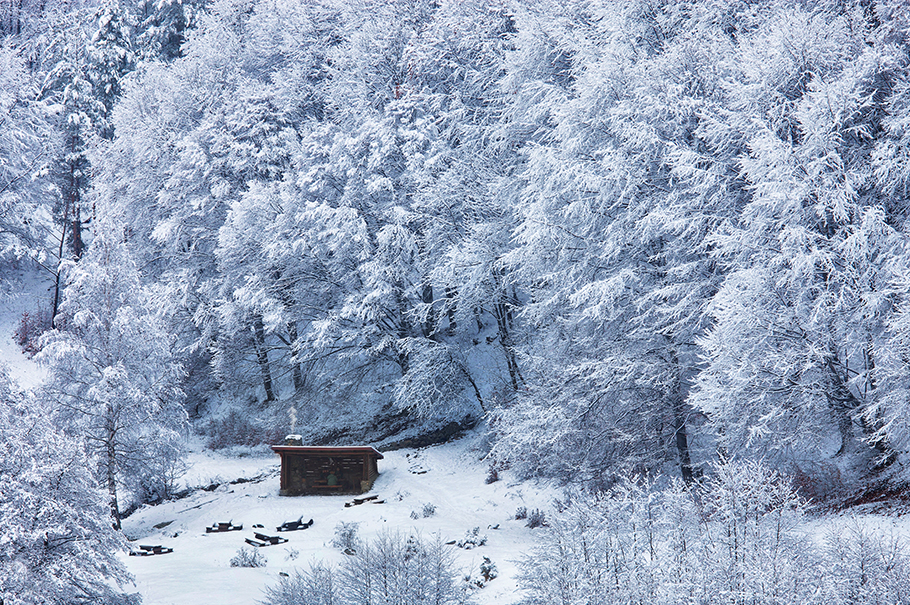}
    \put(5,5){\normalsize \color{white}{\bf Target}}
    \end{overpic}
    \begin{overpic}[width=4.4cm]{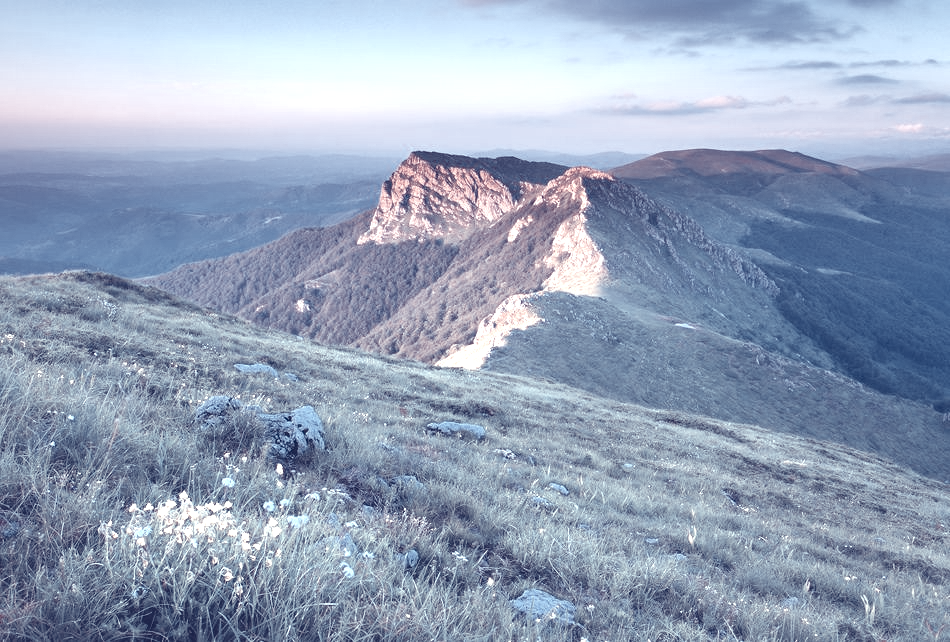}
    \put(5,5){\normalsize \color{white}{\bf IDG}}
    \end{overpic}
    \begin{overpic}[width=4.4cm]{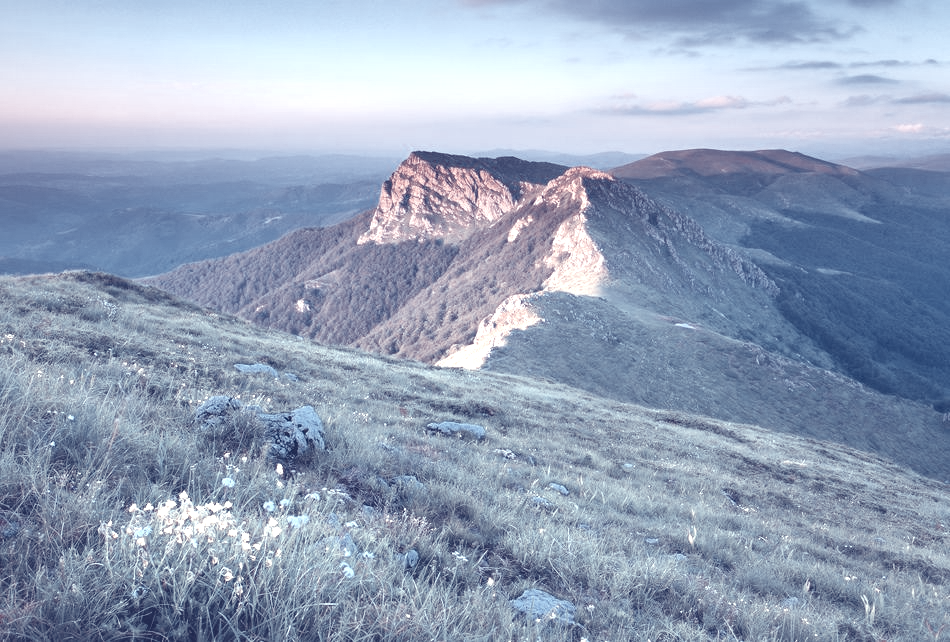}
    \put(5,5){\normalsize \color{white}{\bf ISG}}
    \end{overpic}
    \caption{5D transformation of images that have moderate size}
    \label{fig:mountain}
\end{figure}

Then, gradient-field information was included, so the transformation came to involve 5D, which consist in three color components (of CIELAB) and two components of the image spatial gradient field ($\mathrm{d}x$ and $\mathrm{d}y$). For this application, the shape parameter of the MGGD model was supposed to fixed. Figure \ref{fig:mountain} presents the four different implementations. It can be observed that the output of the three MLE methods is significantly better than that of MM. In the transformed result of MM, the hue is darker and greener. MLE results are better, since the frost on the grass is whiter and appears more natural, and the forest on the mountain in the image also appears darker. The two images in Figure \ref{fig:mountain} have more than $1.2\times 10^6$ pixels (i.e. $1.2\times 10^6$ samples). The FP and IDG need more than $4$ hours to run, on the these two images. The ISG method needs only $21$ seconds.
\begin{figure}[!htbp]
    \centering
    \begin{overpic}[width=4.4cm]{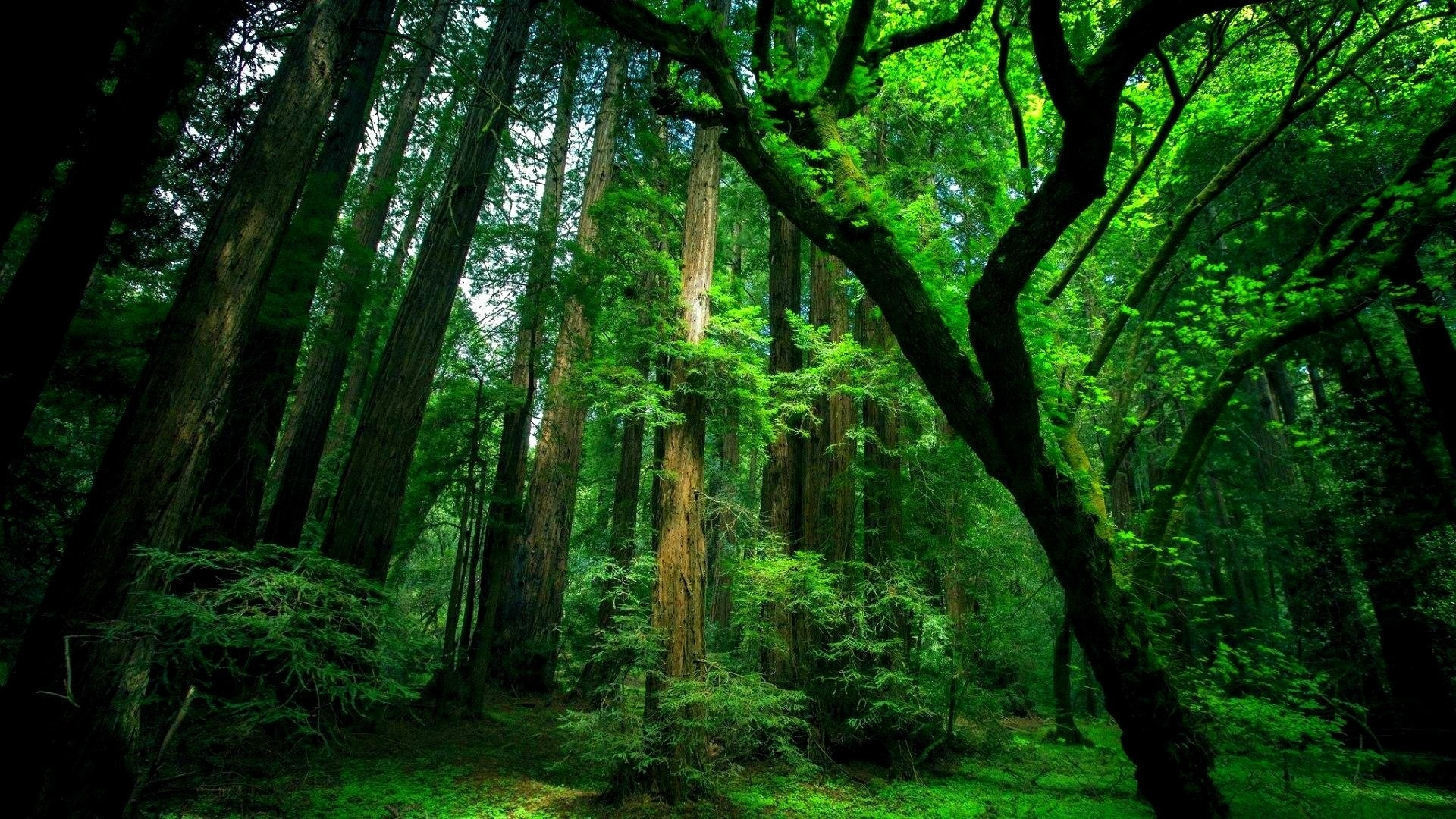}
    \put(5,5){\normalsize \color{white}{\bf Input}}
    \end{overpic}
    \begin{overpic}[width=4.4cm]{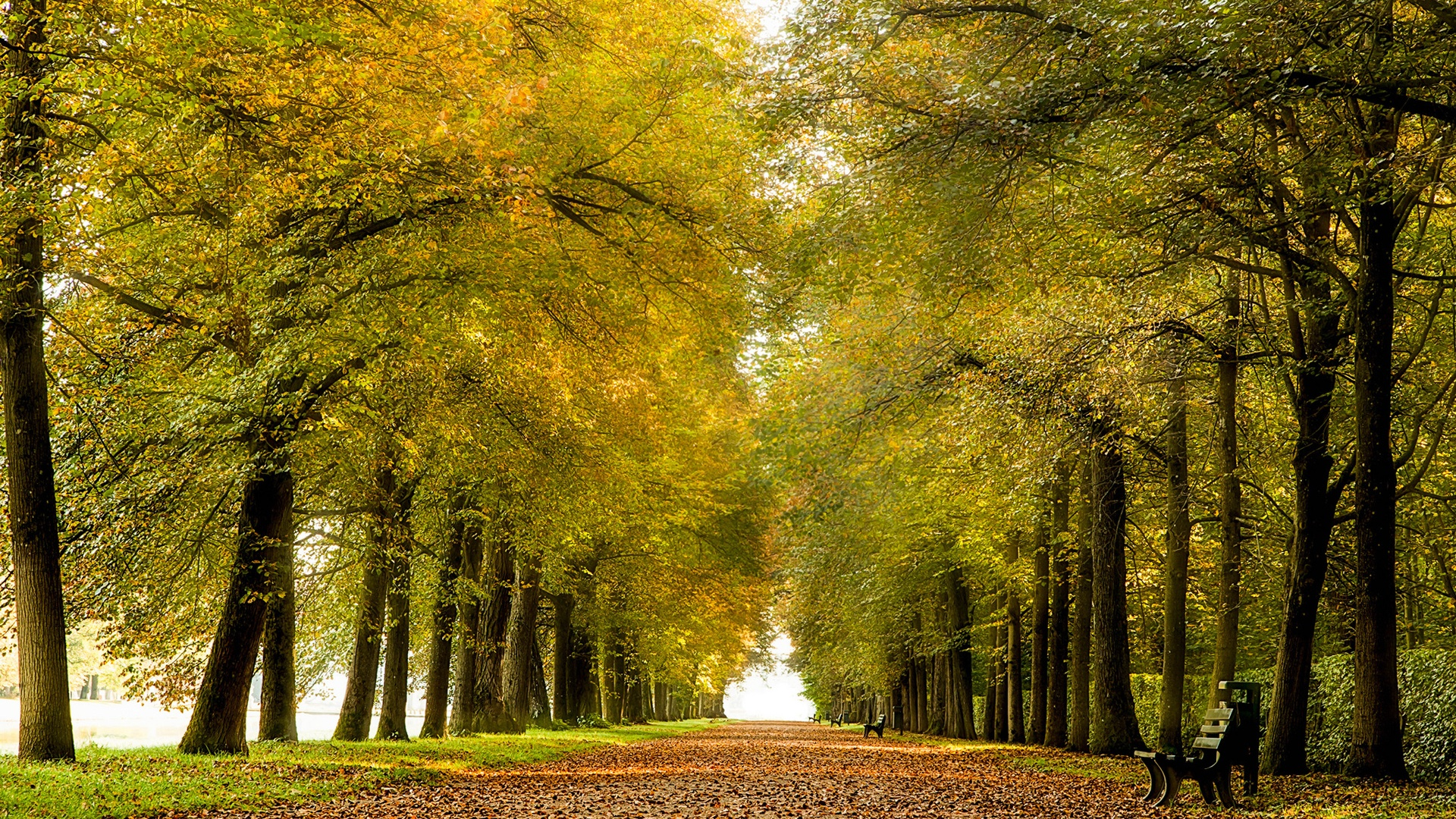}
    \put(5,5){\normalsize \color{white}{\bf Target}}
    \end{overpic}
    \begin{overpic}[width=4.4cm]{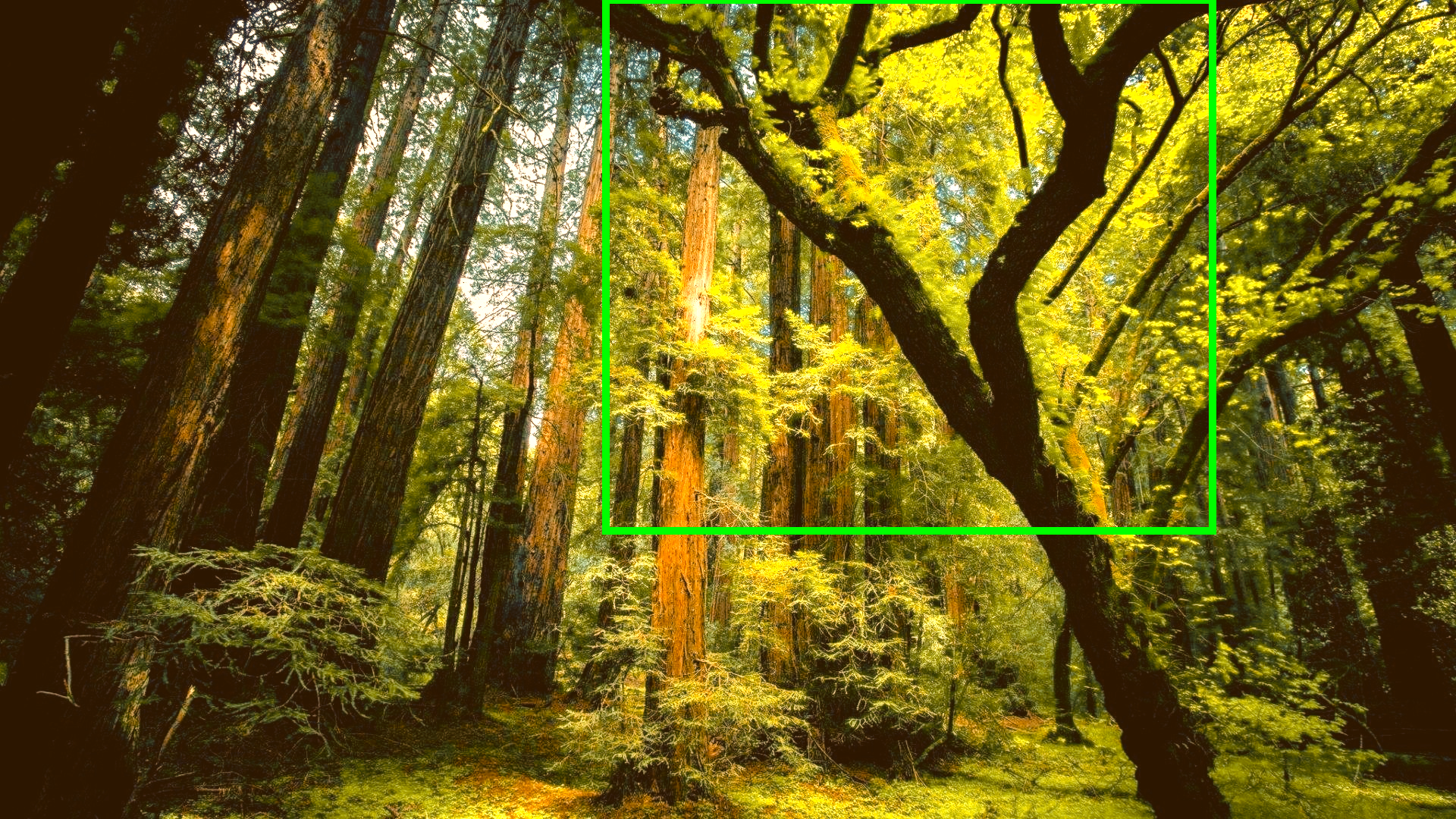}
    \put(5,5){\normalsize \color{white}{\bf MM}}
    \put(45,45){\normalsize \color{white}{\bf c1}}
    \end{overpic}
    
    \vspace*{0.08cm}
    \begin{overpic}[width=4.4cm]{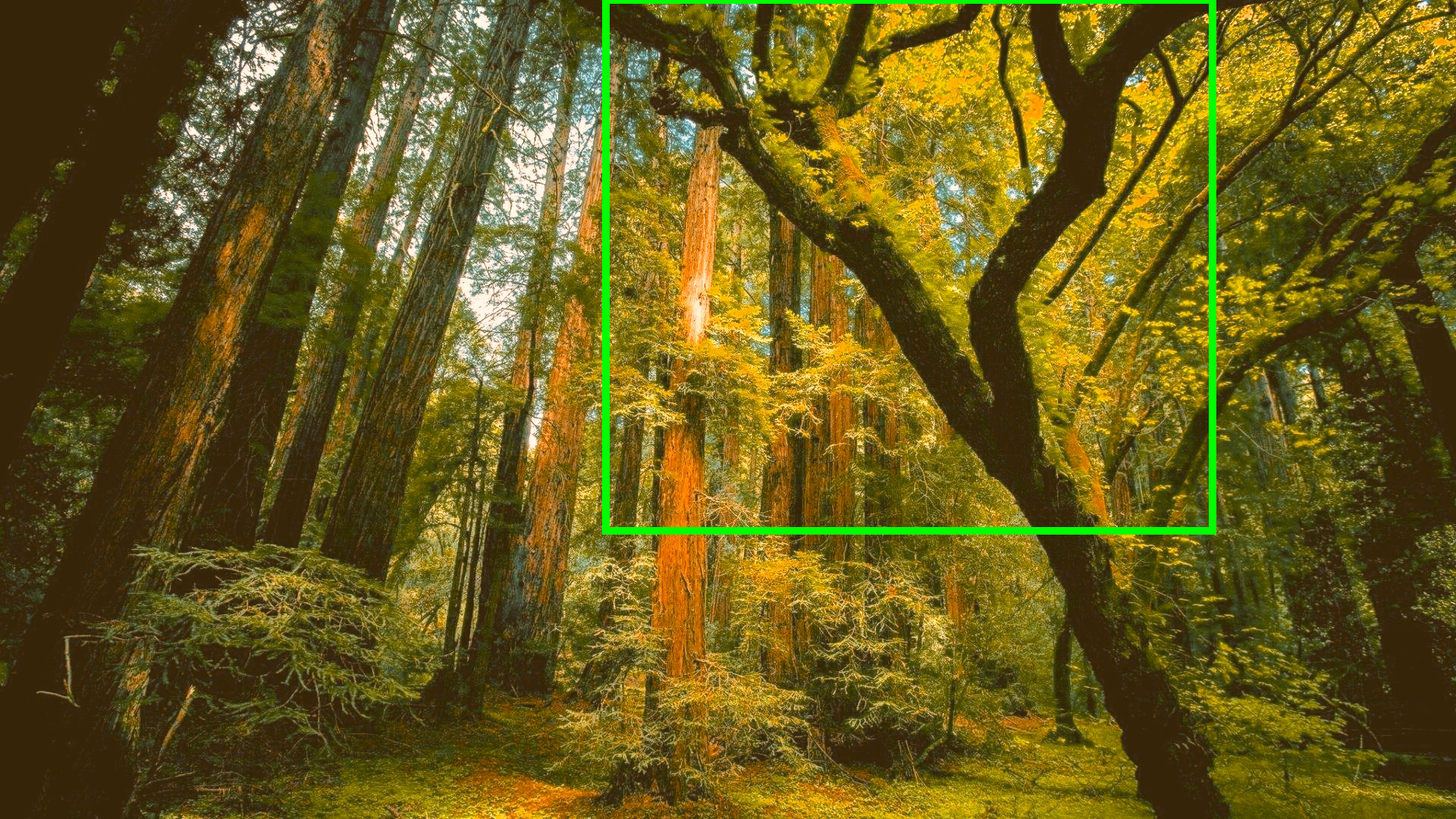}
    \put(5,5){\normalsize \color{white}{\bf FP}}
    \put(45,45){\normalsize \color{white}{\bf c2}}
    \end{overpic}
    \begin{overpic}[width=4.4cm]{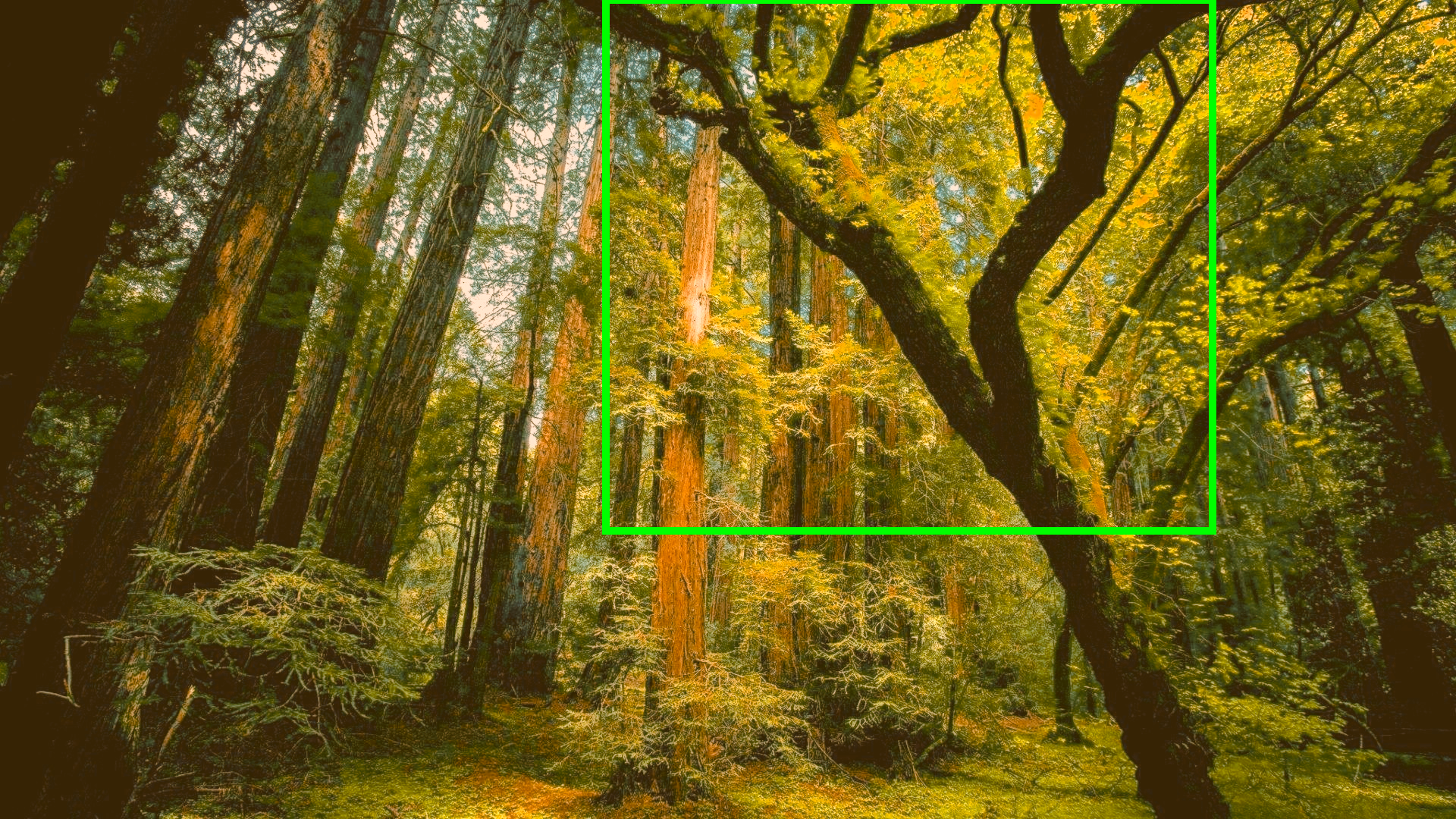}
    \put(5,5){\normalsize \color{white}{\bf IDG}}
    \put(45,45){\normalsize \color{white}{\bf c3}}
    \end{overpic}
    \begin{overpic}[width=4.4cm]{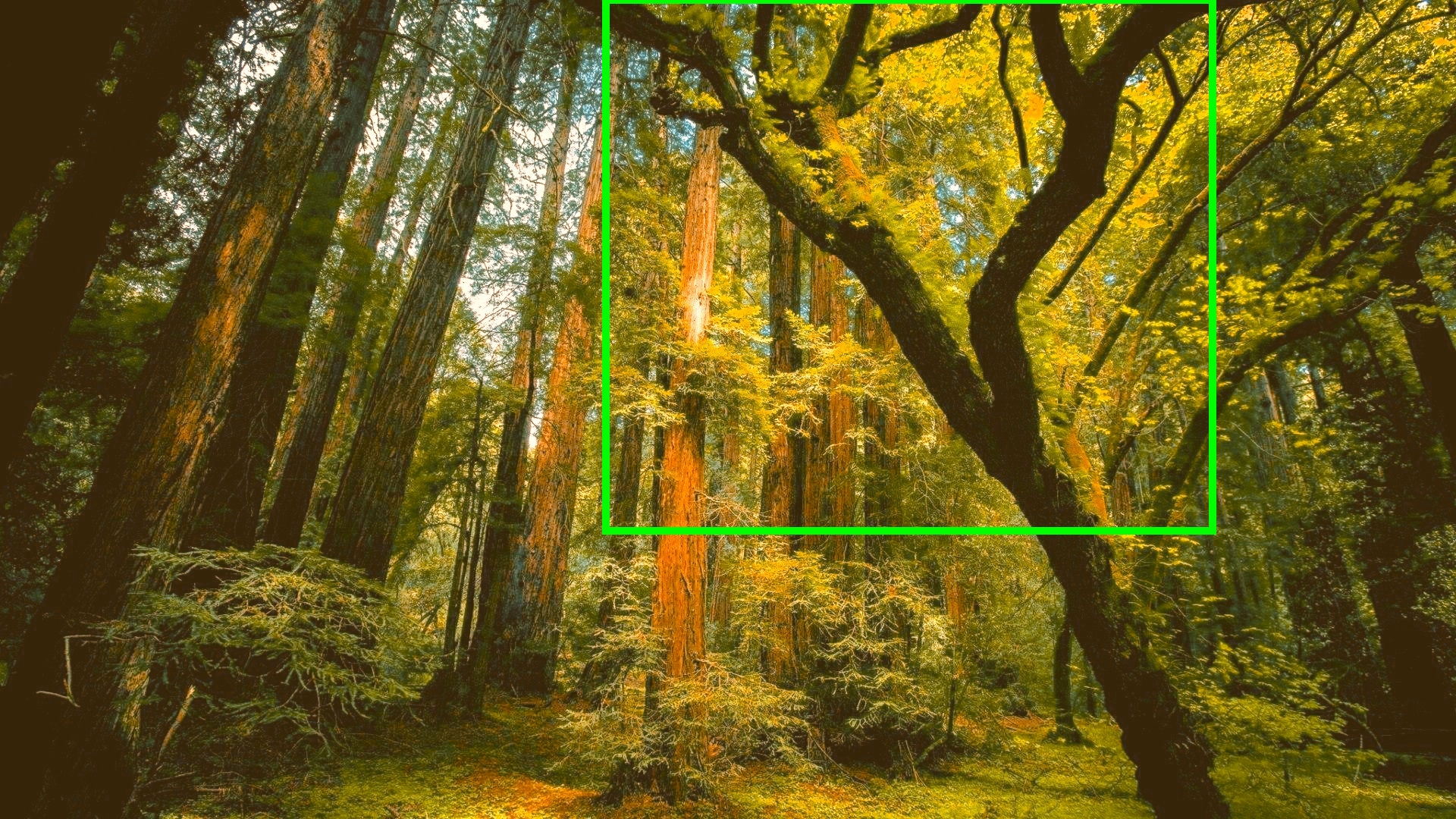}
    \put(5,5){\normalsize \color{white}{\bf ISG}}
    \put(45,45){\normalsize \color{white}{\bf c4}}
    \end{overpic}
    
    \vspace*{0.08cm}
    \begin{overpic}[width=6.65cm]{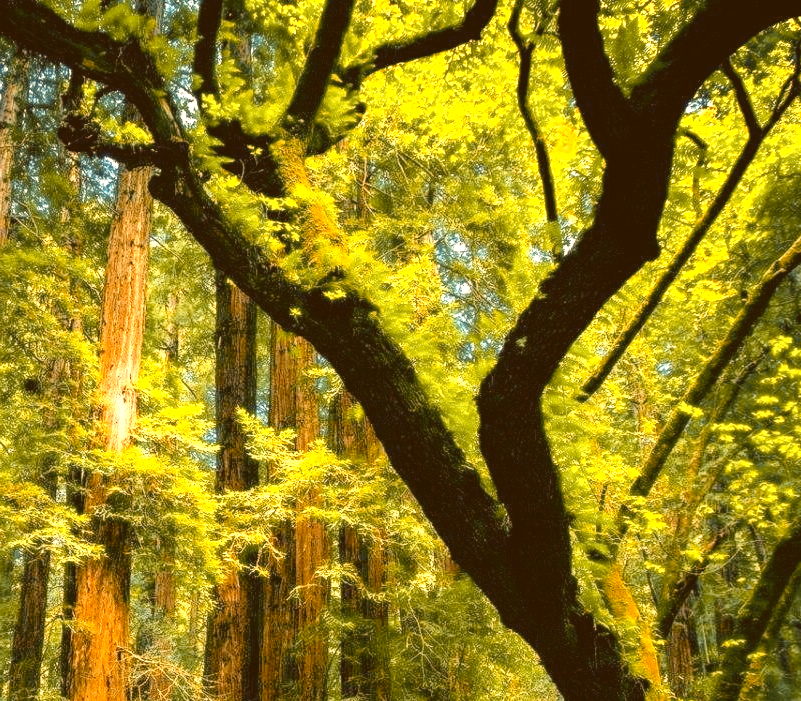}
    \put(5,5){\normalsize \color{white}{\bf c1}}
    \end{overpic}
    \begin{overpic}[width=6.65cm]{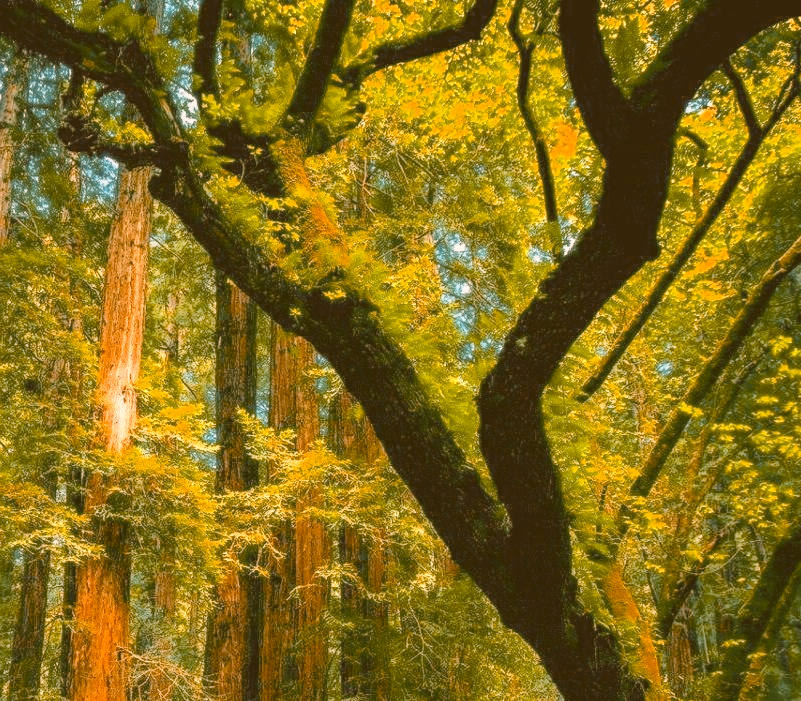}
    \put(5,5){\normalsize \color{white}{\bf c2}}
    \end{overpic}
    
    \vspace*{0.08cm}
    \begin{overpic}[width=6.65cm]{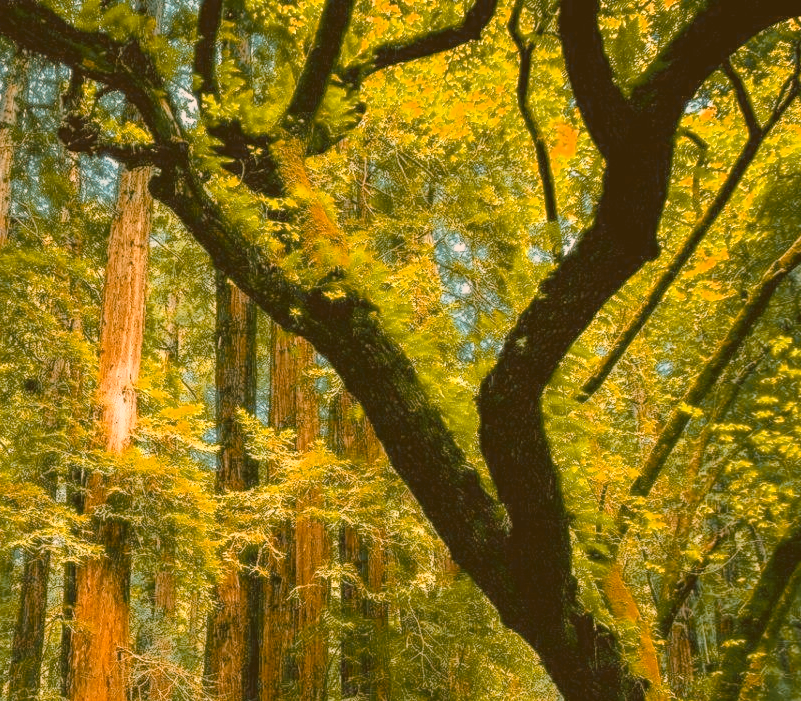}
    \put(5,5){\normalsize \color{white}{\bf c3}}
    \end{overpic}
    \begin{overpic}[width=6.65cm]{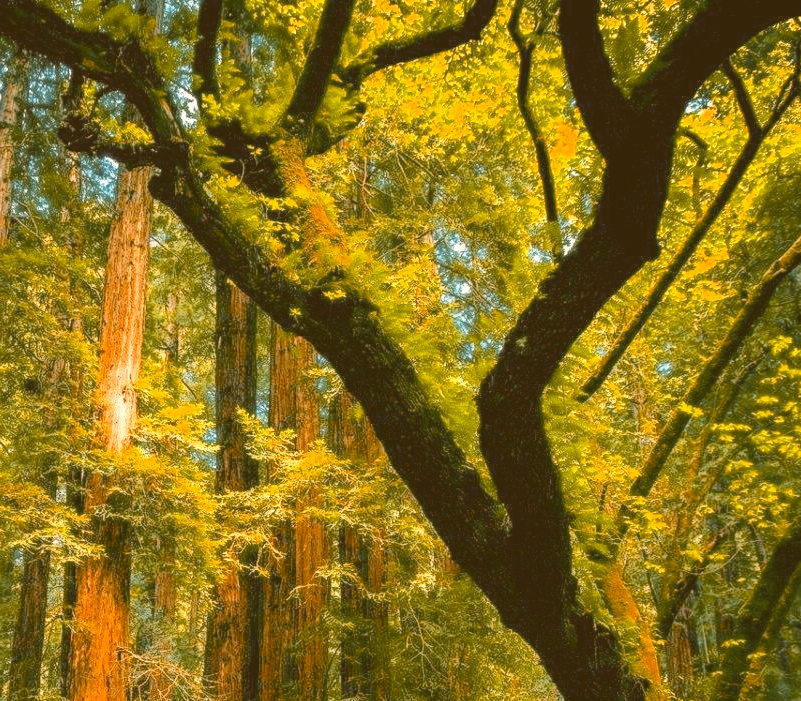}
    \put(5,5){\normalsize \color{white}{\bf c4}}
    \end{overpic}
    \caption{5D transformation of images that have large size}
    \label{fig:forest}
\end{figure}

We also considered an application to full HD images. In this case, as demonstrated in Figure \ref{fig:forest}, the advantages of the ISG algorithm were significant. The result of MM failed to achieve the color of the autumn leaves in the target image, showing cyan instead of yellow. Since the input image and the target image have more than $4 \times 10^6$ pixels (that is $4 \times 10^6$ samples), it was not feasible to run FP and IDG, with the entire dataset. Rather, the estimation was done on subsets of the complete dataset. These two subsets have $4 \times 10^5$ samples, that are randomly taken from the original images. In the autumn leaves obtained using FP and IDG, the yellow color has obviously been smeared. ISG is more natural, in which the yellow color is more uniform, and it is closer to the style of the target image.

\begin{figure}[!htbp]
\centering
\subfigure[FP]{
\label{fig:texclassFP}
\includegraphics[width=12cm]{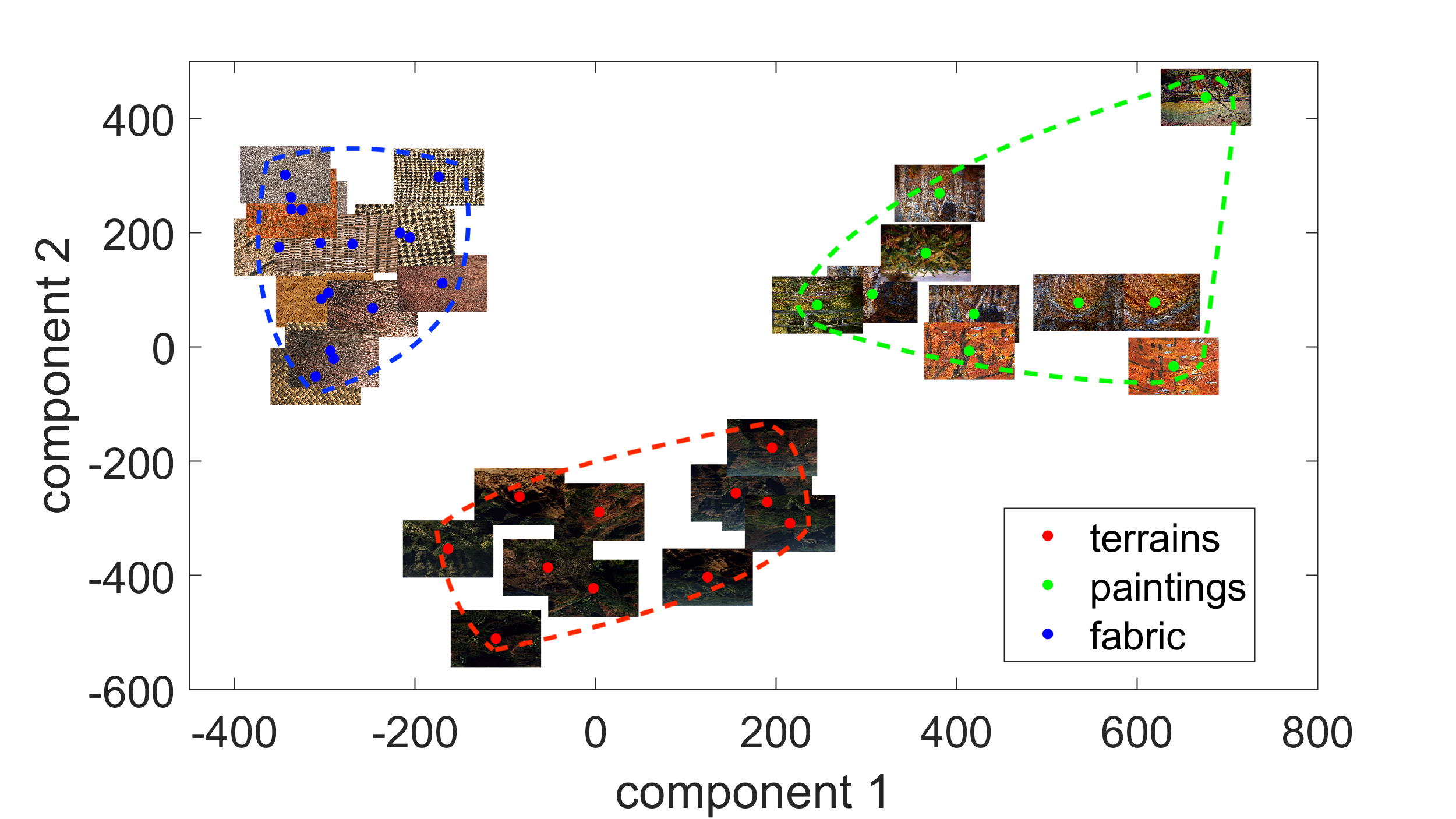}
}
\subfigure[ISG]{
\label{fig:texclassISG}
\includegraphics[width=12cm]{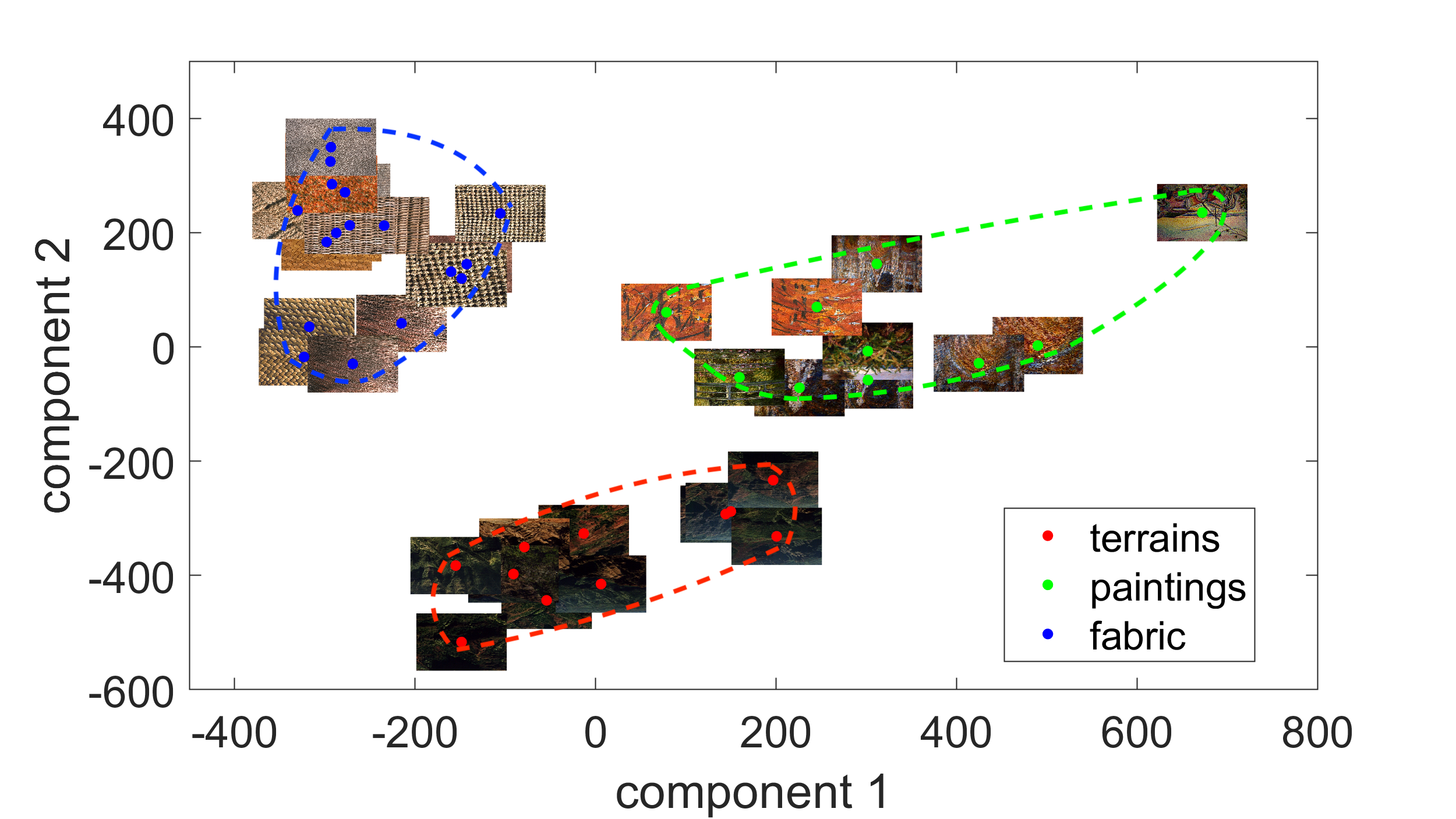}
}
\caption{Texture Classification using MGGD \label{fig:classification}}
\end{figure}

\subsection{Classification}
MGGD are also used for texture modeling~\cite{kwitt2011testing,verdoolaege2011geodesics}. Without going into the details of presently existing classification methods, we attempted to use an MGGD representation, in order to distinguish between different groups of textures. Three groups of textures are selected from the VisTex database~\cite{VisTex}, $11$ paintings, $18$ fabrics, and $11$ terrains. Each texture is considered as an RGB $3$-dimensional image, modeled by an MGGD, whose parameters $\theta = (\mu,\Sigma,\beta)$ are estimated by two MLE methods, i.e. FP and ISG. Then, the scatter matrices $\Sigma$ are normalized by their trace, i.e. $M = \frac{p}{\mathrm{tr}(\Sigma)}\Sigma$ (in order to avoid the elements of $\Sigma$ being too small). Afterwards, for each texture, a new vector is constituted by the eigenvalues of $M$, $\mu$ and $\beta$. This $7$-dimensional vector is projected onto a 2D plane, via a PCA operation. A visual (2D) representation is given in Figure \ref{fig:classification}. These two figures are obtained using FP and ISG, respectively. Each texture contains $512 \times 512 = 262144$ pixels, FP expended $210$ seconds for each image. In stark contrast, ISG expended $4.6$ seconds in average. And for both these methods, the boundary between these $3$ clouds of points is quite sharp, and the distinction is quite clear. We have reason to believe that, in this scenario, ISG has achieved the same performance as FP and simultaneously it used less time.

\section*{Acknowledgement}
At the end, we thank the support from the ANR MAR- GARITA under Grant ANR-17-ASTR-0015 for our works.

\appendix
\section{Proof of proposition \ref{prop:conv_idg}}\label{prove:conv_idg}
Let $(\theta_k)_{k\geq 0}$ be an infinite sequence generated by Algorithm \ref{algo:idg}. Recall the retraction $R_{\theta}$ defined in (\ref{eq:prod_retraction}). Consider the sequence of tangent vectors $(\eta_k)_{k\geq 0}$ where $\eta_k$ belongs to  $T_{\theta_k} \Theta$, and
$\eta_{\scriptscriptstyle 0} = -\nabla_{\scriptscriptstyle \mu} \hat{D}( \mu_{\scriptscriptstyle 0}, \Sigma_{\scriptscriptstyle 0}, \beta_{\scriptscriptstyle 0} )$, $\eta_{\scriptscriptstyle 1} = -\nabla_{\scriptscriptstyle \Sigma} \hat{D}( \mu_{\scriptscriptstyle 1}, \Sigma_{\scriptscriptstyle 0}, \beta_{\scriptscriptstyle 0} )$,
$\eta_{\scriptscriptstyle 2} = - \nabla_{\scriptscriptstyle \beta} \hat{D}( \mu_{\scriptscriptstyle 1}, \Sigma_{\scriptscriptstyle 1}, \beta_{\scriptscriptstyle 0})$, and so on. 

Then, the sequence $(\theta_k)$ is given as in Algorithm 1 of~\cite{absil2009optimization}, $\theta_{k+1} = R_{\theta_{\scriptscriptstyle k}}(t_{\scriptscriptstyle k}\,\eta_{\scriptscriptstyle k})$ with step-size $t_{\scriptscriptstyle k}$ chosen according to Armijo-Goldstein rule (note that $t_{\scriptscriptstyle 0} = \alpha_{\scriptscriptstyle \mu}$, $t_{\scriptscriptstyle 1} = \alpha_{\scriptscriptstyle \Sigma}$, and $t_{\scriptscriptstyle 2} = \alpha_{\scriptscriptstyle \beta}$, \textit{etc.}).

The sequence $(\theta_k)$ remains within the neighborhood $S_0$ of $\theta^*$. Without loss of generality, assume this neighborhood is compact.  
According to Corollary 4.3.2 in \cite{absil2009optimization}, if the sequence $(\eta_k)_{k\geq 0}$ is gradient-related,  
\begin{equation}
\lim_{k \rightarrow \infty} \| \mathrm{grad} \hat{D}(\theta_k) \| = 0
\end{equation}
Then, since $\theta^*$ is the only stationary point of the cost function (\ref{eq:pb_idg}) in $S_0$, it follows that $\lim_{k \rightarrow \infty} \theta_k = \theta^*$, as required. To show that the sequence $(\eta_k)_{k\geq 0}$ is gradient-related, note that 
$$
\left< \eta_{\scriptscriptstyle 0},\nabla_\theta \hat{D}(\theta_{\scriptscriptstyle 0}) \right> = - \| \nabla_{\scriptscriptstyle \mu} \hat{D}( \mu_{\scriptscriptstyle 0}, \Sigma_{\scriptscriptstyle 0}, \beta_{\scriptscriptstyle 0} )\|^2
$$
$$
\left< \eta_{\scriptscriptstyle 1},\nabla_\theta \hat{D}(\theta_{\scriptscriptstyle 1}) \right> = - \| \nabla_{\scriptscriptstyle \Sigma} \hat{D}( \mu_{\scriptscriptstyle 1}, \Sigma_{\scriptscriptstyle 0}, \beta_{\scriptscriptstyle 0} )\|^2
$$
$$
\left< \eta_{\scriptscriptstyle 2},\nabla_\theta \hat{D}(\theta_{\scriptscriptstyle 2}) \right> = - \| \nabla_{\scriptscriptstyle \beta} \hat{D}( \mu_{\scriptscriptstyle 1}, \Sigma_{\scriptscriptstyle 1}, \beta_{\scriptscriptstyle 0} )\|^2
$$
and so on, for $k \geq 3$. In other words, the scalar product between $\eta_k$ and $\nabla_\theta \hat{D}(\theta_{\scriptscriptstyle k})$ is always strictly negative. Therefore,  the sequence $(\eta_k)_{k\geq 0}$ is gradient-related.

\section{Proof of proposition \ref{prop:conv_isg}}\label{prove:conv_isg}
The proof is a direct application of Remark 2, concerning Proposition 1, in~\cite{zhou2019fast}. According to this remark, if $u(\theta_n,x)$ denotes the direction of descent, and if
\begin{equation}\label{eq:assumpt_alt}
\mathbb{E}\left<u(\theta_n,x),\nabla_{\theta}D(\theta_n)\right> < 0, \text{ for } n > 0
\end{equation}
then $\lim \theta_n = \theta^*$ almost surely. Here (compare to the proof of Proposition \ref{prop:conv_idg}), the direction of descent is given by
$u(\theta_{\scriptscriptstyle 0},x) = \nabla_{\scriptscriptstyle \mu} \ell( \mu_{\scriptscriptstyle 0}, \Sigma_{\scriptscriptstyle 0}, \beta_{\scriptscriptstyle 0} )$, $u(\theta_{\scriptscriptstyle 1},x) = \nabla_{\scriptscriptstyle \Sigma} \ell( \mu_{\scriptscriptstyle 1}, \Sigma_{\scriptscriptstyle 0}, \beta_{\scriptscriptstyle 0} )$, $u(\theta_{\scriptscriptstyle 2},x) = \nabla_{\scriptscriptstyle \beta} \ell( \mu_{\scriptscriptstyle 1}, \Sigma_{\scriptscriptstyle 1}, \beta_{\scriptscriptstyle 0} )$ and so on. Therefore, the expectation in (\ref{eq:assumpt_alt}) is equal to
$$
\mathbb{E}\left<u(\theta_{\scriptscriptstyle 0},x),\nabla_{\theta}D(\theta_0)\right>= - \| \nabla_{\scriptscriptstyle \mu} D( \mu_{\scriptscriptstyle 0}, \Sigma_{\scriptscriptstyle 0}, \beta_{\scriptscriptstyle 0} )\|^2
$$
$$
\mathbb{E}\left<u(\theta_{\scriptscriptstyle 1},x),\nabla_{\theta}D(\theta_1)\right>= - \| \nabla_{\scriptscriptstyle \Sigma} D( \mu_{\scriptscriptstyle 1}, \Sigma_{\scriptscriptstyle 0}, \beta_{\scriptscriptstyle 0} )\|^2
$$
$$
\mathbb{E}\left<u(\theta_{\scriptscriptstyle 2},x),\nabla_{\theta}D(\theta_2)\right>= - \| \nabla_{\scriptscriptstyle \beta} D( \mu_{\scriptscriptstyle 1}, \Sigma_{\scriptscriptstyle 1}, \beta_{\scriptscriptstyle 0} )\|^2
$$
and so on, for $k \geq 3$. This shows that (\ref{eq:assumpt_alt}) is verified. 

\section{Proof of propositions \ref{prop:mean_square_rate} and \ref{prop:asymptotic_normality}}\label{prove:msr_&_an}
As for Proposition \ref{prop:conv_isg}, this is an application of Remark 2 in~\cite{zhou2019fast}. According to this remark, in order to obtain the mean-square rate and the asymptotic normality, it is enough to show the mean vector field $X(\theta)=\mathbb{E}_{\theta^*}[u(\theta,x)]$ has an attractive stationary point at $\theta=\theta^*$. Since $u(\theta,x)=\nabla_{\theta} \ell(\theta,x)$
\begin{equation}
\mathbb{E}_{\theta^*} [u(\theta;x)] = \left[
\begin{matrix}
\nabla_{\mu} D(\theta)\\
\nabla_{\Sigma} D(\theta)\\
\nabla_{\beta} D(\theta)
\end{matrix}
\right]
\end{equation}
The covariant derivative of this vector field at the point $\theta=\theta^*$ is equal to the Hessian $\mathcal{H}(\theta^*)$, which is positive definite. Therefore, the results of Propositions \ref{prop:mean_square_rate} and \ref{prop:asymptotic_normality} follow by Remark 2 in~\cite{zhou2019fast}.

\section{Proof of proposition \ref{prop:convex_Sigma}}\label{prove:condition_convex_Sigma}
For the case of $\theta=(\Sigma)$, the geodesic convexity of the cost function $D(\theta)$ (or of $\hat{D}(\theta)$) follows by proving $-\ell(\theta;x)$ is geodesically strictly convex in $\theta = (\Sigma)$ for any $x$.

To do this, for any fixed $x$, denote $g(\theta) = -\ell(\theta;x)$. Recall that, geodesic curves on $\mathcal{P}_m$ are of the form~\cite{pennec2006riemannian}
\begin{equation}
\begin{aligned}
\gamma : & \mathbb{R} & \rightarrow & \hspace*{0.5cm} \mathcal{P}_m \\
 & t & \mapsto & A \exp(t r) A^{\dagger}
\end{aligned}
\end{equation}
where $\exp$ denotes the matrix exponential map, $A$ is an invertible matrix, and $r$ is a diagonal matrix, both of same size as $\Sigma$. Then, $g(\theta)$ is geodesically convex if and only if the composition $(g \circ \gamma)(t)$ is always a convex function with respect to $t$. Moreover, geodesic strict convexity is defined in exactly the same way. The composition $(g \circ \gamma)(t)$ can be expressed 
\begin{equation}\label{eq:g_gamma}
(g\circ \gamma)(t) = \log \det (A) + \mathrm{tr}(r)\frac{t}{2} + \log \left[ (f\circ\varphi)(t) \right]
\end{equation}
where
\begin{equation}\label{eq:varphi_Sigma}
\varphi(t) = \sum_{i=1}^{p} u^2_i \exp(-r_i t) 
\end{equation}
$u=A^{-1}x$ has components $u_i$, and $r_i$ are the diagonal elements of $r$. The function $\varphi:\mathbb{R}\rightarrow\mathbb{R}_+$ is strictly log-convex, because it is the Laplace transform of a positive measure~\cite{shiryayev1984probability}
\begin{equation}
\varphi(t) = \int_{0}^{\infty} \exp(-tx) \mu(\mathrm{d} x)
\end{equation}
where $\mu = \sum_{i=1}^{m} u_i^{2} \delta_{r_i}$, and $\delta_{r_i}$ is the Dirac measure concentrated at $r_i$.

Assume that the function $f$ verifies Condition (\ref{eq:convex_condition_Sigma}). Then, since $\varphi$ is strictly log-convex, $f \circ \varphi$ is strictly log-convex. Thus, the term $\log \left[ (f\circ\varphi)(t) \right]$ of (\ref{eq:g_gamma}) is a
strictly convex function of the real variable $t$. Since the term $\mathrm{tr}(r)\frac{t}{2}$ of (\ref{eq:g_gamma}) amounts to an affine function of $t$, it is now clear that $(g \circ \gamma)(t)$ is a strictly convex function of the real variable $t$, for any geodesic curve $\gamma:\mathbb{R} \rightarrow \mathcal{P}_m$. Finally, since $x$ was chosen arbitrarily, $-\ell(\theta;x)$ is geodesically strictly convex in $\theta = (\Sigma)$ for each $x$. Therefore,  $D(\theta)$ and $\hat{D}(\theta)$ are both geodesically strictly convex.

\section{Proof of corollary \ref{coro:mggd_stud_convex_Sigma} and \ref{coro:mggd_stud_convex_mS}}
For the case of $\theta=(\Sigma)$, note that $\varphi : \mathbb{R} \rightarrow \mathbb{R}_+$ is strictly log-convex if and only if $\varphi(t) = \exp( \psi(t) )$ where $\psi:\mathbb{R} \rightarrow \mathbb{R}$ is strictly convex.

\noindent
1) plugging (\ref{eq:fct_f_mggd}) into (\ref{eq:convex_condition_Sigma}),
\begin{equation}\label{eq:f_phi_mggd}
\log (f \circ \varphi) (t) = \frac{1}{2} \exp \left( \beta \left( \psi(t) \right) \right)
\end{equation}
Therefore, condition (\ref{eq:convex_condition_Sigma}) is verified since $\beta>0$.

\noindent
2) plugging (\ref{eq:fct_f_student}) into (\ref{eq:convex_condition_Sigma}),
\begin{equation}\label{eq:f_phi_mvt}
\log \left( f \circ \varphi \right)(t) = \frac{\beta+m}{2} \log \left( 1+\frac{\exp(\psi(t))}{\beta} \right) 
\end{equation}
Therefore, condition (\ref{eq:convex_condition_Sigma}) is verified since $\beta+m>0$.

For the case of $\theta=(\mu,\Sigma)$, as mentioned above, the function $\tilde{f}$ is reformulated. Then, the same strategy is applied for this reformulated $\tilde{f}$.

\noindent
1) For MGGD, recall the geodesic curve for reformulated matrix $S(t)$,
\begin{equation}
S(t) = B \exp(s t) B^{\dagger}
\end{equation}
where $\exp$ denotes the matrix exponential map, $B$ is an invertible matrix, and $s$ is a diagonal matrix, both of same size as $S$.
\begin{equation}
\delta_y(t) = y^{\dagger} S^{-1} y = \sum_{i=1}^{p+1} v_i^2  e^{-s_i t} \qquad \text{ with } v = B^{-1} y
\end{equation}
According to equation (\ref{eq:delta_y_delta_x}), we have $\delta_y > 1$. Therefore, $\exists w \in \mathbb{R}^{p+1}$ and $\exists q \in (0,+\infty)^{p+1}$ (e.g. $w = (u, 0)$ and $q = (r,1)$ ) such that
\begin{equation}
\sum_{i=1}^{p+1} v_i^2  e^{-s_i t} = \sum_{i=1}^{p+1} w_i^2  e^{-q_i t} + 1
\end{equation}
Plugging $\sum_{i=1}^{p+1} w_i^2  e^{-q_i t} + 1$ into the reformulated $\tilde{f}$
\begin{equation}
\tilde{f} \circ \delta_y (t) = \exp \left\{ \frac{1}{2} \left( \sum_{i=1}^{p+1} w_i^2  e^{-q_i t} \right)^{\beta} \right\}
\end{equation}
This function is proved to be log-convex in equation (\ref{eq:varphi_Sigma}). Therefore, condition (\ref{eq:convex_condition_Sigma}) is verified since $\beta > 0$ for MGGD model.

\noindent 2) For Student-T, plugging (\ref{eq:reformed_f_mS_stud}) into (\ref{eq:convex_condition_Sigma}),
\begin{equation}
\log \left( \tilde{f} \circ \varphi \right)(t) = \frac{\beta + m}{2} \left[ 1 - \frac{1}{\beta} + \frac{1}{\beta} \exp(\psi(t)) \right]
\end{equation}
Therefore, condition (\ref{eq:convex_condition_Sigma}) is verified since $\beta > 0$.

\bibliographystyle{elsarticle-num-names} 
\bibliography{ref}

\begin{thebibliography}{40}
\expandafter\ifx\csname natexlab\endcsname\relax\def\natexlab#1{#1}\fi
\providecommand{\url}[1]{\texttt{#1}}
\providecommand{\href}[2]{#2}
\providecommand{\path}[1]{#1}
\providecommand{\DOIprefix}{doi:}
\providecommand{\ArXivprefix}{arXiv:}
\providecommand{\URLprefix}{URL: }
\providecommand{\Pubmedprefix}{pmid:}
\providecommand{\doi}[1]{\href{http://dx.doi.org/#1}{\path{#1}}}
\providecommand{\Pubmed}[1]{\href{pmid:#1}{\path{#1}}}
\providecommand{\bibinfo}[2]{#2}
\ifx\xfnm\relax \def\xfnm[#1]{\unskip,\space#1}\fi
\bibitem[{Kelker(1970)}]{kelker1970distribution}
\bibinfo{author}{D.~Kelker},
\newblock \bibinfo{title}{Distribution theory of spherical distributions and a
  location-scale parameter generalization},
\newblock \bibinfo{journal}{Sankhy{\=a}: The Indian Journal of Statistics,
  Series A}  (\bibinfo{year}{1970}) \bibinfo{pages}{419--430}.
\bibitem[{Fang and Zhang(1990)}]{fang1990generalized}
\bibinfo{author}{K.~Fang}, \bibinfo{author}{Y.~Zhang},
  \bibinfo{title}{Generalized multivariate analysis},
  \bibinfo{publisher}{Science Press}, \bibinfo{year}{1990}.
\bibitem[{Fang(2018)}]{fang2018symmetric}
\bibinfo{author}{K.~W. Fang}, \bibinfo{title}{Symmetric multivariate and
  related distributions}, \bibinfo{publisher}{CRC Press}, \bibinfo{year}{2018}.
\bibitem[{G{\'o}mez et~al.(1998)G{\'o}mez, Gomez-Viilegas, and
  Marin}]{gomez1998multivariate}
\bibinfo{author}{E.~G{\'o}mez}, \bibinfo{author}{M.~Gomez-Viilegas},
  \bibinfo{author}{J.~Marin},
\newblock \bibinfo{title}{A multivariate generalization of the power
  exponential family of distributions},
\newblock \bibinfo{journal}{Communications in Statistics-Theory and Methods}
  \bibinfo{volume}{27} (\bibinfo{year}{1998}) \bibinfo{pages}{589--600}.
\bibitem[{S{\'a}nchez-Manzano et~al.(2002)S{\'a}nchez-Manzano, Gomez-Villegas,
  and Mar{\'\i}n-Diazaraque}]{sanchez2002matrix}
\bibinfo{author}{E.~G. S{\'a}nchez-Manzano}, \bibinfo{author}{M.~A.
  Gomez-Villegas}, \bibinfo{author}{J.-M. Mar{\'\i}n-Diazaraque},
\newblock \bibinfo{title}{A matrix variate generalization of the power
  exponential family of distributions},
\newblock \bibinfo{journal}{Communications in Statistics-Theory and Methods}
  \bibinfo{volume}{31} (\bibinfo{year}{2002}) \bibinfo{pages}{2167--2182}.
\bibitem[{Kotz and Nadarajah(2004)}]{kotz2004multivariate}
\bibinfo{author}{S.~Kotz}, \bibinfo{author}{S.~Nadarajah},
  \bibinfo{title}{Multivariate t-distributions and their applications},
  \bibinfo{publisher}{Cambridge University Press}, \bibinfo{year}{2004}.
\bibitem[{Boubchir and Fadili(2005)}]{boubchir2005multivariate}
\bibinfo{author}{L.~Boubchir}, \bibinfo{author}{J.~M. Fadili},
\newblock \bibinfo{title}{Multivariate statistical modeling of images with the
  curvelet transform},
\newblock in: \bibinfo{booktitle}{Proceedings of the Eighth International
  Symposium on Signal Processing and Its Applications},
  volume~\bibinfo{volume}{2}, \bibinfo{organization}{IEEE},
  \bibinfo{year}{2005}, pp. \bibinfo{pages}{747--750}.
\bibitem[{Cho et~al.(2009)Cho, Bui, and Chen}]{cho2009image}
\bibinfo{author}{D.~Cho}, \bibinfo{author}{T.~D. Bui},
  \bibinfo{author}{G.~Chen},
\newblock \bibinfo{title}{Image denoising based on wavelet shrinkage using
  neighbor and level dependency},
\newblock \bibinfo{journal}{International journal of wavelets, multiresolution
  and information processing} \bibinfo{volume}{7} (\bibinfo{year}{2009})
  \bibinfo{pages}{299--311}.
\bibitem[{Verdoolaege et~al.(2008)Verdoolaege, De~Backer, and
  Scheunders}]{verdoolaege2008multiscale}
\bibinfo{author}{G.~Verdoolaege}, \bibinfo{author}{S.~De~Backer},
  \bibinfo{author}{P.~Scheunders},
\newblock \bibinfo{title}{Multiscale colour texture retrieval using the
  geodesic distance between multivariate generalized gaussian models},
\newblock in: \bibinfo{booktitle}{2008 15th IEEE International Conference on
  Image Processing}, \bibinfo{organization}{IEEE}, \bibinfo{year}{2008}, pp.
  \bibinfo{pages}{169--172}.
\bibitem[{Bazi et~al.(2007)Bazi, Bruzzone, and Melgani}]{bazi2007image}
\bibinfo{author}{Y.~Bazi}, \bibinfo{author}{L.~Bruzzone},
  \bibinfo{author}{F.~Melgani},
\newblock \bibinfo{title}{Image thresholding based on the em algorithm and the
  generalized gaussian distribution},
\newblock \bibinfo{journal}{Pattern Recognition} \bibinfo{volume}{40}
  (\bibinfo{year}{2007}) \bibinfo{pages}{619--634}.
\bibitem[{Scharcanski(2006)}]{scharcanski2006wavelet}
\bibinfo{author}{J.~Scharcanski},
\newblock \bibinfo{title}{A wavelet-based approach for analyzing industrial
  stochastic textures with applications},
\newblock \bibinfo{journal}{IEEE Transactions on Systems, Man, and
  Cybernetics-Part A: Systems and Humans} \bibinfo{volume}{37}
  (\bibinfo{year}{2006}) \bibinfo{pages}{10--22}.
\bibitem[{Gupta et~al.(2018)Gupta, Moorthy, Soundararajan, and
  Bovik}]{GUPTA201887}
\bibinfo{author}{P.~Gupta}, \bibinfo{author}{A.~K. Moorthy},
  \bibinfo{author}{R.~Soundararajan}, \bibinfo{author}{A.~C. Bovik},
\newblock \bibinfo{title}{Generalized gaussian scale mixtures: A model for
  wavelet coefficients of natural images},
\newblock \bibinfo{journal}{Signal Processing: Image Communication}
  \bibinfo{volume}{66} (\bibinfo{year}{2018}) \bibinfo{pages}{87 -- 94}.
\bibitem[{Desai and Mangoubi(2003)}]{desai2003robust}
\bibinfo{author}{M.~N. Desai}, \bibinfo{author}{R.~S. Mangoubi},
\newblock \bibinfo{title}{Robust gaussian and non-gaussian matched subspace
  detection},
\newblock \bibinfo{journal}{IEEE Transactions on Signal Processing}
  \bibinfo{volume}{51} (\bibinfo{year}{2003}) \bibinfo{pages}{3115--3127}.
\bibitem[{Elguebaly and Bouguila(2010)}]{elguebaly2010bayesian}
\bibinfo{author}{T.~Elguebaly}, \bibinfo{author}{N.~Bouguila},
\newblock \bibinfo{title}{Bayesian learning of generalized gaussian mixture
  models on biomedical images},
\newblock in: \bibinfo{booktitle}{IAPR Workshop on Artificial Neural Networks
  in Pattern Recognition}, \bibinfo{organization}{Springer},
  \bibinfo{year}{2010}, pp. \bibinfo{pages}{207--218}.
\bibitem[{Laus and Steidl(2019)}]{laus2019multivariate}
\bibinfo{author}{F.~Laus}, \bibinfo{author}{G.~Steidl},
\newblock \bibinfo{title}{Multivariate myriad filters based on parameter
  estimation of student-t distributions},
\newblock \bibinfo{journal}{SIAM Journal on Imaging Sciences}
  \bibinfo{volume}{12} (\bibinfo{year}{2019}) \bibinfo{pages}{1864--1904}.
\bibitem[{{Bombrun} et~al.(2011){Bombrun}, {Vasile}, {Gay}, and
  {Totir}}]{5570982}
\bibinfo{author}{L.~{Bombrun}}, \bibinfo{author}{G.~{Vasile}},
  \bibinfo{author}{M.~{Gay}}, \bibinfo{author}{F.~{Totir}},
\newblock \bibinfo{title}{Hierarchical segmentation of polarimetric sar images
  using heterogeneous clutter models},
\newblock \bibinfo{journal}{IEEE Transactions on Geoscience and Remote Sensing}
  \bibinfo{volume}{49} (\bibinfo{year}{2011}) \bibinfo{pages}{726--737}.
\bibitem[{{Fernández-Michelli} et~al.(2017){Fernández-Michelli}, {Hurtado},
  {Areta}, and {Muravchik}}]{7887730}
\bibinfo{author}{J.~I. {Fernández-Michelli}}, \bibinfo{author}{M.~{Hurtado}},
  \bibinfo{author}{J.~A. {Areta}}, \bibinfo{author}{C.~H. {Muravchik}},
\newblock \bibinfo{title}{Unsupervised polarimetric sar image classification
  using $\mathcal {G}_{p}^{0}$ mixture model},
\newblock \bibinfo{journal}{IEEE Geoscience and Remote Sensing Letters}
  \bibinfo{volume}{14} (\bibinfo{year}{2017}) \bibinfo{pages}{754--758}.
\bibitem[{{Chen} et~al.(2020){Chen}, {Yang}, {Li}, and {Liu}}]{8930947}
\bibinfo{author}{Q.~{Chen}}, \bibinfo{author}{H.~{Yang}},
  \bibinfo{author}{L.~{Li}}, \bibinfo{author}{X.~{Liu}},
\newblock \bibinfo{title}{A novel statistical texture feature for sar building
  damage assessment in different polarization modes},
\newblock \bibinfo{journal}{IEEE Journal of Selected Topics in Applied Earth
  Observations and Remote Sensing} \bibinfo{volume}{13} (\bibinfo{year}{2020})
  \bibinfo{pages}{154--165}.
\bibitem[{Tyler(1987)}]{tyler1987}
\bibinfo{author}{D.~E. Tyler},
\newblock \bibinfo{title}{A distribution-free m-estimator of multivariate
  scatter},
\newblock \bibinfo{journal}{The annals of Statistics}  (\bibinfo{year}{1987})
  \bibinfo{pages}{234--251}.
\bibitem[{{Ollila} et~al.(2012){Ollila}, {Tyler}, {Koivunen}, and
  {Poor}}]{6263313}
\bibinfo{author}{E.~{Ollila}}, \bibinfo{author}{D.~E. {Tyler}},
  \bibinfo{author}{V.~{Koivunen}}, \bibinfo{author}{H.~V. {Poor}},
\newblock \bibinfo{title}{Complex elliptically symmetric distributions: Survey,
  new results and applications},
\newblock \bibinfo{journal}{IEEE Transactions on Signal Processing}
  \bibinfo{volume}{60} (\bibinfo{year}{2012}) \bibinfo{pages}{5597--5625}.
\bibitem[{Sra and Hosseini(2013)}]{sra2013geometric}
\bibinfo{author}{S.~Sra}, \bibinfo{author}{R.~Hosseini},
\newblock \bibinfo{title}{Geometric optimisation on positive definite matrices
  for elliptically contoured distributions},
\newblock in: \bibinfo{booktitle}{Advances in Neural Information Processing
  Systems}, \bibinfo{year}{2013}, pp. \bibinfo{pages}{2562--2570}.
\bibitem[{Zhang et~al.(2013)Zhang, Wiesel, and Greco}]{zhang2013multivariate}
\bibinfo{author}{T.~Zhang}, \bibinfo{author}{A.~Wiesel}, \bibinfo{author}{M.~S.
  Greco},
\newblock \bibinfo{title}{Multivariate generalized gaussian distribution:
  Convexity and graphical models},
\newblock \bibinfo{journal}{IEEE Transactions on Signal Processing}
  \bibinfo{volume}{61} (\bibinfo{year}{2013}) \bibinfo{pages}{4141--4148}.
\bibitem[{Sra and Hosseini(2015)}]{sra2015conic}
\bibinfo{author}{S.~Sra}, \bibinfo{author}{R.~Hosseini},
\newblock \bibinfo{title}{Conic geometric optimization on the manifold of
  positive definite matrices},
\newblock \bibinfo{journal}{SIAM Journal on Optimization} \bibinfo{volume}{25}
  (\bibinfo{year}{2015}) \bibinfo{pages}{713--739}.
\bibitem[{Pascal et~al.(2013)Pascal, Bombrun, Tourneret, and
  Berthoumieu}]{pascal2013parameter}
\bibinfo{author}{F.~Pascal}, \bibinfo{author}{L.~Bombrun},
  \bibinfo{author}{J.-Y. Tourneret}, \bibinfo{author}{Y.~Berthoumieu},
\newblock \bibinfo{title}{Parameter estimation for multivariate generalized
  gaussian distributions},
\newblock \bibinfo{journal}{IEEE Transactions on Signal Processing}
  \bibinfo{volume}{61} (\bibinfo{year}{2013}) \bibinfo{pages}{5960--5971}.
\bibitem[{Verdoolaege and Scheunders(2011)}]{verdoolaege2011geodesics}
\bibinfo{author}{G.~Verdoolaege}, \bibinfo{author}{P.~Scheunders},
\newblock \bibinfo{title}{Geodesics on the manifold of multivariate generalized
  gaussian distributions with an application to multicomponent texture
  discrimination},
\newblock \bibinfo{journal}{International Journal of Computer Vision}
  \bibinfo{volume}{95} (\bibinfo{year}{2011}) \bibinfo{pages}{265}.
\bibitem[{Bonnabel(2013)}]{bonnabel2013stochastic}
\bibinfo{author}{S.~Bonnabel},
\newblock \bibinfo{title}{Stochastic gradient descent on riemannian manifolds},
\newblock \bibinfo{journal}{IEEE Transactions on Automatic Control}
  \bibinfo{volume}{58} (\bibinfo{year}{2013}) \bibinfo{pages}{2217--2229}.
\bibitem[{Tripuraneni et~al.(2018)Tripuraneni, Flammarion, Bach, and
  Jordan}]{tripuraneni2018averaging}
\bibinfo{author}{N.~Tripuraneni}, \bibinfo{author}{N.~Flammarion},
  \bibinfo{author}{F.~Bach}, \bibinfo{author}{M.~I. Jordan},
\newblock \bibinfo{title}{Averaging stochastic gradient descent on riemannian
  manifolds},
\newblock \bibinfo{journal}{arXiv preprint arXiv:1802.09128}
  (\bibinfo{year}{2018}).
\bibitem[{Zhou and Said(2019)}]{zhou2019fast}
\bibinfo{author}{J.~Zhou}, \bibinfo{author}{S.~Said},
\newblock \bibinfo{title}{Fast, asymptotically efficient, recursive estimation
  in a riemannian manifold},
\newblock \bibinfo{journal}{Entropy} \bibinfo{volume}{21}
  (\bibinfo{year}{2019}) \bibinfo{pages}{1021}.
\bibitem[{Amari(2016)}]{amari2016information}
\bibinfo{author}{S.-i. Amari}, \bibinfo{title}{Information geometry and its
  applications}, volume \bibinfo{volume}{194}, \bibinfo{publisher}{Springer},
  \bibinfo{year}{2016}.
\bibitem[{Amari(1998)}]{amari1998natural}
\bibinfo{author}{S.-I. Amari},
\newblock \bibinfo{title}{Natural gradient works efficiently in learning},
\newblock \bibinfo{journal}{Neural computation} \bibinfo{volume}{10}
  (\bibinfo{year}{1998}) \bibinfo{pages}{251--276}.
\bibitem[{Verdoolaege and Scheunders(2012)}]{verdoolaege2012geometry}
\bibinfo{author}{G.~Verdoolaege}, \bibinfo{author}{P.~Scheunders},
\newblock \bibinfo{title}{On the geometry of multivariate generalized gaussian
  models},
\newblock \bibinfo{journal}{Journal of mathematical imaging and vision}
  \bibinfo{volume}{43} (\bibinfo{year}{2012}) \bibinfo{pages}{180--193}.
\bibitem[{Fuhrer et~al.(1995)Fuhrer, Moore, and Schuh}]{fuhrer1995estimating}
\bibinfo{author}{J.~C. Fuhrer}, \bibinfo{author}{G.~R. Moore},
  \bibinfo{author}{S.~D. Schuh},
\newblock \bibinfo{title}{Estimating the linear-quadratic inventory model
  maximum likelihood versus generalized method of moments},
\newblock \bibinfo{journal}{Journal of Monetary Economics} \bibinfo{volume}{35}
  (\bibinfo{year}{1995}) \bibinfo{pages}{115--157}.
\bibitem[{Absil et~al.(2009)Absil, Mahony, and
  Sepulchre}]{absil2009optimization}
\bibinfo{author}{P.-A. Absil}, \bibinfo{author}{R.~Mahony},
  \bibinfo{author}{R.~Sepulchre}, \bibinfo{title}{Optimization algorithms on
  matrix manifolds}, \bibinfo{publisher}{Princeton University Press},
  \bibinfo{year}{2009}.
\bibitem[{Berkane et~al.(1997)Berkane, Oden, and Bentler}]{berkane1997geodesic}
\bibinfo{author}{M.~Berkane}, \bibinfo{author}{K.~Oden}, \bibinfo{author}{P.~M.
  Bentler},
\newblock \bibinfo{title}{Geodesic estimation in elliptical distributions},
\newblock \bibinfo{journal}{Journal of Multivariate Analysis}
  \bibinfo{volume}{63} (\bibinfo{year}{1997}) \bibinfo{pages}{35--46}.
\bibitem[{Pennec et~al.(2006)Pennec, Fillard, and
  Ayache}]{pennec2006riemannian}
\bibinfo{author}{X.~Pennec}, \bibinfo{author}{P.~Fillard},
  \bibinfo{author}{N.~Ayache},
\newblock \bibinfo{title}{A riemannian framework for tensor computing},
\newblock \bibinfo{journal}{International Journal of computer vision}
  \bibinfo{volume}{66} (\bibinfo{year}{2006}) \bibinfo{pages}{41--66}.
\bibitem[{Mostajeran and Sepulchre(2018)}]{cyrus}
\bibinfo{author}{C.~Mostajeran}, \bibinfo{author}{R.~Sepulchre},
\newblock \bibinfo{title}{Ordering positive definite matrices},
\newblock \bibinfo{journal}{Information Geometry} \bibinfo{volume}{1}
  (\bibinfo{year}{2018}) \bibinfo{pages}{287--313}.
\bibitem[{Hristova et~al.(2017)Hristova, Le~Meur, Cozot, and
  Bouatouch}]{hristova2017transformation}
\bibinfo{author}{H.~Hristova}, \bibinfo{author}{O.~Le~Meur},
  \bibinfo{author}{R.~Cozot}, \bibinfo{author}{K.~Bouatouch},
\newblock \bibinfo{title}{Transformation of the multivariate generalized
  gaussian distribution for image editing},
\newblock \bibinfo{journal}{IEEE transactions on visualization and computer
  graphics} \bibinfo{volume}{24} (\bibinfo{year}{2017})
  \bibinfo{pages}{2813--2826}.
\bibitem[{Kwitt et~al.(2011)Kwitt, Meerwald, Uhl, and
  Verdoolaege}]{kwitt2011testing}
\bibinfo{author}{R.~Kwitt}, \bibinfo{author}{P.~Meerwald},
  \bibinfo{author}{A.~Uhl}, \bibinfo{author}{G.~Verdoolaege},
\newblock \bibinfo{title}{Testing a multivariate model for wavelet
  coefficients},
\newblock in: \bibinfo{booktitle}{2011 18th IEEE International Conference on
  Image Processing}, \bibinfo{organization}{IEEE}, \bibinfo{year}{2011}, pp.
  \bibinfo{pages}{1277--1280}.
\bibitem[{Vis(95)}]{VisTex}
\bibinfo{title}{Mit vision and modeling group, vision texture}
  (\bibinfo{year}{95}). \URLprefix
  \url{https://vismod.media.mit.edu/pub/VisTex/}.
\bibitem[{Shiryayev(1988)}]{shiryayev1984probability}
\bibinfo{author}{A.~Shiryayev},
\newblock \bibinfo{title}{Probability. 1988},
\newblock \bibinfo{journal}{Springer-Verlag}  (\bibinfo{year}{1988}).

\end{thebibliography}

\end{document}